\documentclass[aps,prb,showpacs,twocolumn,superscriptaddress]{revtex4-1}

\usepackage{amsmath}
\usepackage{amsfonts}
\usepackage{amssymb}
\usepackage{dsfont}
\usepackage{epsfig}
\usepackage{epstopdf}
\usepackage{physics}
\usepackage{microtype}

\usepackage{graphicx}

\usepackage{hyphenat}
\usepackage{hyperref}
\usepackage{placeins}	
\usepackage{tikz}
\usepackage{xcolor}
\usepackage{color}
\usepackage{xspace}
\usepackage{accents}
\usepackage{braket}
\usepackage{subcaption}
\usepackage{graphicx}  
\usepackage{todonotes}
\usepackage{comment}
\usepackage{ragged2e}
\usepackage{soul}
\usepackage{float}
\DeclareCaptionJustification{justified}{\justifying}
\usepackage{array}
\newcolumntype{C}[1]{>{\centering\arraybackslash}m{#1}}
\usepackage{overpic}

\renewcommand{\vec}[1]{\boldsymbol{\mathbf{#1}}}
\renewcommand{\vec}[1]{\boldsymbol{\mathbf{#1}}}

\usepackage[normalem]{ulem}

\usepackage{booktabs}
\usepackage{diagbox}
\usepackage{siunitx}
\AtBeginDocument{
	\heavyrulewidth=.08em
	\lightrulewidth=.05em
	\cmidrulewidth=.03em
	\belowrulesep=.65ex
	\belowbottomsep=0pt
	\aboverulesep=.4ex
	\abovetopsep=0pt
	\cmidrulesep=\doublerulesep
	\cmidrulekern=.5em
	\defaultaddspace=.5em
}
\bibpunct{[}{]}{;}{n}{}{,~}

\begin{document}

\title{Dynamic structure factor of Heisenberg bilayer dimer phases in the presence of quenched disorder and frustration}
\author{Max H\"ormann}
\affiliation{Lehrstuhl f\"ur Theoretische Physik I, Staudtstra{\ss}e 7, FAU Erlangen-N\"urnberg, D-91058 Erlangen, Germany}
\author{Kai Phillip Schmidt}
\affiliation{Lehrstuhl f\"ur Theoretische Physik I, Staudtstra{\ss}e 7, FAU Erlangen-N\"urnberg, D-91058 Erlangen, Germany}

\begin{abstract}
  We investigate the influence of quenched disorder on the dynamic structure factor of Heisenberg bilayers on the square, triangular, and kagome lattice in the quantum paramagnetic phase. Perturbative continuous unitary transformations and white graphs are employed to calculate the one-triplon contribution up to high orders in perturbation about the dimer limit for bimodal and continuous disorder. For the square lattice we find that the lifetime of the gap mode is increased by stronger quantum correlations while stronger disorder effects are observed for the triangular lattice due to geometric frustration. For intra-dimer disorder, in-band energy gaps are observed for both lattices which can be understood in terms of a level repulsion on dimers with low and high intra-dimer exchange that are close in energy at the momentum where the in-band gap opens. For the highly frustrated kagome lattice disorder even allows to decrease the gap energy. In addition, the localization length of the low-energy flat band is increased up to order $7$ in perturbation theory. The interplay of quenched disorder, geometric frustration, and strong correlations leads therefore to rich structures in the dynamical structure factor of two-dimensional quantum magnets.
\end{abstract}

\maketitle

\section{Introduction} 
Disorder is an inevitable ingredient of any condensed matter, but most of the times perfect translational symmetry is still a very good approximation. However, there are situations where this approximation breaks down and effects of disorder play a dominant role so that the physical behavior of clean systems can be fundamentally altered. Effects can range from the famous disorder-driven metal-insulator transition of excitations in electronic systems \cite{anderson1958absence} to the appearance of new phases of matter like cluster spin glasses \cite{andrade2018cluster} or random singlet phases \cite{melin2002strongly} in strongly correlated systems. Furthermore, if the Harris criterion \cite{harris74} is violated, the critical properties of physical systems can be governed by the behavior in rare regions of the system \cite{vojta2006rare}.

The effect of quenched disorder on low-lying excitations is of particular interest. On the one hand, experimentally, these can be probed by various measurement techniques as for example ineleastic neutron scattering \cite{povarov2015dynamics,smirnov2017order,mannig2018spin}. On the other hand, theoretically, to predict the effect of disorder on the shape of  eigenfunctions is challenging. 
In this respect the interplay of strong correlations and disorder as well as the effect on quasi-particles is of particular interest. Clearly, to gain such an understanding is crucial for interpreting experiments in disordered quantum materials with strong correlations.\par
Quantum magnetism is a great playground to examine strongly-correlated quantum matter. For quantum magnets with quenched disorder, ground-state properties were examined by various methods \cite{igloi2018strong,lin2001strongly,singh2017critical,thompson2019griffiths} whereas the investigation of the low-lying excitations and their eigenstates is still in its infancy. For coupled-dimer magnets like the square lattice Heisenberg bilayer, the well-known bond-operator method has been generalized to describe the dynamic structure factor (DSF) of both singlet and magnetically ordered phases on the mean-field level \cite{vojta2013excitation}. For the antiferromagnetic Heisenberg spin-1/2 ladder, a quasi one-dimensional building block of a Heisenberg bilayer, high-order series expansions using the method of perturbative continuous unitary transformations (pCUTs) and white graphs allow to evaluate the DSF beyond the mean-field level in a wide window of coupling ratios \cite{hormann2018dynamic}. In particular, the effect of quenched disorder on elementary triplon excitations \cite{schmidt03} but also two-triplon continua and bound states has been determined.\par
Here, we advance from these calculations for the quasi one-dimensional Heisenberg ladder \cite{hormann2018dynamic} and calculate the one-triplon properties of the DSF for two-dimensional Heisenberg bilayers within the gapped quantum paramagnetic dimer phase. A primary focus is laid on the interplay of quenched disorder and geometric frustration, which we study by comparing the unfrustrated square lattice Heisenberg bilayer with the frustrated cases on the triangular and kagome lattice. Furthermore, the kagome bilayer has a more complex triplon band structure originating from the three-dimer unit cell and we investigate the influence of quenched disorder on such triplon bands.\par
The article is organized as follows. In Sect.~\ref{sec:model} we introduce the model and the observables we focus on. In Sect.~\ref{sec:methods} we explain the methods that we use to do the calculations.
Finally, in Sect.~\ref{Sect::DSF} the results for the one-triplon contribution to the disorder-averaged DSF are given and we conclude in Sect.~\ref{sec:conclusion}.

\section{Model and observables} 
\label{sec:model}
The Hamiltonian of the disordered Heisenberg bilayer for a fixed disorder configuration $\{J\}$ is given by
%
\begin{equation}
\mathcal{H}\left(\{J\}\right) = \sum_\nu J_{\nu}^{\perp}\;\vec{S}_{\nu,1}\cdot \vec{S}_{\nu,2}+\sum_{<\nu,\nu'>}\sum_{n=1}^2 J_{\nu,n}^{\parallel}\;\vec{S}_{\nu,n}\cdot \vec{S}_{\nu',n}\, , 
\label{Eq::Ham_QSB}
\end{equation}
%
where $\vec{S}_{\nu,n}$ with $n\in\{1,2\}$ represent spin-1/2 operators on the upper ($n=1$) and lower ($n=2$) site of dimer $\nu$. The first sum therefore runs over all intra-dimer couplings $J_{\nu}^{\perp}$ while the second term includes all inter-dimer interactions $J_{\nu,n}^{\parallel}$. The disorder configuration $\{J\}$ is given by the set of antiferromagnetic couplings $J_{\nu}^{\perp}$ and $J_{\nu,n}^{\parallel}$ and depends on the type of quenched disorder. Here our main focus is on bimodal disorder, i.e.~the intra- and inter-dimer exchanges can take either the value $J_{1}^{\kappa}$ with probability $p$ or $J_{2}^{\kappa}$ with probability $1-p$ for $\kappa\in\{\perp,\parallel\}$. However, our technical treatment is more general and allows to consider any stationary quenched dimer disorder distribution. Specifically, we will compare bimodal disorder with Gaussian, box and bimodal Gaussian disorder. Gaussian disorder is given by a Gaussian distribution with mean $\mu$ given as the mean value of intra- and inter-dimer couplings and standard deviation $\sigma$ given by the standard deviation of the intra- and inter-dimer couplings. For box disorder the couplings are drawn uniformly from a box with lower intra-dimer (inter-dimer) bound $J^\perp_1$ ($J^\parallel_{1}$) and upper intra-dimer (inter-dimer) bound $J^\perp_2$ ($J^\parallel_{2}$). The bimodal Gaussian distribution is made up of two Gaussian distributions. With a probability $p$ the inter- or intra-dimer coupling is either chosen of one or the other of these two distributions.\par 
Clean Heisenberg bilayers with $J_{\nu}^{\perp}\equiv J^{\perp}$ and $J_{\nu,n}^{\parallel}\equiv J^{\parallel}$ have a quantum paramagnetic singlet ground state and gapped triplon excitations for values of $J^{\parallel}/J^{\perp}$ small enough. This disordered phase is adiabatically connected to the isolated dimer limit $J^{\parallel}=0$. In this limit the ground state is a product state of singlets \mbox{$|s\rangle$=$(|\uparrow\downarrow\rangle - |\downarrow\uparrow\rangle )/\sqrt{2}$} and excitations are local triplets $|t^{+1}\rangle$=$|\uparrow\uparrow\rangle$, \mbox{$|t^{+0}\rangle$=$(|\uparrow\downarrow\rangle + |\downarrow\uparrow\rangle )/\sqrt{2}$}, and \mbox{$|t^{-1}\rangle$=$|\downarrow\downarrow\rangle$}. 

We then can reformulate the disordered Heisenberg bilayer \eqref{Eq::Ham_QSB} exactly in terms of triplet creation and annihilation operators $t^{(\dagger )}_{\nu,\alpha}$ with \mbox{$t^{\dagger}_{\nu,\alpha }|s\rangle \equiv |t^{\alpha}\rangle $} and \mbox{$\alpha\in\{\pm 1,0\}$} on rung $\nu$. Setting the average rung exchange \mbox{$\bar{J}^\perp\equiv (J_{1}^{\perp} +J_{2}^{\perp})/2\equiv 1$} and introducing the deviations $\Delta\bar{J}^\perp_{\pm}$ from it, allows to express \eqref{Eq::Ham_QSB} as
%
\begin{equation}
\mathcal{H}\left(\{J\}\right) = E_0+\mathcal{Q}+\sum_{n=-2}^2 \hat{T}_n\left(\{J\}\right)\quad , 
\label{Eq::Ham_QSL_Trp}
\end{equation}
%
where $E_0=-3/4\,(\mathcal{N}_{\rm d}+\sum_{\nu}\Delta J^{\perp}_{\nu}/\bar{J}^\perp)$ with $\mathcal{N}_{\rm d}$ the number of dimers, the counting operator \mbox{$\mathcal{Q}=\sum_{\nu,\alpha}\hat{n}_{\nu,\alpha}$} with $\hat{n}_{\nu,\alpha}=t^\dagger_{\nu,\alpha}t^{\phantom{\dagger}}_{\nu,\alpha}$, and the \mbox{$\hat{T}_n$} with $[\hat{T}_n,\mathcal{Q}]=n\hat{T}_n$ change the triplet number by $n$. The $\hat{T}_n$ depend explicitly on $\Delta\bar{J}^\perp_{\pm}$ as well as $J^\parallel_{1,2}$ and therefore on the disorder configuration $\{J\}$. Here $\hat{T}_{\pm 2}$ correspond to pair creation and annihilation processes, $\hat{T}_{0}$ contains triplet hopping as well as quartic triplet-triplet interactions, and $\hat{T}_{\pm 1}$ represent decay processes of one triplet into two or vice versa. Note that $\hat{T}_{\pm 1}=0$ holds for the clean case where the Heisenberg bilayers possess an exact reflection symmetry about the centerline giving rise to a conserved parity quantum number $\pm 1$.

The central quantity for inelastic neutron scattering on disordered Heisenberg bilayers is the disorder averaged DSF
%
\begin{equation}
S_{\pm}(\vec{k},\omega) \equiv \lim_{\mathcal{N}_{\rm dc}\rightarrow\infty}\frac{1}{\mathcal{N}_{\rm dc}} S_\pm (\vec{k},\omega,\{J\}) 
\label{Eq::DSF_Averaged}
\end{equation}
%
with momentum $\vec{k}$, frequency $\omega$, number of disorder configurations $\mathcal{N}_{\rm dc}$, and 
%
\begin{equation}
S_{\pm}(\vec{k},\omega,\{J\}) \equiv -\frac{1}{\pi}{\rm Im}\langle 0| \mathcal{O}^{\dagger}_\pm \frac{1}{\omega-\mathcal{H}+{\rm i}0^+} \mathcal{O}^{\phantom{\dagger}}_\pm|0\rangle,\label{Eq::DSF}
\end{equation}
%
where $\mathcal{O}_\pm(\vec{k})\equiv\sum_{\vec{\nu}} e^{{\rm i}\vec{k}\vec{\nu}} \mathcal{O}_\pm(\vec{\nu})/\sqrt{\mathcal{N}_{\rm d}}$ with \mbox{$\mathcal{O}_\pm(\vec{\nu})\equiv (S_{\vec{\nu},1}^z\pm S_{\vec{\nu},2}^z)/2$}. The index $\pm $ reduces to the parity quantum number for the clean Heisenberg bilayers.  

\begin{figure}
	\centering
	\includegraphics[width=0.85\linewidth]{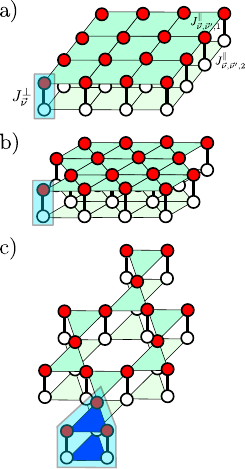}
	\caption[]{Illustration of the Heisenberg bilayer on the (a) square, (b) triangular, and (c) kagome lattice. Blue-shaded boxes illustrate the unit cell in each case.
	}
	\label{fig:Square_Bilayers}
\end{figure}

Although all Heisenberg bilayers realize a singlet dimer phase for sufficiently small $J^{\parallel}/J^{\perp}$ and amount of disorder, the physical properties of Heisenberg bilayers depend strongly on the specific lattice under consideration. This concerns the ground-state properties but also the properties of excited states. Here we investigate the disordered Heisenberg bilayer \eqref{Eq::Ham_QSB} on the square, triangular, and kagome lattice as illustrated in Fig. \ref{fig:Square_Bilayers}.

\subsection{Ground-state properties} 
\label{ssec:pd}

In the absence of disorder, the Heisenberg bilayer on the unfrustrated hypercubic lattice including the square lattice displays for $d>1$ a second-order phase transition between the gapped dimer singlet phase for $J^{\parallel}/J^{\perp}$ small enough and a gapless N\'eel phase \cite{weihong1997various,coester2016linked} if the two hypercubic lattices are coupled sufficiently weak. The gapless modes are Goldstone bosons originating from the spontaneous breaking of the SU(2) symmetry. 
    For the square lattice Heisenberg bilayer the quantum critical point is located at $J^{\parallel}/J^{\perp}\approx 0.3969$ \cite{wenzel08}. In the presence of dimer dilution, several quantum Monte Carlo simulations come to the conclusion that this model gains new critical exponents which fulfill the Harris criterion \cite{sknepnek2004exotic,sandvik2006quantum,vajk2002quantum}. Only recently it was found that for a model with a quantum phase transition between similar phases as those of the square lattice Heisenberg bilayer the gapless Goldstone modes remain delocalized in the presence of disorder \cite{puschmann2019collective}. This rises the question how the gap mode in the disordered square lattice Heisenberg bilayer behaves when approaching the quantum critical point from the disordered phase.\par

In contrast to the square lattice Heisenberg bilayer, the triangular and kagome Heisenberg bilayer are geometrically frustrated. That disorder and the lifting of the ground-state degeneracy by it can have an influence on order by disorder transitions was shown in the frustrated pyrochlore lattice in Ref.~\onlinecite{andrade2018cluster}. In case of the clean triangular Heisenberg bilayer high-order series expansions indicate that the universality class of the phase transition at $J^{\parallel}/J^{\perp}\approx 1.2$  changes from $O(3)$ Heisenberg for the unfrustrated square lattice bilayer to the chiral universality class \cite{singh1998quantum} known from quantum Monte-Carlo simulations of Heisenberg antiferromagnets on stacked-triangular lattices \cite{Kawamura_1992}.
In contrast, for the kagome Heisenberg bilayer, the situation is unclear already in the clean case. Even in the limit of vanishing inter-dimer coupling $J^{\perp}=0$ and although there is consensus concerning the absence of magnetic long-range order, the ground state is either believed to be a gapped topologically-ordered or a gapless algebraic quantum spin liquid \cite{singh2007ground,evenbly2010frustrated,depenbrock2012nature}.
The fate of this quantum spin liquid for finite $J^{\perp}$ and the quantum phase transition to the dimer singlet phase for large $J^{\perp}$ is not yet understood to the best of our knowledge.

\subsection{Triplon excitations of the dimer singlet phase} 
\label{ssec:triplon}
The elementary excitations of dimer singlet phases are triplon excitations \cite{schmidt03}, i.e.~dressed triplet excitations with total spin one and finite energy gap $\Delta$ which reduce to local triplets centered on dimers $\nu$ for vanishing inter-dimer interactions $J^{\parallel}_{\nu,n}=0$. In the absence of disorder one has 
\begin{equation}
 \Delta=\min_{\tilde{\vec{k}},a}\left[\omega_a(\tilde{\vec{k}})\right]\quad ,
\end{equation} 
where $\omega_a(\tilde{\vec{k}})$ denotes the one-triplon band $a$ as a function of wave vector $\tilde{\vec{k}}$. Note that for the kagome lattice $\tilde{\vec{k}}$ differs from the momentum $\vec{k}$ used in the DSF due to the three-dimer unit cell. The Heisenberg bilayer on the square and triangular lattice has a single-dimer unit cell and therefore only a single band with $a=1$. In contrast, the unit cell for the kagome Heisenberg bilayer contains three dimers resulting in three one-triplon bands $\omega_a(\tilde{\vec{k}})$ with $a\in\{1,2,3\}$. For second-order phase transitions between dimer singlet phases and magnetically ordered phases, like for the clean Heisenberg bilayer on the square and triangular lattice, the one-triplon gap $\Delta$ vanishes with critical exponent $z\nu$ of the underlying universality class and at an wave vector $\tilde{\vec{k}}$ associated with the magnetically ordered state, i.e.~$\tilde{\vec{k}}=(\pi,\pi)$ ($\tilde{\vec{k}}=(\pm 2\pi/,\mp 2\pi/3)$) for the square (triangular) Heisenberg bilayer. In contrast, the one-triplon gap $\Delta$ is not expected to close for the highly frustrated Heisenberg bilayer on the kagome lattice due to the absence of magnetic order. In fact, this model hosts an almost flat lowest energy one-triplon band, which is a consequence of the frustration. In general and already for the triangular Heisenberg bilayer, geometric frustration results in negative interferences and a reduced kinetic energy of triplon excitations.\par
It is another aim of this work to examine the interplay between geometric frustration and quenched disorder on the physical properties of these triplon excitations. For the latter a special focus is laid on the effect of disorder on the (almost) flat one-triplon band of the kagome Heisenberg bilayer. To study all these questions, we calculate the one-triplon contribution to the disorder-averaged DSF with the method of pCUTs \cite{knetter2000perturbation,knetter2003structure} employing white graphs \cite{coester2015optimizing,hormann2018dynamic}.

\section{Methods}\label{sec:methods}

Technically, we apply the method of pCUTs with the help of white graphs along the same lines as done for the disordered quasi one-dimensional Heisenberg ladder in Ref.~\onlinecite{hormann2018dynamic}. In order to be self-contained we nevertheless summarize the most important steps in the following.

\subsection{pCUT}\label{ssec:high_field_expansion}

Here we perform pCUTs to obtain effective one-triplon Hamiltonians and the associated one-triplon contribution to the DSF for the Heisenberg bilayer in the dimer singlet phase and in the presence of quenched disorder.\par
The major target of a pCUT is to unitarily transform \eqref{Eq::Ham_QSL_Trp}, order by order in $J_{\nu,n}^{\parallel}$ and $\Delta\bar{J}^\perp_{\pm}$, to an effective Hamiltonian $\mathcal{H}_{\rm eff}$ which conserves the number of triplons so that $[\mathcal{H}_{\rm eff},\mathcal{Q}]=0$ holds. As a consequence, the complicated quantum many-body system is mapped to an effective few-body problem which is easier to solve. A pCUT application has a model-independent step, which expresses $\mathcal{H}_{\rm eff}$ in a sum of operator product sequences of the $\hat{T}_n$ operators with exactly known rational coefficients. The most efficient way of performing the second model-dependent step, which amounts to a normal ordering of $\mathcal{H}_{\rm eff}$, is a full-graph decomposition using the linked-cluster theorem (see also next subsection). To model the disorder every bond has an individual perturbation parameter assigned to. This idea is essential, since the pCUT calculation is only done once and then can be used for any realization of disorder $\{J\}$ \cite{coester2015optimizing,hormann2018dynamic}.\par 
The effective one-triplon Hamiltonian $\mathcal{H}_{\rm eff}^{1{\rm qp}}$ and the one-triplon contribution to the effective observable $\mathcal{O}^{1{\rm qp}}_{\pm,{\rm eff}}$ can be written as
%
\begin{align}
  \mathcal{H}_{\rm eff}^{\rm 1qp} &= \sum_{\vec{\mu},\vec{\delta}}\sum_{\kappa,\kappa^\prime} \sum_\alpha a^{\kappa,\kappa^\prime}_{\vec{\mu},\vec{\delta}} \left( t^{\dagger}_{\vec{\mu}+\vec{\delta},\kappa^\prime,\alpha }t^{\phantom\dagger}_{\vec{\mu},\kappa,\alpha} + {\rm h.c.}\right)\\
  \mathcal{O}_{\pm, \rm eff}^{\rm 1qp}(\vec{\mu},\kappa)  &= \sum_{\vec{\delta}}\sum_{\kappa^\prime } \sum_\alpha w^{\kappa,\kappa^\prime}_{\vec{\mu},\vec{\delta}} \left( t^{\dagger}_{\vec{\mu}+\vec{\delta},\kappa^\prime,\alpha }+ {\rm h.c.}\right)                                                                                               \quad , 
\label{Eq::Ham_eff}
\end{align}
%
where $\vec{\nu}=(\vec{\mu},\kappa )$ so that $\vec{\mu}$ denotes the unit cell and $\kappa$ the supersite (dimer) within $\vec{\mu}$. Further, $a^{\kappa,\kappa^\prime}_{\vec{\mu},\vec{\delta}}$ and $w^{\kappa,\kappa^\prime}_{\vec{\mu},\vec{\delta}}$ represent the one-triplon hopping amplitudes and the amplitudes for the transformed observable in the one-triplon basis, respectively. Using the pCUT method, we calculated these coefficients up to order $8$ in perturbation for the square and kagome lattice, while we reached order $7$ for the triangular lattice.\par
The effective one-triplon problem is then diagonalized for finite Heisenberg bilayers with $\mathcal{N}_{\rm d}\approx 10000$ dimers and the DSF for a fixed disorder configuration is obtained using bins of width $2\bar{J}^\perp/1000$. Averaging over \mbox{$\mathcal{N}_{\rm dc}=100$} disorder configurations results in the averaged DSF \eqref{Eq::DSF_Averaged}. From an experimental viewpoint all relevant aspects of the DSF discussed below are converged because a finite broadening decreases the finite-size effects. A more detailed analysis of finite-size effects are given in appendix B.

\subsection{Graph decomposition}\label{ssec:pade}
One important advantage of a pCUT is that it obeys the linked-cluster theorem \cite{knetter2003structure}. Due to that it is sufficient to do the pCUT calculations on finite graphs and embed the graph contributions afterwards on the lattice. We calculated all embeddings of all relevant graphs of a given order on the lattice with the boost graph library \cite{boost_graph}. Then, knowing the disorder of one realization on the lattice, we determined the one-triplon effective Hamiltonian using these embeddings.\par
Since a single triplon is centered on a dimer, it is natural to consider a dimer as the elementary building block which we call supersite. The Heisenberg bilayer is then reduced to a single layer of supersites and we perform the graph expansion in terms of these supersites.\par
It is instructive to illustrate the procedure with most simple graphs, which are the point graph containing only one supersite (dimer) and the graph with two supersites and one edge. For the first graph there is only one embedding per supersite and hence for the whole lattice there are $\mathcal{N}_{\rm d}$ embeddings. The point graph only contributes for intra-dimer disorder and only in first-order perturbation theory. Depending on the specific disorder configuration $\{J\}$ the local contribution of this graph to the one-triplon hopping amplitude $a^{\kappa,\kappa}_{\vec{\mu},\vec{\delta}=\vec{0}}$ at supersite $(\vec{\mu},\kappa)$ is either $J^\perp_1-\bar{J}^\perp$ or $J^\perp_2-\bar{J}^\perp$ for bimodal disorder. For the next larger graph with two supersites and one edge there are two embeddings per supersite and hence there are $2\mathcal{N}_{\rm d}$ ($3\mathcal{N}_{\rm d}$) embeddings for the square and kagome (triangular) lattice. Locally, the contribution of that graph depends on the inter-dimer exchanges between two neighboring supersites $\vec{\nu}$ and $\vec{\nu}'$ and on the intra-dimer disorder at these supersites. This graph represents therefore the (leading) first-order contribution for inter-dimer disorder which is given by $J^\parallel_1/2$, $(J^\parallel_1+J^\parallel_2)/4$, and $J^\parallel_2/2$ for bimodal disorder.\par
In general, for a given order $k$ in perturbation theory, graphs with at most $k$ edges can contribute due to the linked-cluster theorem. The number of these graphs in a given order depends on the lattice under consideration. The number of local embeddings depends exponentially on the number of edges of the graphs and hence scales exponentially with the order $k$ of the perturbation theory. This makes the computational cost strongly dependent on the lattice geometry and on the order $k$ of the perturbative expansion. On the one hand for the triangular lattice almost all graphs with $8$ edges match on the lattice while on the other hand the number of those graphs on the kagome lattice is significantly smaller. Also the high coordination number of the triangular lattice leads to a much larger number of embeddings compared to the square or kagome lattice. Because of that calculations on the triangular lattice were only performed up to order $7$ and on the other two lattices they were done up to order $8$.

\section{Dynamic structure factor} 
\label{Sect::DSF}
For the standard Anderson model all states are localized and only in three dimensions a localization-delocalization transition can occur \cite{eilmes1998two}. A crucial difference to these first-order models is that in the Heisenberg bilayers the disorder acts at the distance of the correlation length approximately given by the perturbative order. Furthermore, the disorder in the different matrix elements is correlated and this correlation depends on the geometry and the quantum fluctuations. Intra-dimer disorder in first order resembles the Anderson model while inter-dimer disorder in first order leads to random nearest-neighbor one-triplon hopping amplitudes. Such models are also sometimes referred to as Anderson models but show larger localization lengths for similar disorder strengths than the usual Anderson models with on-site disorder \cite{eilmes1998two}. Though, almost all eigenstates are still localized for such models except the ones in the band center \cite{theodorou1976extended,eilmes1998two}. Differences between intra- and inter-dimer disorder in one dimension were already discussed in detail for antiferromagnetic Heisenberg spin ladders in \cite{hormann2018dynamic}.\par 
Another characteristic of the disorder is the probability distribution of the exchange couplings. Most of the times we will investigate bimodal disorder since it is relevant for the modeling of various disordered quantum materials synthesized by intentional doping. However, to find out how sensitive the DSF is to the moments of the probability distribution we will also compare with continuous distributions that either agree up to the first two or four moments with the bimodal distribution.\par
To really compare experimental results with the theoretical calculations of the DSF on Heisenberg bilayers it is essential to incorporate higher order perturbative effects as we do here. It is thus another aim of the calculations to show that it is possible to give quantitative calculations for the DSF of disordered quantum magnets. 
\subsection{Intra-dimer disorder}
For intra-dimer disorder the reflection symmetry at the centerline of the dimers is not broken and hence the observable $S_{+}(k,\omega)$ is zero. In all Heisenberg bilayers the first-order effect of inter-dimer disorder is a random on-site hopping term $a^{\kappa,\kappa}_{\vec{\mu},\vec{\delta}=\vec{0}}$, i.e.~a disordered chemical potential, in a nearest-neighbor one-triplon hopping Hamiltonian. 
For intra-dimer disorder in one dimension we found very characteristic structures in the DSF that had their origin in the fragmentation of the ladder into finite ladder segments \cite{hormann2018dynamic}. This fragmentation could be understood by the fact that there is no percolation in one dimension. For the two-dimensional Heisenberg bilayers with intra-dimer disorder this is different. On all of the three lattices we study in this work there is a threshold $p_c$ for site percolation. To understand the connection between percolation and intra-dimer disorder better it is instructive to go to the limit of bimodal intra-dimer disorder with an extremely small nearest-neighbor one-triplon hopping amplitude, i.e.~sufficiently small $J^\parallel_{\nu,n}$. In this limit the eigenstates are either solely with all their weight on dimers with intra-dimer-exchange $J^\perp_{1}$ or $J^\perp_{2}$. Hence, if $p$ is the probability for the value $J^\perp_{1}$ and $p<p_c$, the eigenstates can not be extended since there is no percolating cluster. Furthermore, there is no smooth density of states (DOS) in these energy regions since the probability for a finite contributing cluster decays exponentially with size. If one increases the one-triplon hopping amplitudes this picture is obviously not correct anymore. However, it still serves as a good approximation if the hopping is small compared to $J^\perp_{1}-J^\perp_{2}$. Because percolation properties are dependent on the specific lattice it will be interesting to see if we can explain differences in the DSF on the different lattices within the just described limit.

\subsubsection{Square lattice}
The disorder setup considered for the square lattice is an inter-dimer coupling of $J^\parallel=0.2$ and two different intra-dimer exchanges $J^\perp_1=0.8$ and $J^\perp_2=1.2$ and \mbox{$\mathcal{N}_{\rm d}=10000$} and  $\mathcal{N}_{\rm dc}=100$. The value $J^\perp_1$ was chosen with probabilities $p=0.1,0.3,0.5,0.7,0.9$. In Fig. \ref{fig:Rung_Square_Minus} the DSF is shown for $p=0.1,0.5,0.9$. For $p=0.3,0.7$ the plots are given in appendix C in Fig.~\ref{fig:Rung_Square_p0307}.\par
Highest intensities are found at the gap momentum $\vec{k}=(\pi,\pi)$ for $p=0.9$. This has two reasons. First, the effective ratio between inter-dimer and intra-dimer couplings $J^\parallel/[p\,J_1^\perp+(1-p)\,J_2^\perp ]$ is close to \mbox{$0.2/0.8=1/4$} and highest in comparison to $p=0.1,0.5$. Since the static structure factor at $\vec{k}=(\pi,\pi)$ increases upon approaching the critical point larger coupling ratios lead to larger intensities. The other reason is that the effect of disorder depends on the ratio of the standard deviation of the intra-dimer disorder and the effective coupling ratio. Consequently also the effect of disorder is weakest for $p=0.9$ leading to longer lifetimes that result in higher intensities.\par

The description in terms of bands with a finite lifetime breaks down at certain momenta that depend on $p$. At these momenta we see the remarkable presence of small in-band energy gaps. We see these in-band gaps close to the momenta $\vec{k}=(0,\pi)$ for $p=0.5$ while for $p=0.3$ the momentum is closer to $\vec{k}=(\pi,\pi)$ and for $p=0.7$ it is further away. For $p=0.9$ it is hard to observe because of the small disorder and for $p=0.1$ it opens in the surrounding of $\vec{k}=(\pi,\pi)$. The mechanism responsible for this effect shall be described briefly in the following:\\

\begin{figure}
	\centering
		\includegraphics[width=1\linewidth]{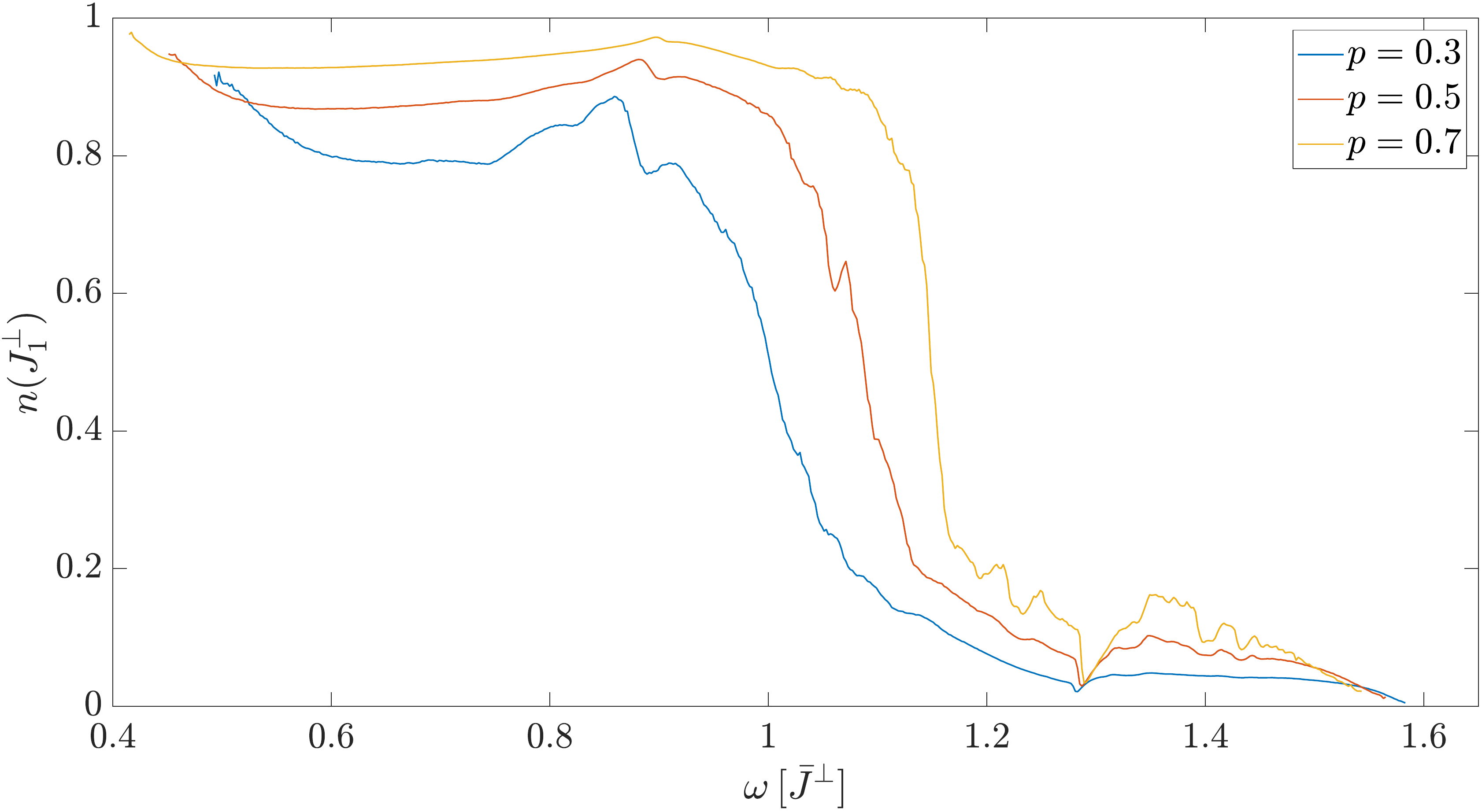}
	\caption[]{For the bimodal intra-dimer disorder in the square lattice and for the probabilities $p=0.3,0.5,0.7$ the overlap $n$ of the eigenfunctions with dimer states of value $J^\perp_{1}=0.8$ is shown as a function of $\omega$. Up to the transition energy region the overlap is larger than $0.8$ for $p=0.3,0.5,0.7$. In contrast, for larger $\omega$ the overlap with $J^\perp_{1}=0.8$ dimer states is smaller than $0.2$. 
	}
	\label{fig:Square_Rung1_Density}
\end{figure}

On the one hand for zero intra-dimer disorder the eigenstates are momentum states. Inducing a small disorder leads to a finite lifetime of the momentum states, but a description in terms of them is still possible. On the other hand, for infinite intra-dimer disorder corresponding to hopping amplitudes of zero the eigenstates are purely local and the system has a density of $p$ with the first intra-dimer exchange $J^\perp_{1}$ as eigenvalue and $1-p$ with the second intra-dimer exchange $J^\perp_{1}$ as eigenvalue. Increasing the coupling ratio in that situation leads in first order to a hopping between the blocks of dimer states with $J^\perp_{1}$ or $J^\perp_{2}$. When one increases the hopping further a small density of  $J^\perp_{1}$-states mixes into the $J^\perp_{2}$-eigenstates and vice versa. The disorder we are dealing with here is in between those two limiting regimes. Analyzing the eigenstates shows that below the in-band gap the density of $J^\perp_{1}$-sites in the eigenstates is almost 80\% and that of $J^\perp_{1}$-sites 20\% and vice versa above the in-band gap as can be seen in Fig.~\ref{fig:Square_Rung1_Density}. Though, apparent when looking at the DSF using momentum with a finite broadening as descriptor seems not too bad in regions not close to the in-band gaps. The regions with the lower intra-dimer exchange occupy predominantly the momenta with lowest energy and vice versa for the regions with the larger intra-dimer exchange. The in-band gap results from a repulsion of states on dimers with low and high intra-dimer exchanges that are close in energy at the momentum where the in-band gap opens.

Next we address the issue of percolation. For zero inter-dimer coupling one has a density of $p$ states with energy $0.8$ and a density of $1-p$ with energy $1.2$. If the inter-dimer coupling is small compared to $J^\perp_2-J^\perp_1$ the states remain separated in regions of the lattice with $J^\perp_\nu=J^\perp_1$ or $J^\perp_\nu=J^\perp_2$ respectively. For such an situation the DOS in the regions with $p<p_{\rm pc}\approx 0.59$ \cite{derrida1985corrections}, where $p_{\rm pc}$ is the bond percolation threshold of the lattice, consists of states that can only interact with a finite part of the lattice and hence the DOS for $p<p_{\rm pc}$ should not show smooth behavior. As one increases the inter-dimer coupling the probability for hopping processes over more sites increases and one expects a possible percolation of the states for $p<p_{\rm pc}$. For the behavior of the DOS (see Fig.~\ref{fig:Rung_DOS} in appendix C) and also of the DSF in the momentum region around the momentum $\vec{k}=(\pi,\pi)$ in the $p=0.1$-case and for the momentum region around $\vec{k}=(0,0)$ in the $p=0.9$-case (hardly visible) one finds roughly a peaked shape. This is in accordance with the percolation picture.

\begin{figure} 
	\centering
	\begin{subfigure}[b]{.9\columnwidth}
		\caption{}
		\includegraphics[width=1\linewidth]{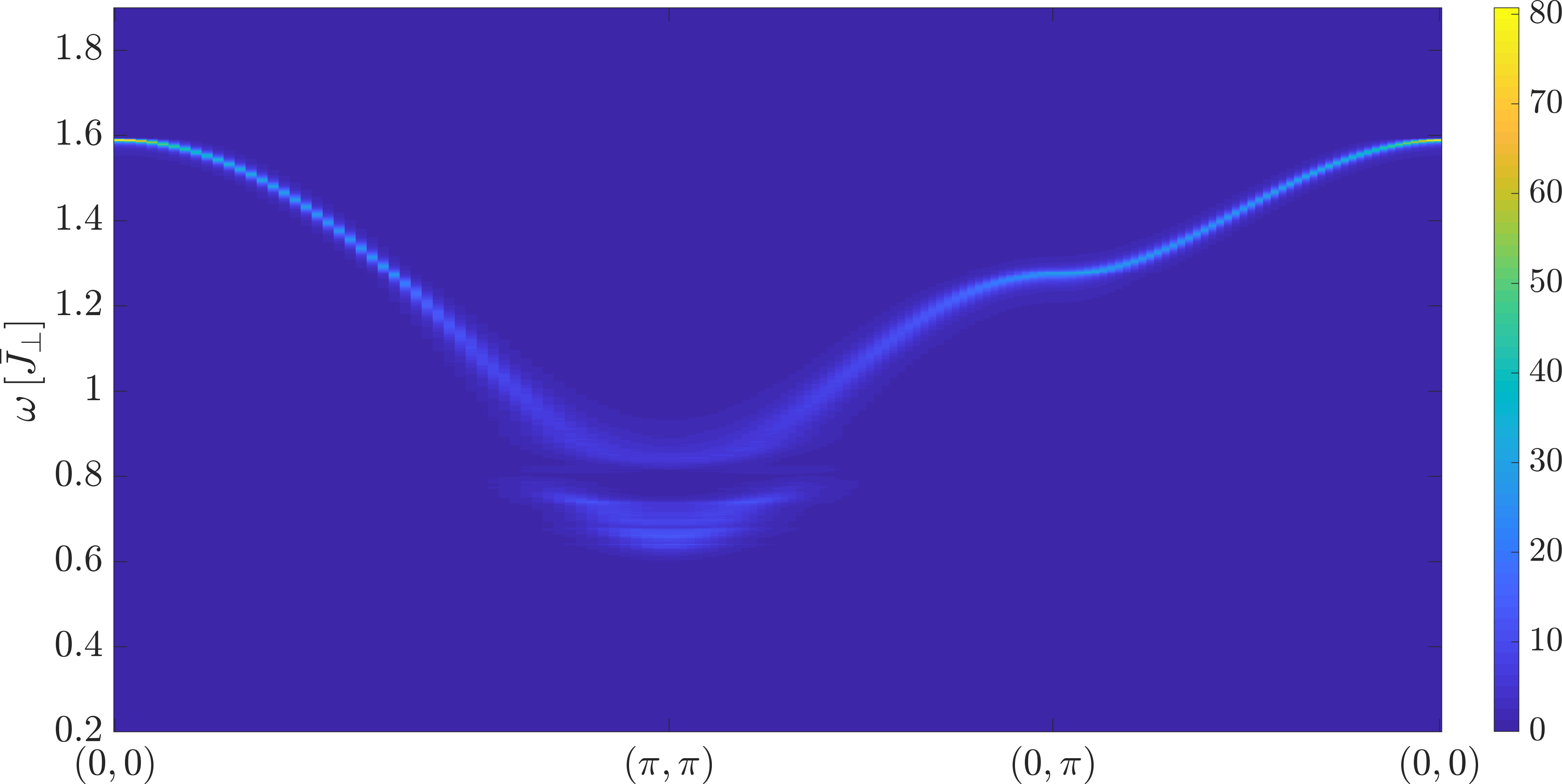}
		\label{fig:Ng1} 
	\end{subfigure}
	
	\begin{subfigure}[b]{.9\columnwidth}
		\caption{}
		\includegraphics[width=1\linewidth]{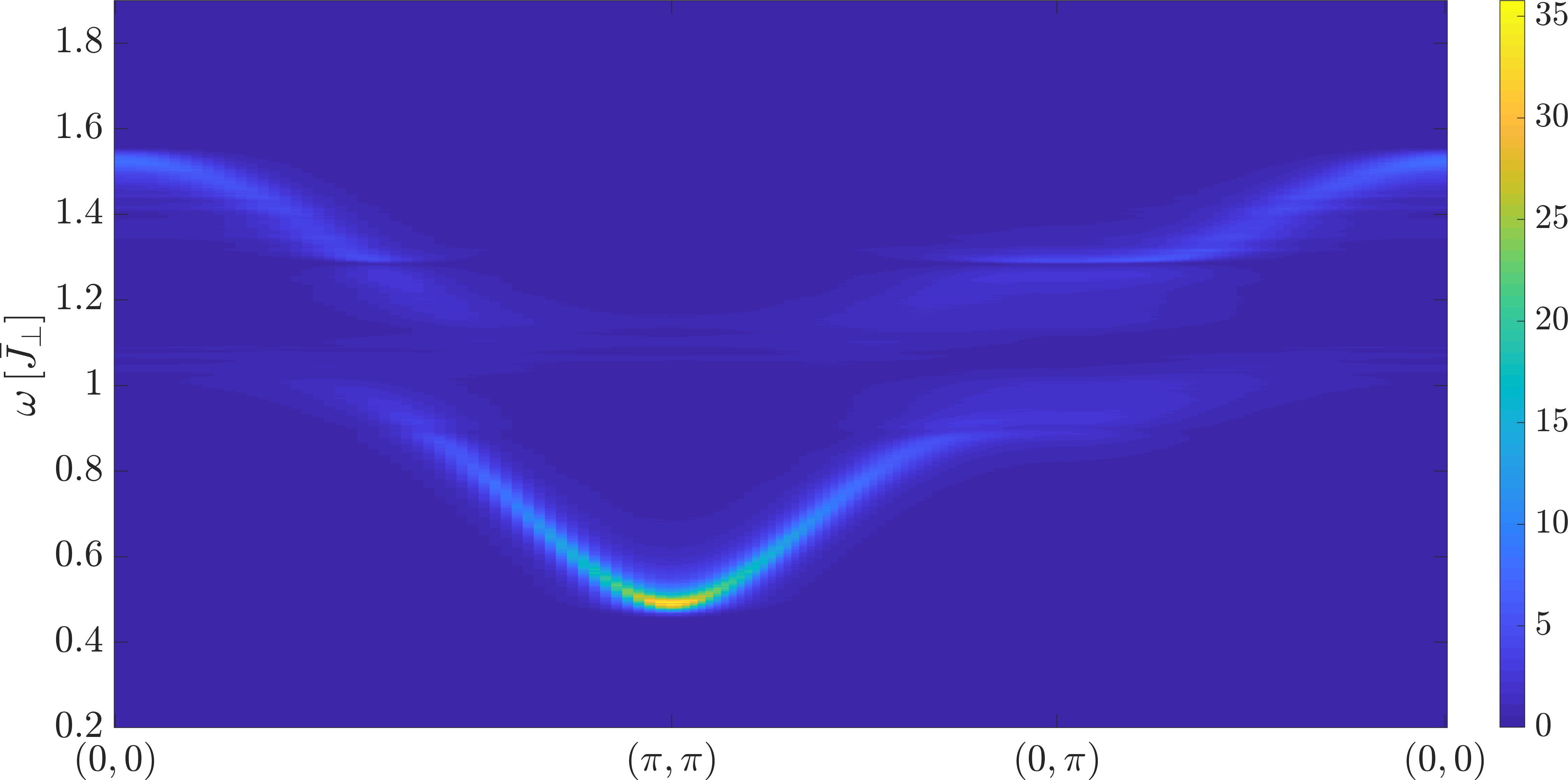}
		\label{fig:Ng2}
	\end{subfigure}
	
	\begin{subfigure}[b]{.9\columnwidth}
		\caption{}
		\includegraphics[width=1\linewidth]{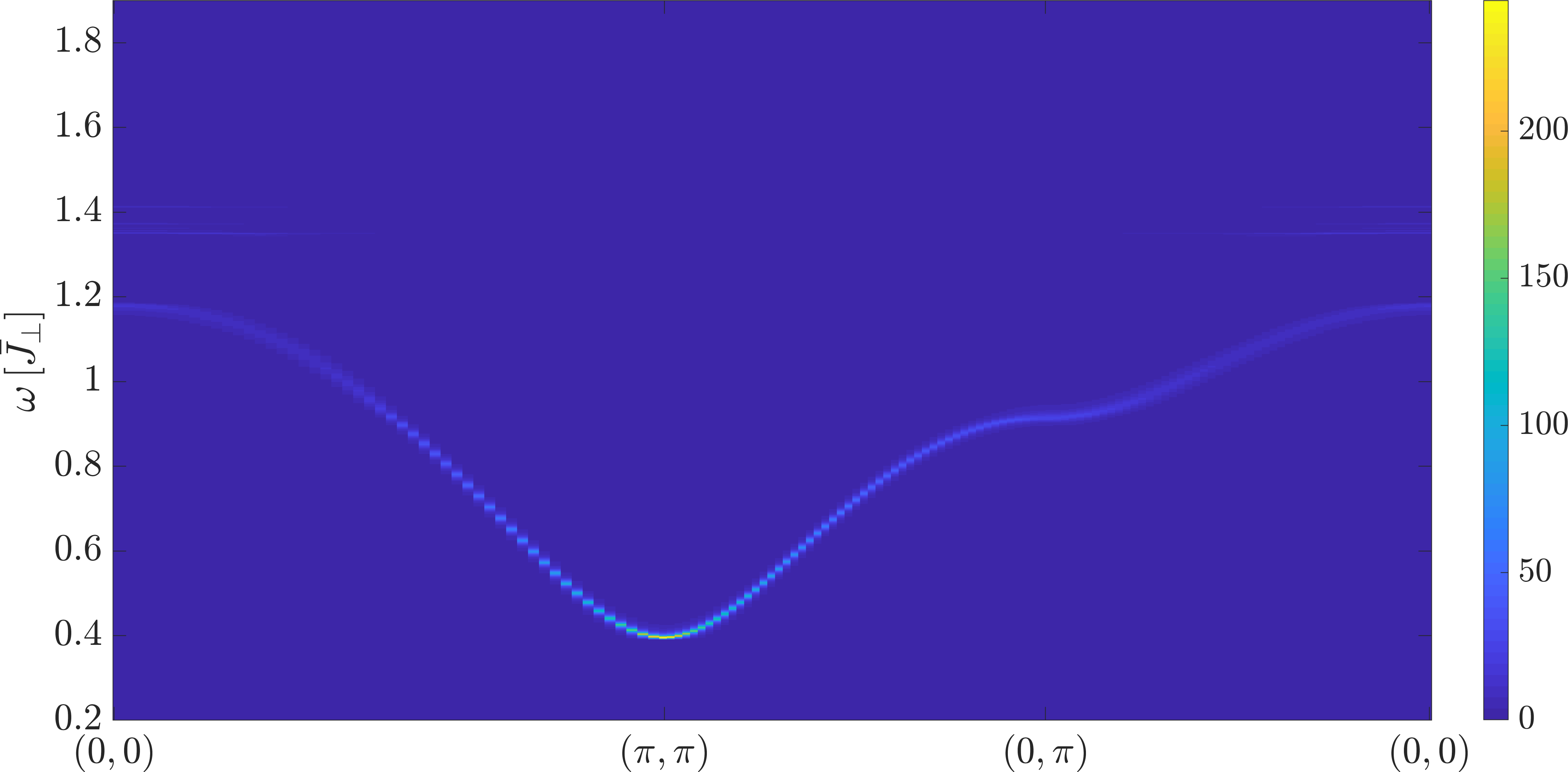}
		\label{fig:Ng2}
	\end{subfigure}
	
	\caption[]{The DSF $\mathcal{S}_-(k,\omega)$ is shown for bimodal intra-dimer disorder for the square lattice with $J^\parallel=0.2$ and $\Delta J^\perp=\pm 0.2$ for 
		$p=0.1$ in (a), $p=0.5$ in (b) and $p=0.9$ in (c).
	}
	\label{fig:Rung_Square_Minus}
\end{figure}

\subsubsection{Triangular lattice}
\begin{figure}
	\centering
	\begin{subfigure}[b]{.9\columnwidth}
		\caption{}		
		\includegraphics[width=1\linewidth]{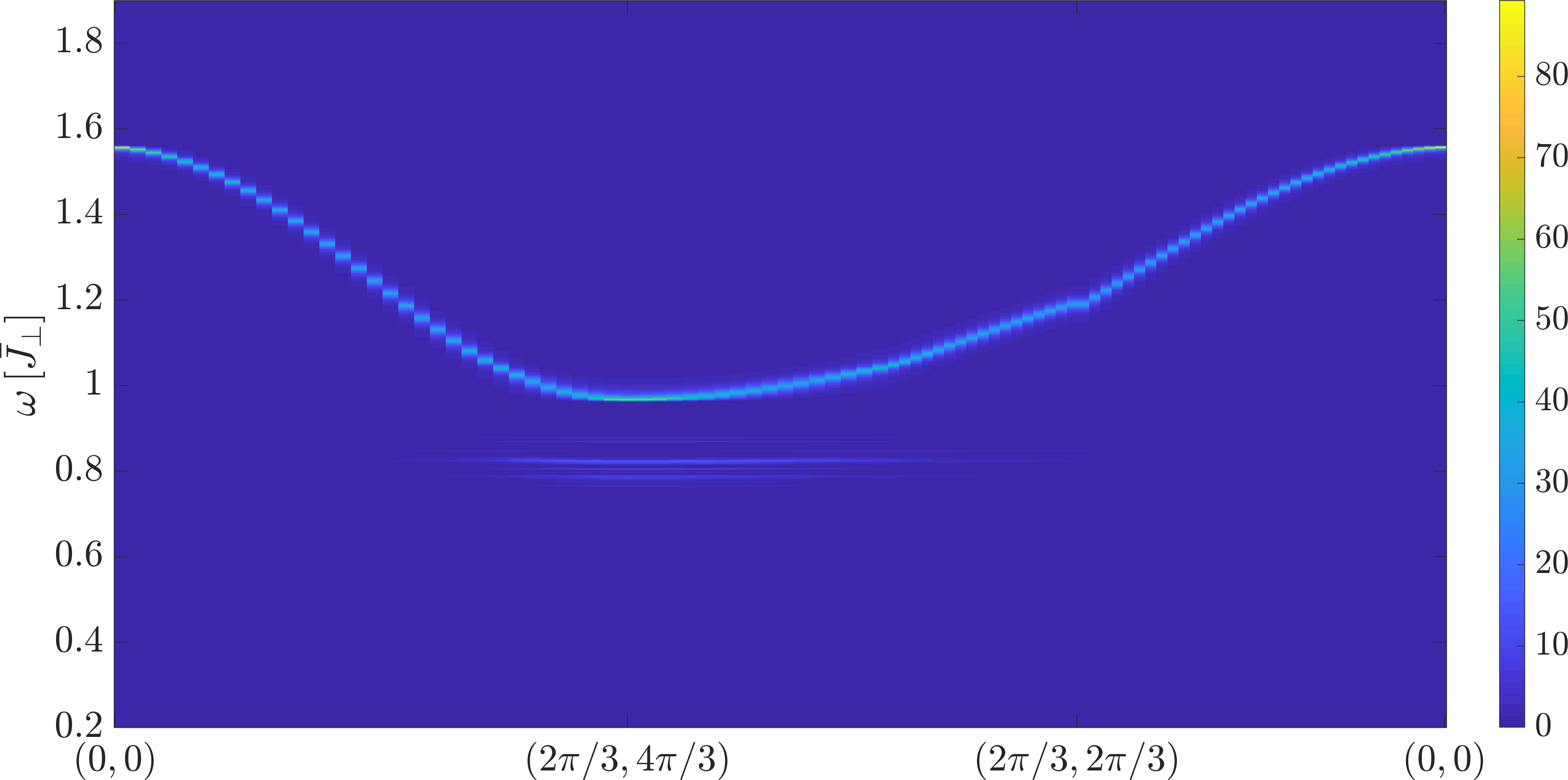}
		\label{fig:Ng1} 
	\end{subfigure}
	
	\begin{subfigure}[b]{.9\columnwidth}
		\caption{}		
		\includegraphics[width=1\linewidth]{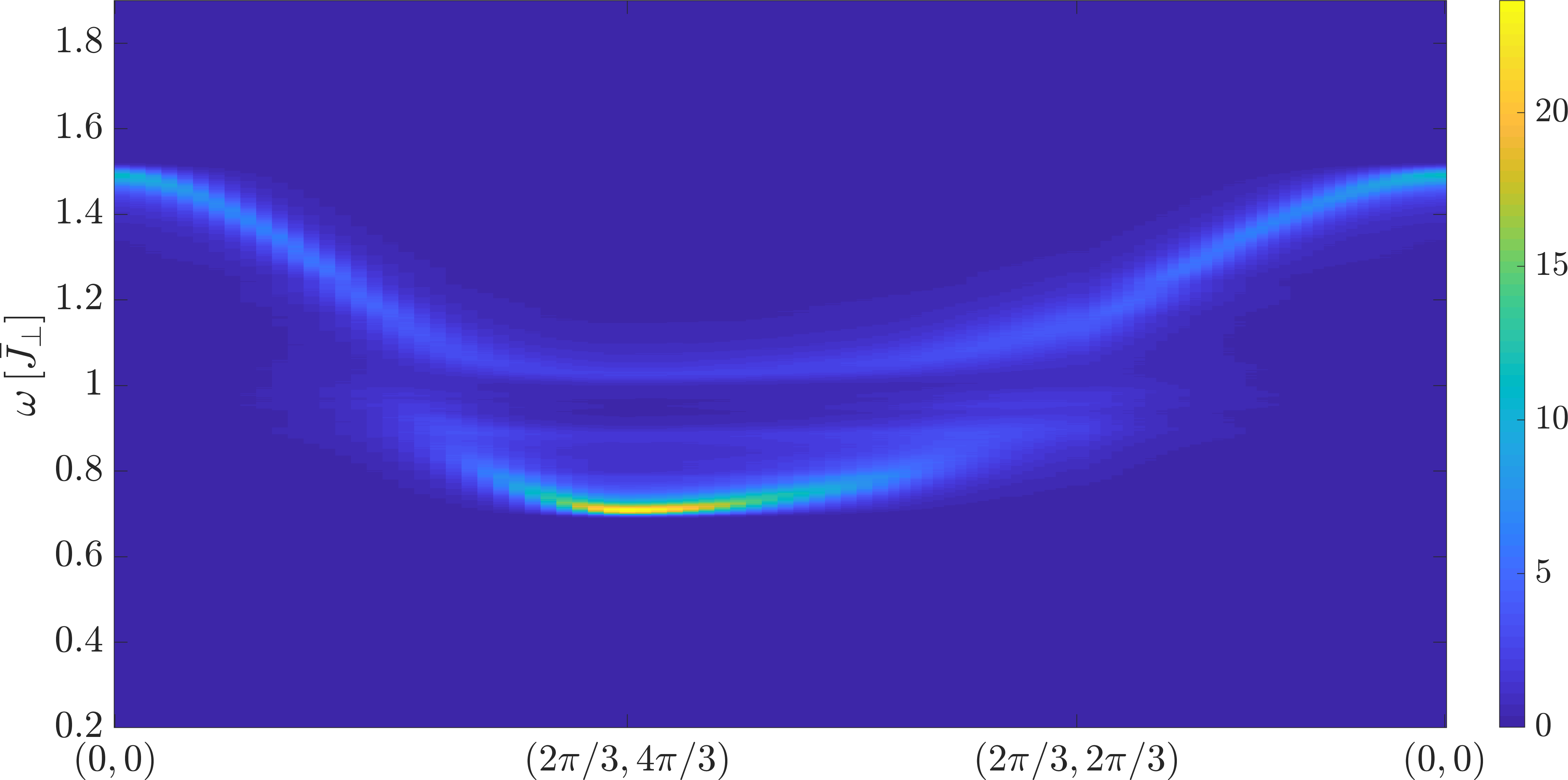}
		\label{fig:Ng2}
	\end{subfigure}
	
	\begin{subfigure}[b]{.9\columnwidth}
		\caption{}
		\includegraphics[width=1\linewidth]{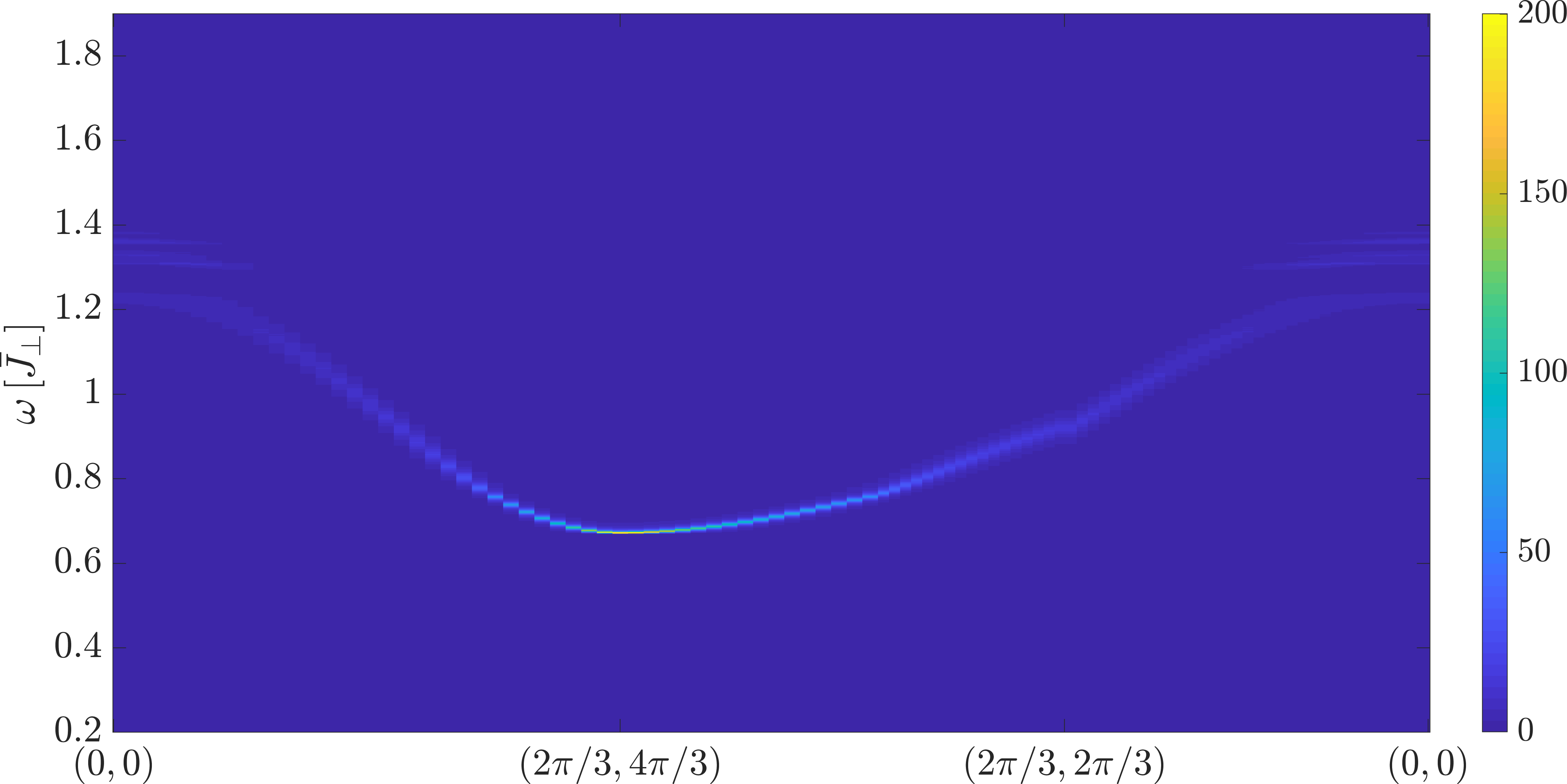}
		\label{fig:Ng2}
	\end{subfigure}
	
	\caption[]{
		The DSF $\mathcal{S}_-(k,\omega)$ is shown for bimodal intra-dimer disorder in the triangular lattice with $J^\parallel=0.15$ and $\Delta J^\perp=\pm 0.15$ for 
		$p=0.1$ in (a), $p=0.5$ in (b) and $p=0.9$ in (c).
	}
	\label{fig:Triangular_Square_Minus}
\end{figure}

In the triangular lattice the disorder setup is an inter-dimer coupling of $J^\parallel=0.15$ and two different intra-dimer exchanges $J^\perp_1=0.85$ and $J^\perp_2=1.15$ and $\mathcal{N}_{\rm d}=9801$ and  $\mathcal{N}_{\rm dc}=100$. The probabilities chosen for  $J^\perp_1=0.85$ are again $p=0.1,0.3,0.5,0.7,0.9$. In Fig.~\ref{fig:Triangular_Square_Minus} the DSF is displayed for $p=0.1,0.5,0.9$ while the cases $p=0.3,0.7$ are shown in appendix C in Fig.~\ref{fig:Rung_Square_p0307}. The calculations were only done up to order 7 because of the high coordination number of the triangular lattice and the resulting large number of embeddings.\par
Similar to the square lattice we see that a small in-band gap opens and the energy of this in-band gap as well as the dominant momenta depend on the probability $p$. The mechanism responsible for that is the same as in the square lattice.\par
On the one hand the site percolation threshold for the triangular lattice of $p_{\rm pc}=0.5$ is smaller compared to the square and kagome lattice. On the other hand frustration leads to a suppression of kinetic terms. Furthermore, the inter-dimer coupling used for the triangular lattice is slightly smaller than the one used for the square lattice due to convergence issues of the perturbative expansion. Comparing the DSF for $p=0.5$ we find stronger localization effects and claim that these are mainly due to the suppression of kinetic terms by frustration. Also for $p=0.1$ the states below the gap are flatter and hence more localized than in the square lattice. Hence even though the states in these regions have more neighbours than in the square lattice the effect of suppression of kinetic terms is stronger and the states are more localized.\par
For the DOS (see Fig.~\ref{fig:Rung_DOS} in appendix C) one finds similar effects as in the square lattice in the non-percolating regions. However, it is not possible to relate these effects in a quantitative way to the percolation threshold of the lattice since the percolation picture is only valid exactly in the limit $J^\parallel/J^\perp \rightarrow 0$.

\subsubsection{Kagome lattice}
We chose an inter-dimer coupling of $J^\parallel=0.2$ and two intra-dimer exchanges $J^\perp_1=0.8$ and $J^\perp_2=1.2$ and $\mathcal{N}_{\rm d}=9747$ and  $\mathcal{N}_{\rm dc}=100$. In Fig.~\ref{fig:kagome_Square_Minus} the DSF is shown for $p=0.1,0.5,0.9$ while the cases $p=0.3,0.7$ are given in appendix C in Fig.~\ref{fig:Rung_kagome_p0307}.\par
Without disorder the lowest band of the kagome lattice is completely flat up to order 7 in perturbation theory similar to the one-particle spin-flip excitation in the transverse-field Ising model \cite{powalski2013disorder}. The states of that band can be chosen as completely localized states on the hexagons of the kagome lattice. In the cases $p=0.1,0.9$ there remains a strong intensity in the flat band coming from hexagons with solely \mbox{$J^\perp=J^\perp_{2,1}$-couplings}. This is not surprising since $0.9^6 \approx 0.5$ and hence there are still approximately $0.5\,\mathcal{N}_{\rm d}/3$ states with the energy of the flat band of the pure cases. For $p=0.1$ below and for $p=0.9$ above this band one sees further contributions of other very localized states. These are linked to other intra-dimer coupling configurations on a hexagon or are very localized states evolving out of the dispersive bands. For the kagome lattice the site percolation threshold is $p_{\rm pc} \approx 0.65$ \cite{sykes1964exact}.Even without disorder the DOS of the flat band shows just one peak up to order 7 in perturbation theory because the band is completely flat. With disorder the DOS in the flat band region is not smooth and shows peaks belonging to the hexagon states of different disorder configurations (see Fig.~\ref{fig:Rung_DOS} in appendix C). For $p=0.1$ we see a more peaked structure for energies smaller than $1$ while for $p=0.9$ we see it for energies larger than $1$. These energies come from the dispersive bands and hence show a similar behaviour in context of percolation as seen for the other two lattices.\par
    The effects of disorder on the dispersive bands are similar to the ones on the other lattices for $p=0.1,0.9$ and lead to a small broadening and the lower energetic dispersive band splits depending on the momentum. However, for $p=0.5$ things change drastically and the DSF shows a wild appearance. One can not distinguish clearly anymore between the three bands since parts evolving out of the three bands of the clean cases contribute to the DSF at the same momenta and energies. This is reflected in the small value of the inverse participation ratio (IPR) in Fig.~\ref{fig:Rung_IPR} in appendix C. We claim that stronger interactions between the flat and dispersive bands for $p=0.5$ are responsible for this behavior.\par
The overall gap gets decreased compared to the pure cases $p=0$ and $p=1$. The reason for that is the larger localization length of the flat band states in the presence of disorder up to order $7$ in perturbation. This is different in higher orders where the local hexagon modes are no eigenstates anymore.

\begin{figure}[H]
	\centering
	\begin{subfigure}[b]{.9\columnwidth}
		\caption{}		
		\includegraphics[width=1\linewidth]{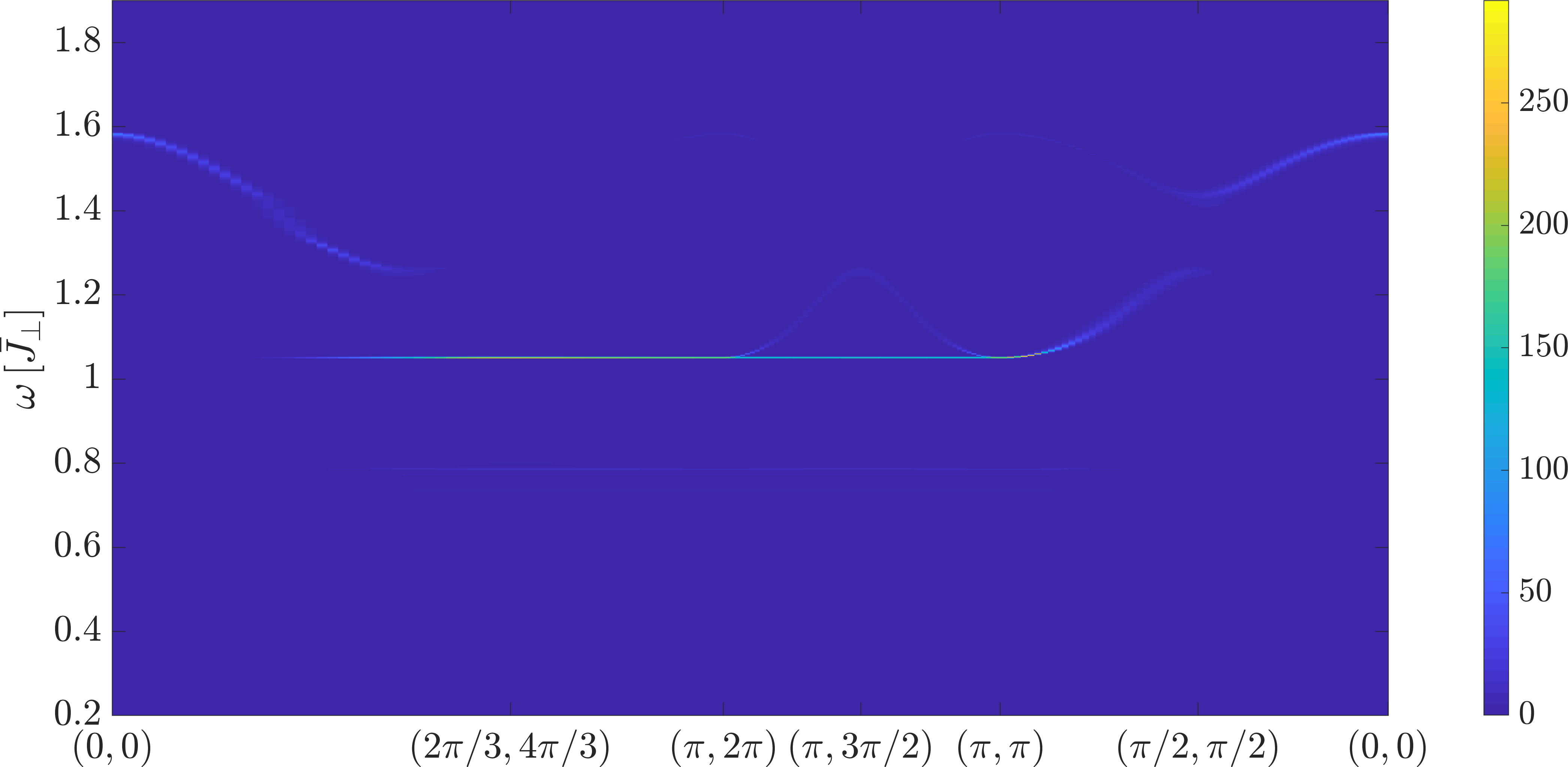}
		\label{fig:Ng1} 
	\end{subfigure}
	
	\begin{subfigure}[b]{.9\columnwidth}
		\caption{}		
		\includegraphics[width=1\linewidth]{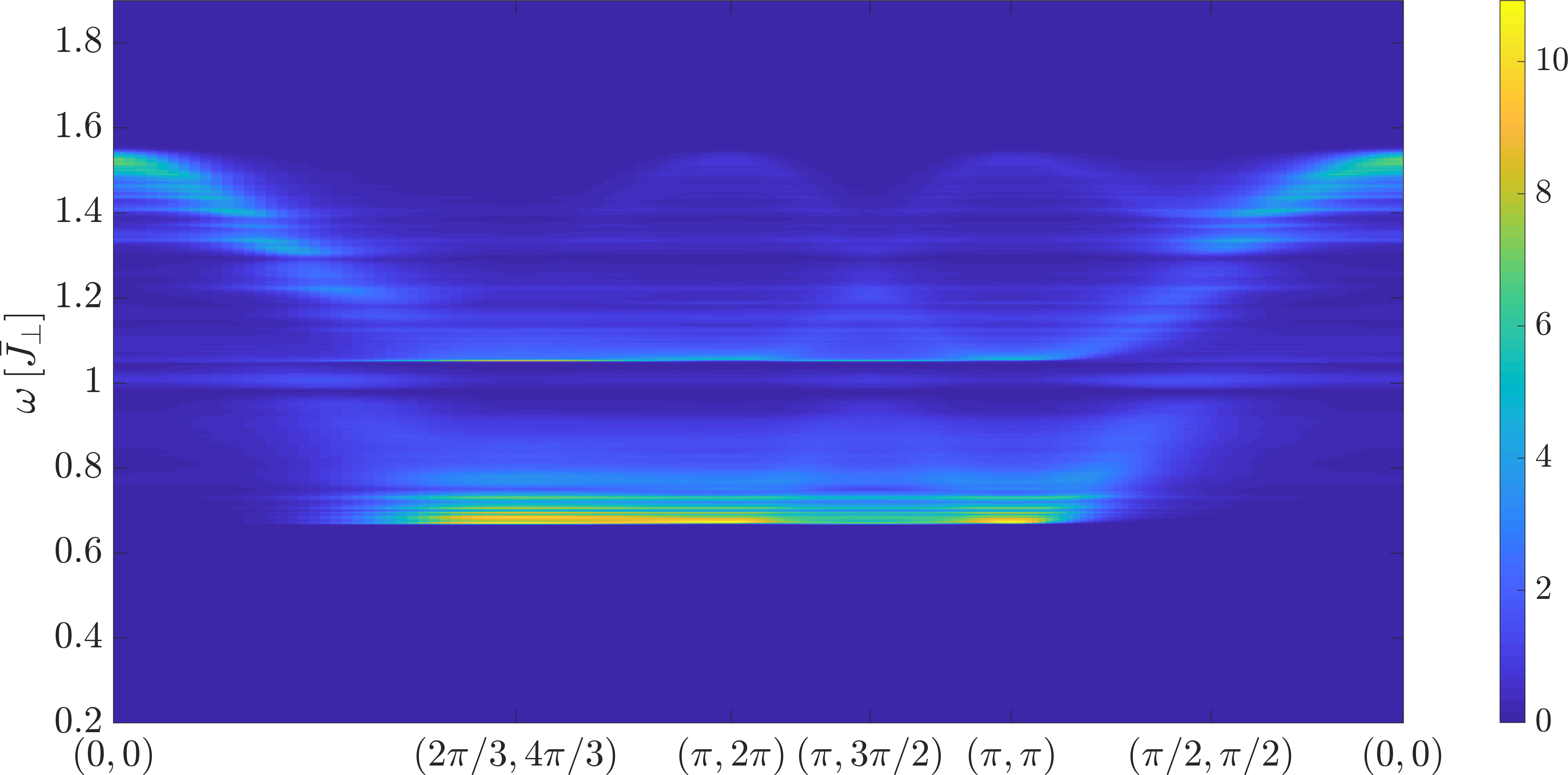}
		\label{fig:Ng2}
	\end{subfigure}
	
	\begin{subfigure}[b]{.9\columnwidth}
		\caption{}		
		\includegraphics[width=1\linewidth]{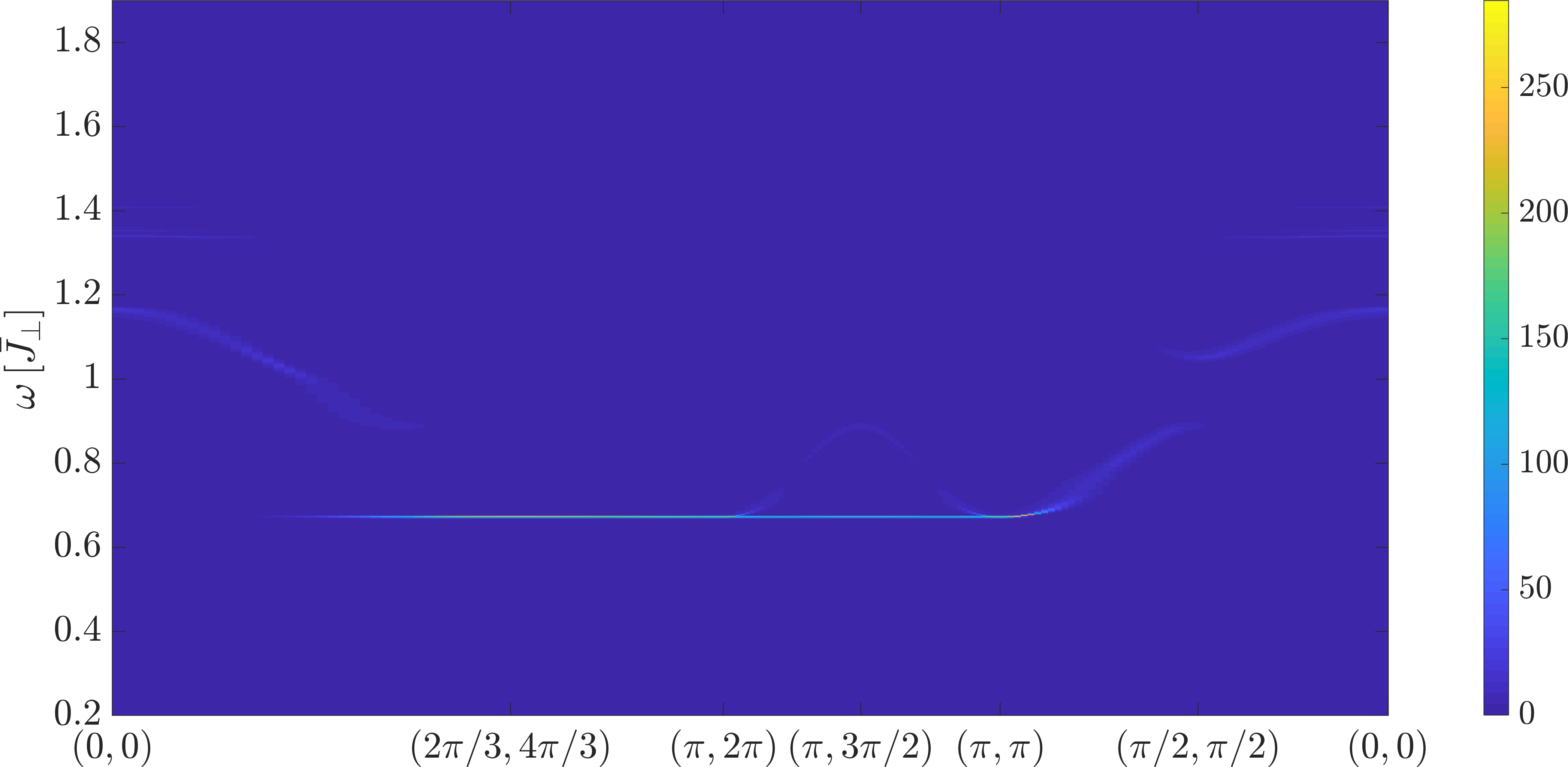}
		\label{fig:Ng2}
	\end{subfigure}
	
	\caption[]{
		The DSF $\mathcal{S}_-(k,\omega)$ is shown for bimodal intra-dimer disorder in the kagome lattice with $J^\parallel=0.2$ and $\Delta J^\perp=\pm 0.2$ for 
		$p=0.1$ in (a), $p=0.5$ in (b) and $p=0.9$ in (c).}
	\label{fig:kagome_Square_Minus}
\end{figure}

\subsubsection{Continuous disorder}
 Continuous distributions without bimodal character can not reproduce the DSF we find for bimodal intra-dimer disorder. This is not surprising since the bimodal intra-dimer disorder has a huge effect on the local energy distribution. For better comparisons a bimodal Gaussian distribution was chosen. This distribution was obtained by choosing with probability $p=0.5$ a value of a Gaussian distribution with mean $J^\perp_1$ and width $\Delta$ and with probability $1-p=0.5$ a value of a Gaussian distribution with mean $J^\perp_2$ and width $\Delta$. This width was chosen such that the second and fourth moment of this distribution and the bimodal distribution coincide. For the continuous disorder we again took $\mathcal{N}_{\rm d}=10000$ and  $\mathcal{N}_{\rm dc}=100$ but the observable was only calculated up to order 7. 
 \begin{figure}
 	\centering
 	\begin{subfigure}[b]{.9\columnwidth}
 		\caption{}		
 		\includegraphics[width=1\linewidth]{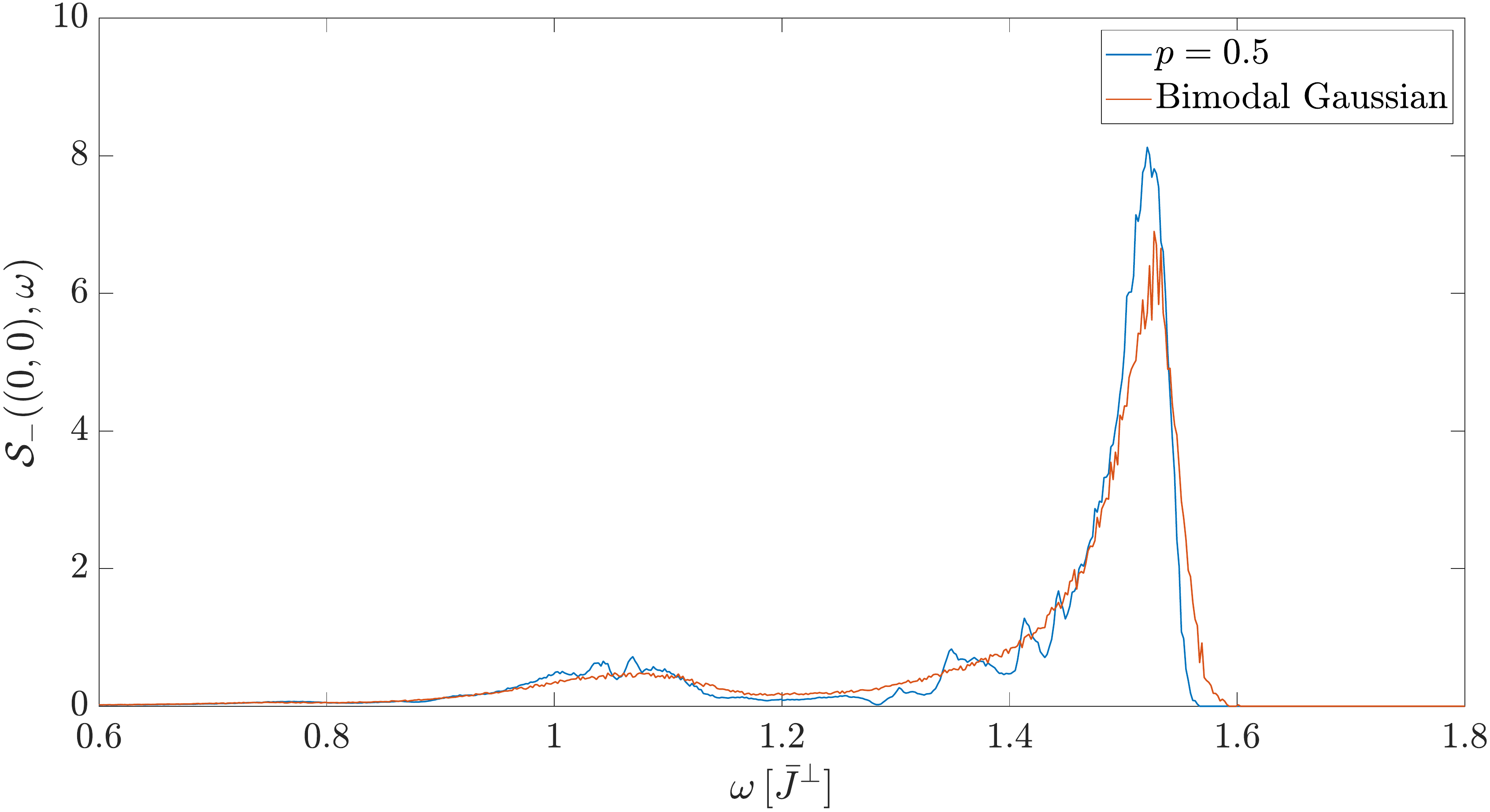}
 		\label{fig:Ng1} 
 	\end{subfigure}
 	
 	\begin{subfigure}[b]{.9\columnwidth}
 		\caption{}		
 		\includegraphics[width=1\linewidth]{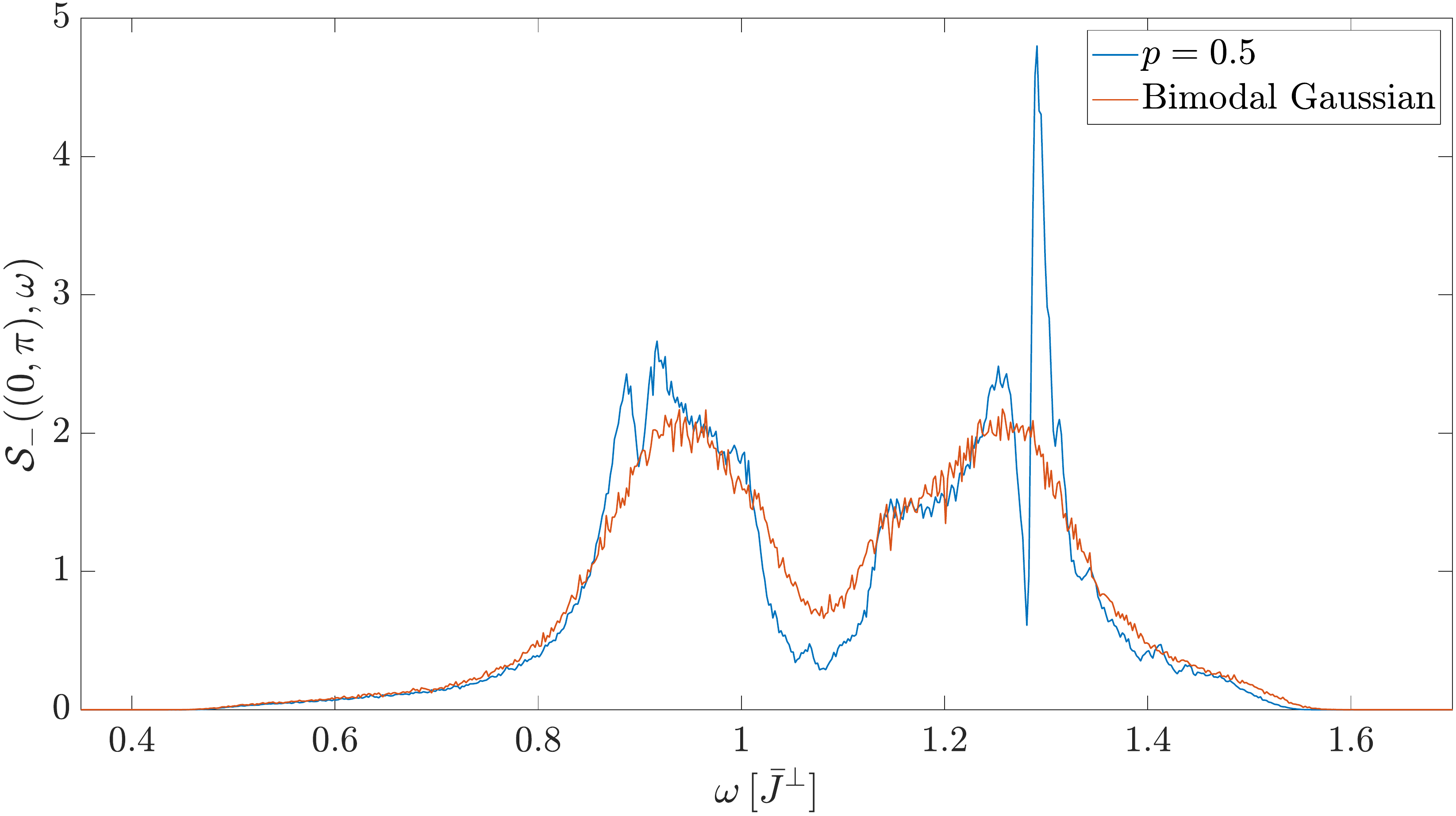}
 		\label{fig:Ng2}
 	\end{subfigure}
 	
 	\begin{subfigure}[b]{.9\columnwidth}
 		\caption{}		
 		\includegraphics[width=1\linewidth]{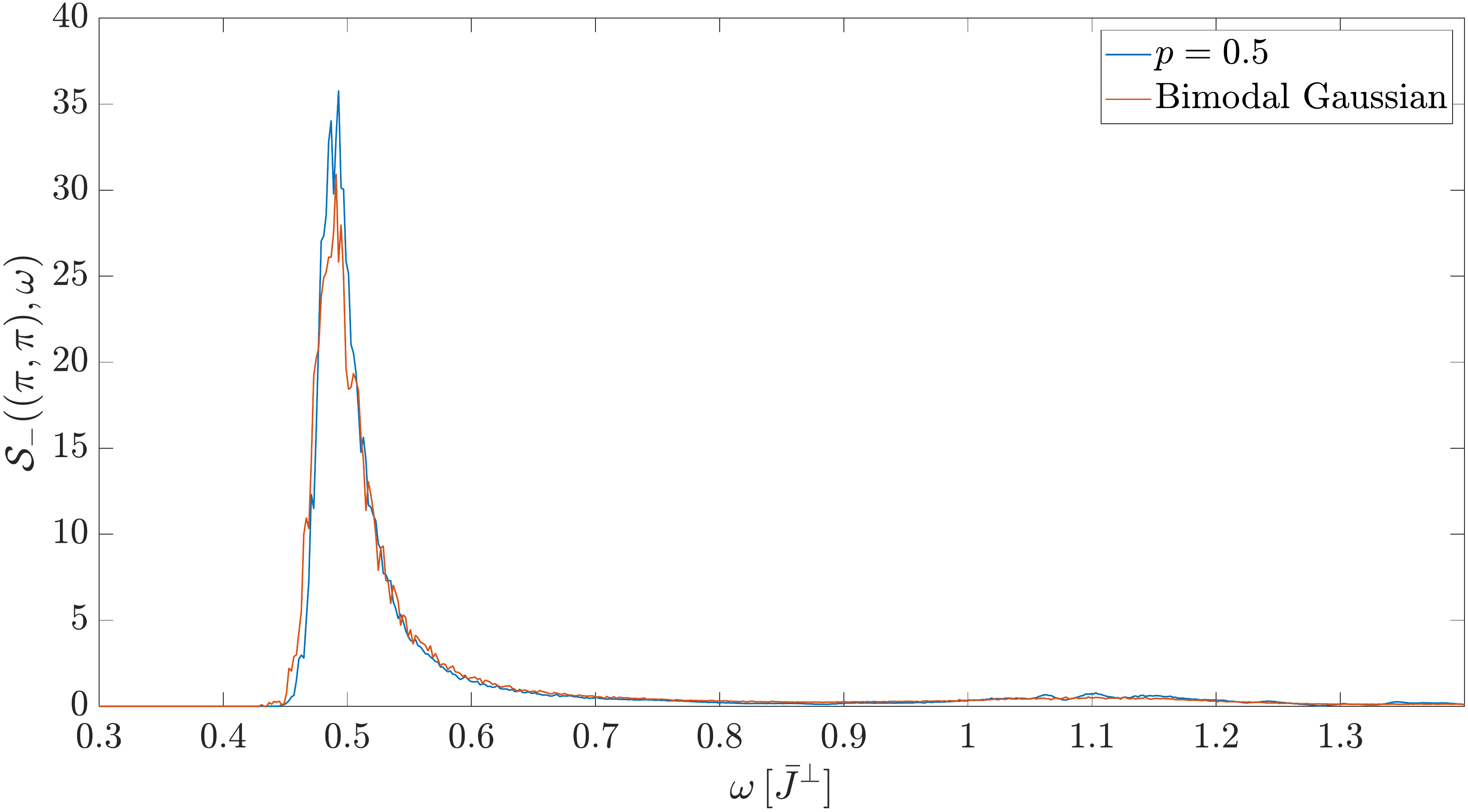}
 		\label{fig:Ng2}
 	\end{subfigure}
 	
 	\caption[]{
 		The DSF $\mathcal{S}_-(k,\omega)$ is shown for bimodal intra-dimer disorder in the square lattice with $J^\parallel=0.2$ and $\Delta J^\perp=\pm 0.2$ for 
 		$p=0.5$ in blue and in red for a bimodal Gaussian distribution with agreement in the first four moments with the bimodal distribution. In (a) $\vec{k}=(0,0)$, in (b) $\vec{k}=(0,\pi)$ and in (c) $\vec{k}=(\pi,\pi)$.}
 	\label{fig:Rung_ContinuousDisorder}
 \end{figure}
 In Fig.~\ref{fig:Rung_ContinuousDisorder} the DSF of the intra-dimer disorder on the square lattice for $p=0.5$ is compared with this bimodal Gaussian distribution. One sees that the agreement is decent. If one chooses just a distribution that agrees with the bimodal one up to the first two moments such an agreement can in general not be obtained. It is quite interesting to note that the continuous distribution only agrees up to the first four moments with the bimodal distribution. The question arises how important moments higher than four are for the DSF in the case of intra-dimer disorder.
\subsection{Inter-dimer disorder}
A limiting case of inter-dimer disorder is that $J^\parallel_{1}-J^\parallel_{2}$ is small compared to the average one-triplon hopping amplitude $\bar{J}^\parallel$.
Then one can treat $J^\parallel_{\nu}-\bar{J}^\parallel$ as a perturbation that will give rise to a finite lifetime of momentum states. Obviously the quantum correlations of the Heisenberg bilayer on the corresponding lattice influence this perturbation. Thus, the lifetime and the shape of the DSF in momentum space is not only dependent on the strength or the form of the disorder but also on the strong correlations on the lattice. These effects can enhance the effect of disorder and decrease the lifetime of the momentum modes or can stabilize these modes by increasing the lifetime.

 The symmetric DSF $\mathcal{S}_+(\vec{k},\omega)$ has finite weight for inter-dimer disorder. However, in contrast to $\mathcal{S}_-(\vec{k},\omega)$ this weight is very small for two reasons. First, the weight is only non-vanishing for different inter-dimer couplings $J^\parallel$ on the upper and lower part of the bilayer. The probability for that to occur is $2p(1-p)$. Second, even for the parts of the lattice $\nu$ where the former holds the weight of $\mathcal{O}_{+}(\nu)$ is very small because it is zero up to the first order of perturbation theory. That the probability for $\mathcal{O}_+(\nu)$ to have a contribution is $2p(1-p)\leq 0.5$ and to have a contribution with constructive phase is \mbox{$p(1-p)\leq 0.25$} implies that $\mathcal{O}_+(\vec{k})$ is composed solely of local disconnected terms because no bond percolation can occur on any of the three lattices. This is the reason why $\mathcal{S}_+(\vec{k},\omega)$ can be seen as a local measurement and why its form is similar to the DOS. Nevertheless, especially for high coupling ratios, an inelastic neutron scattering measurement of  $\mathcal{S}_+(\vec{k},\omega)$ might be possible for energies in the one-triplon regime.
 
\begin{figure}[H]
	\centering
	\begin{subfigure}[b]{.9\columnwidth}
		\caption{}		
		\includegraphics[width=1\linewidth]{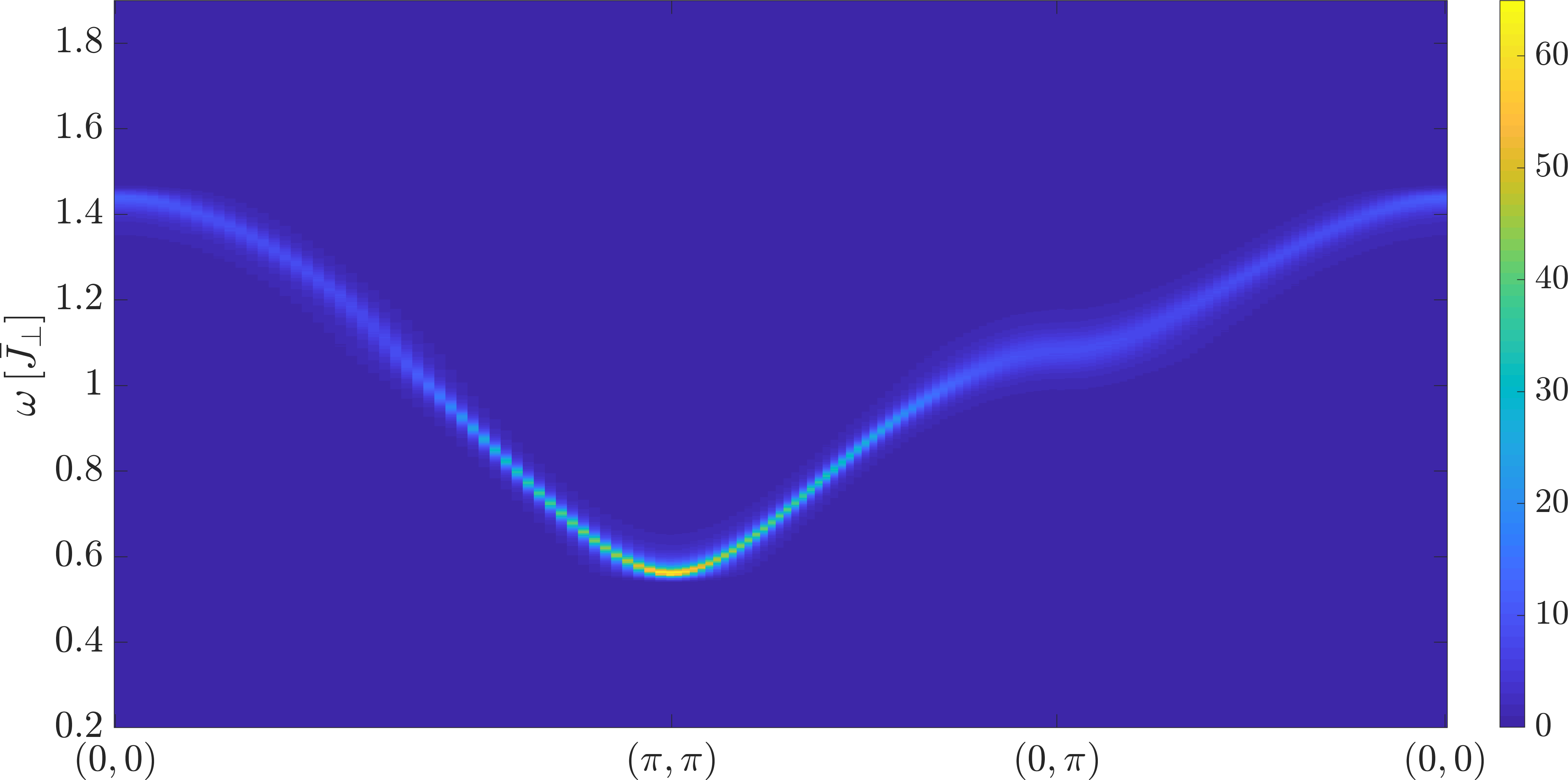}
		\label{fig:Ng1} 
	\end{subfigure}
	\begin{subfigure}[b]{.9\columnwidth}
		\caption{}		
		\includegraphics[width=1\linewidth]{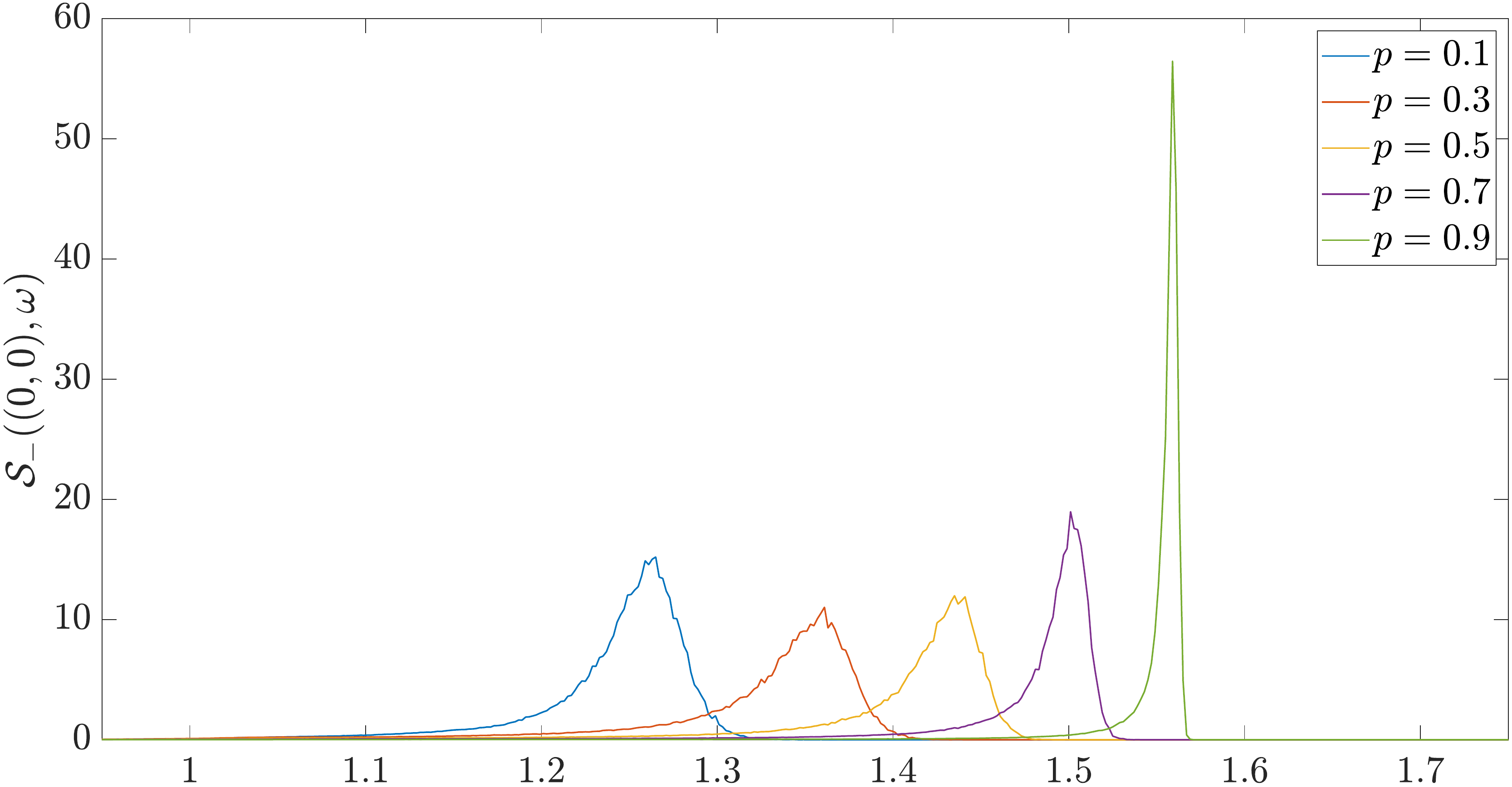}
		\label{fig:Ng2}
	\end{subfigure}
	\begin{subfigure}[b]{.9\columnwidth}
		\caption{}		
		\includegraphics[width=1\linewidth]{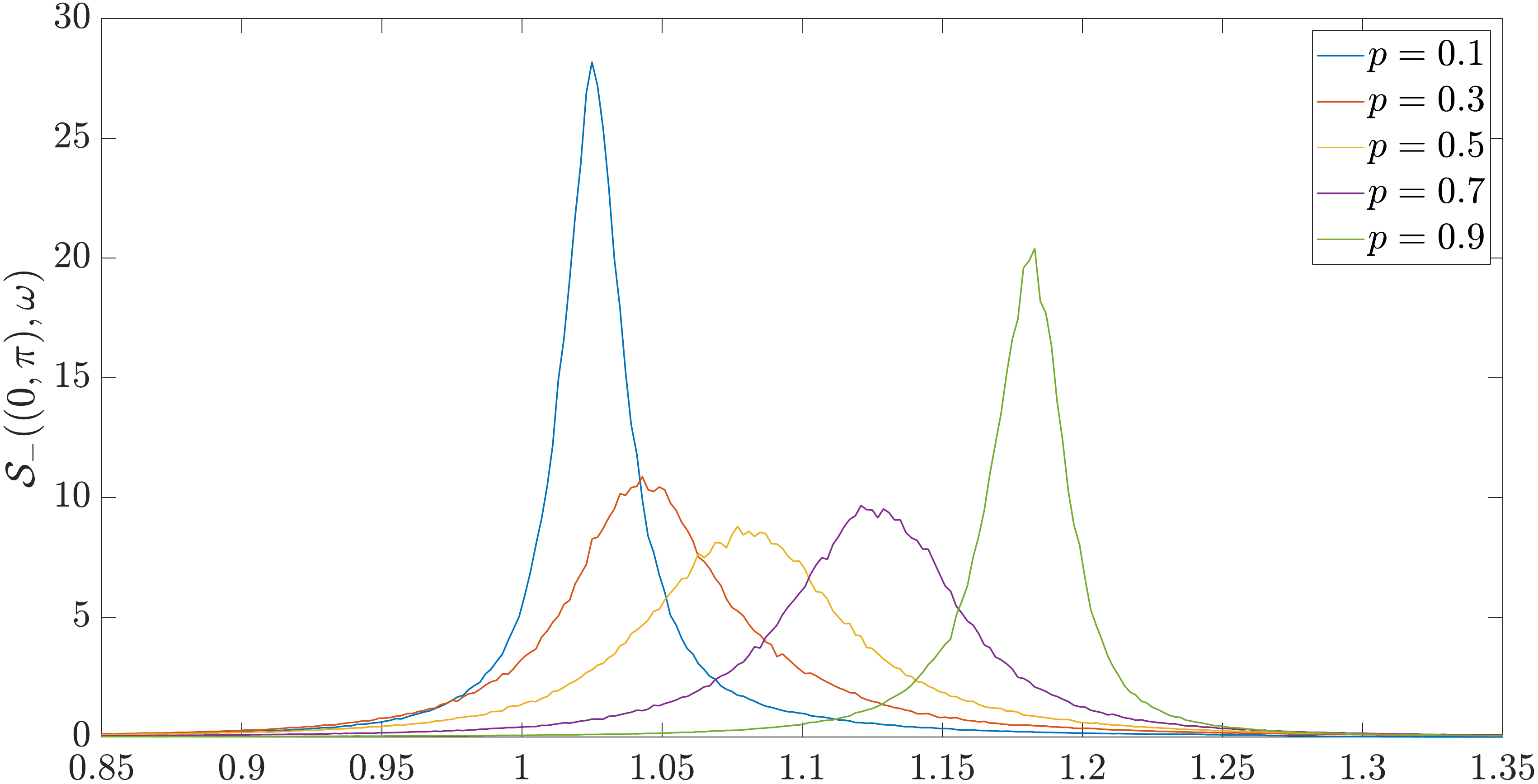}
		\label{fig:Ng2}
	\end{subfigure}
\begin{subfigure}[b]{.9\columnwidth}
	\caption{}	
	\includegraphics[width=1\linewidth]{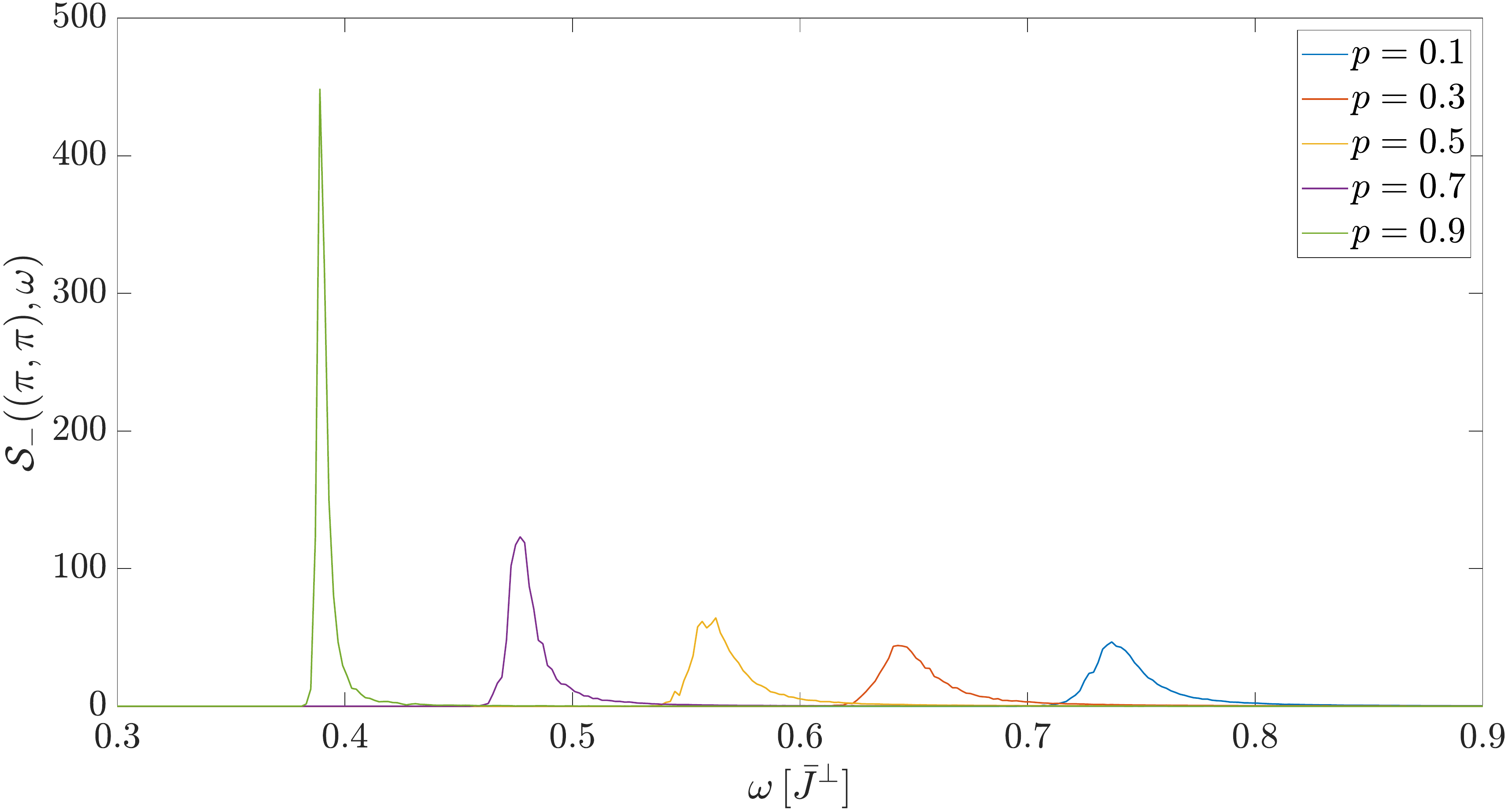}
	\label{fig:Ng2}
\end{subfigure}
	\caption[]{The DSF $\mathcal{S}_-(k,\omega)$ is shown for bimodal inter-dimer disorder in the square lattice with $J^\perp=1$, $J^\parallel_1=0.1$ and $J^\parallel_2=0.3$ for 
		$p=0.5$ in (a) and for $p=0.1,0.3,0.5,0.7,0.9$ and momenta $\vec{k}=(0,0)$ in (b), $\vec{k}=(0,\pi)$ in (c) and $\vec{k}=(\pi,\pi)$ in (d).
	}
	\label{fig:Leg_Square}
\end{figure}

\subsubsection{Square lattice}
The inter-dimer disorder we looked at for the square lattice is bimodal with two different inter-dimer exchanges of $J^\parallel_{1}=0.3$ and $J^\parallel_{2}=0.1$ and $\mathcal{N}_{\rm d}=10000$ and \mbox{$\mathcal{N}_{\rm dc}=100$}. $\mathcal{S}_-(k,\omega)$ was calculated for the probabilities \mbox{$p=0.1,0.3,0.5,0.7,0.9$} and it is shown in Fig.~\ref{fig:Leg_Square}. In appendix C the symmetric structure factor $\mathcal{S}_+(k,\omega)$ is depicted in the Figs.~\ref{fig:Leg_Plus} and \ref{fig:Leg_Plus_k_p}. We stress that the intensities in $\mathcal{S}_+(k,\omega)$ are always very small compared to $\mathcal{S}_-(k,\omega)$.

Interestingly, the lifetimes are highest at the gap momentum $\vec{k}=(\pi,\pi)$. In first-order perturbation theory $\vec{k}=(\pi,\pi)$ can be mapped to $\vec{k}=(0,0)$ and vice versa. Differences in the lifetime at these two momenta are hence caused by higher order perturbations \cite{hormann2018dynamic}. One can thus say that the quantum correlations lead to a stabilization of the lifetime of the mode at $\vec{k}=(\pi,\pi)$. Inter-dimer disorder leads to a disorder potential in $\vec{k}$-space. Apart from the details of this scattering the more states close in energy can be scattered to the lower one expects the lifetime to be. This is an important reason why one finds the low lifetimes at $\vec{k}=(0,\pi)$ and higher ones at the band edges. The high DOS at these energies in Fig.~\ref{fig:Leg_DOS} is in accordance with this reasoning. This also explains why the lifetimes are much higher at the band edges compared to the results for the spin ladder in one dimension \cite{hormann2018dynamic}. 

\subsubsection{Triangular lattice}
For the triangular lattice we considered bimodal inter-dimer disorder with $J^\parallel_{1}=1/15$ and $J^\parallel_{1}=1/5$ and $\mathcal{N}_{\rm d}=9801$ and  $\mathcal{N}_{\rm dc}=100$. We calculated the DSF for \mbox{$p=0.1,0.3,0.5,0.7,0.9$}. It is depicted in Fig.~\ref{fig:Leg_Triangular}. In appendix C the symmetric structure factor $\mathcal{S}_+(k,\omega)$ is shown in the Figs.~\ref{fig:Leg_Plus} and \ref{fig:Leg_Plus_k_p}. As in the square lattice the intensities in $\mathcal{S}_+(k,\omega)$ are very small compared to $\mathcal{S}_-(k,\omega)$.

In contrast to the inter-dimer disorder results for the square lattice here the couplings are chosen further away from the critical point because of the enhancement of the disordered phase by frustration and convergence issues of the perturbative expansion.

The lifetime at the gap momentum $\vec{k}=(4\pi/3,2\pi/3)$ is lower than in the square lattice. As in the intra-dimer disorder we identify frustration and its effect on kinetic terms as the main reason for that. Even in first-order perturbation theory there is no symmetry between the spectral shape at $\vec{k}=(0,0)$ and at the gap momentum. We see that the lifetimes at the gap momentum are shorter than at $\vec{k}=(0,0)$ for all $p$ which is a significant difference to the situation on the square lattice. 

\begin{figure}[H]
	\centering
	\begin{subfigure}[b]{.9\columnwidth}
		\caption{}		
		\includegraphics[width=1\linewidth]{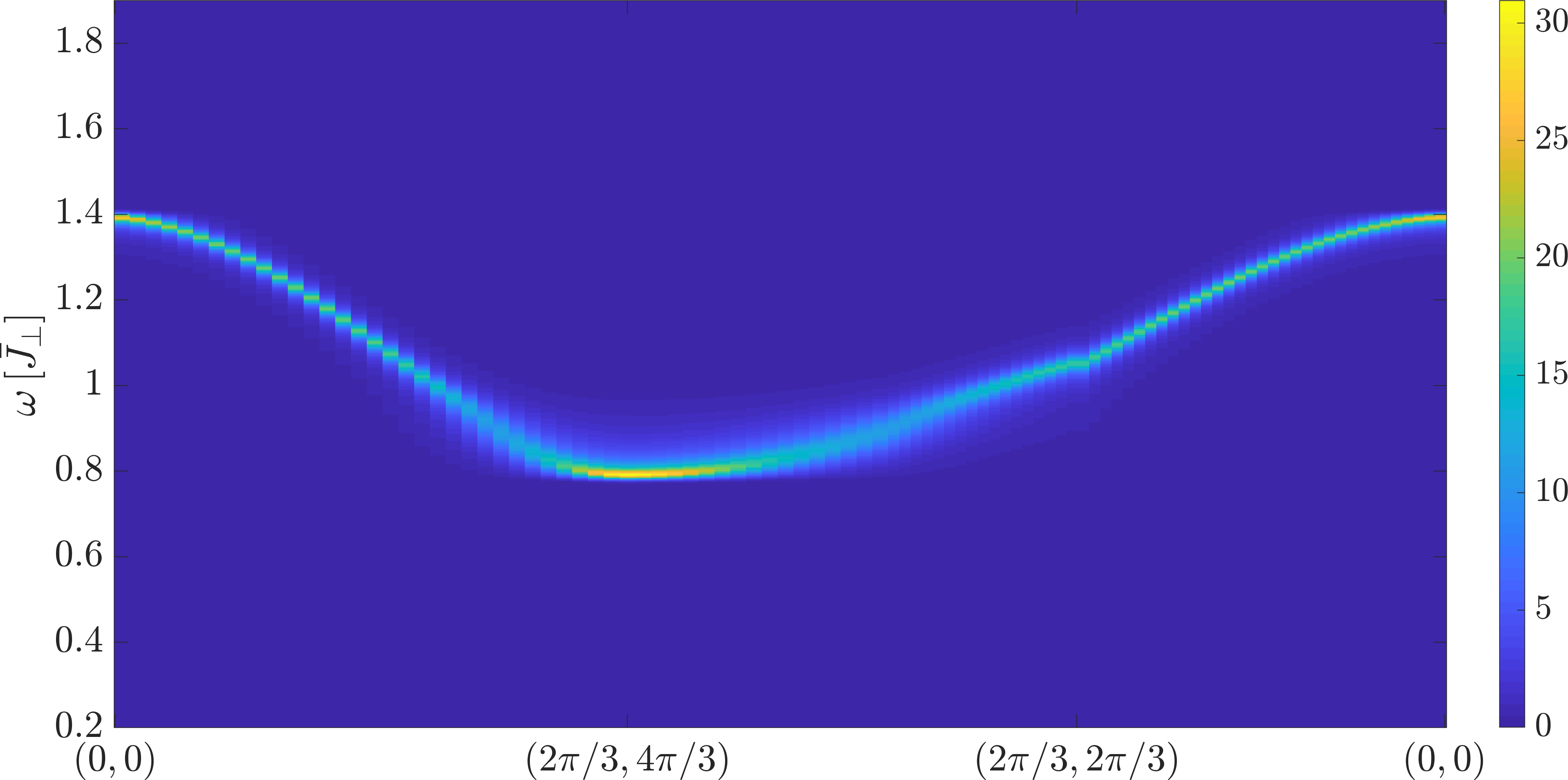}
		\label{fig:Ng1} 
	\end{subfigure}
	\begin{subfigure}[b]{.9\columnwidth}
		\caption{}		
		\includegraphics[width=1\linewidth]{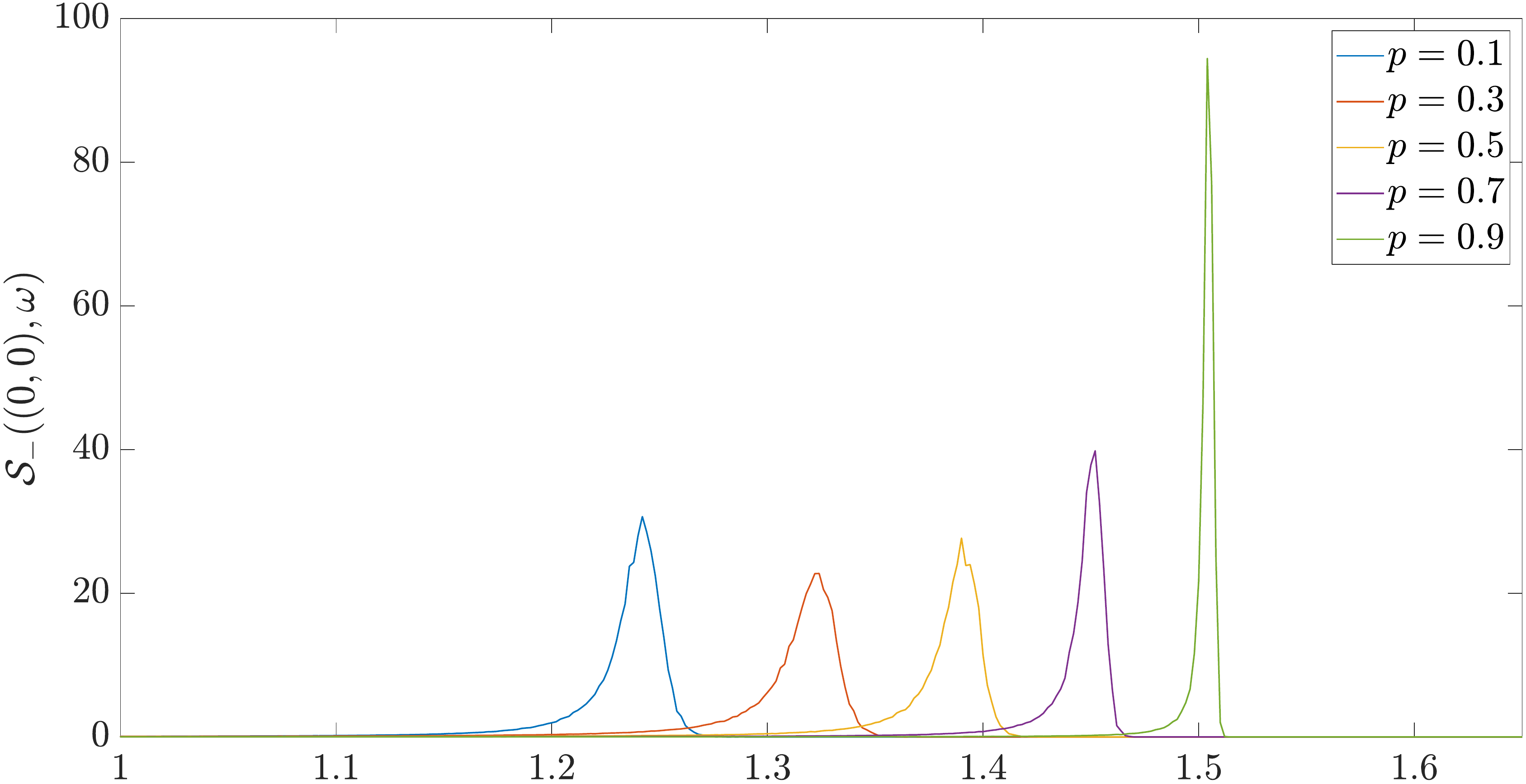}
		\label{fig:Ng2}
	\end{subfigure}
	\begin{subfigure}[b]{.9\columnwidth}
		\caption{}		
		\includegraphics[width=1\linewidth]{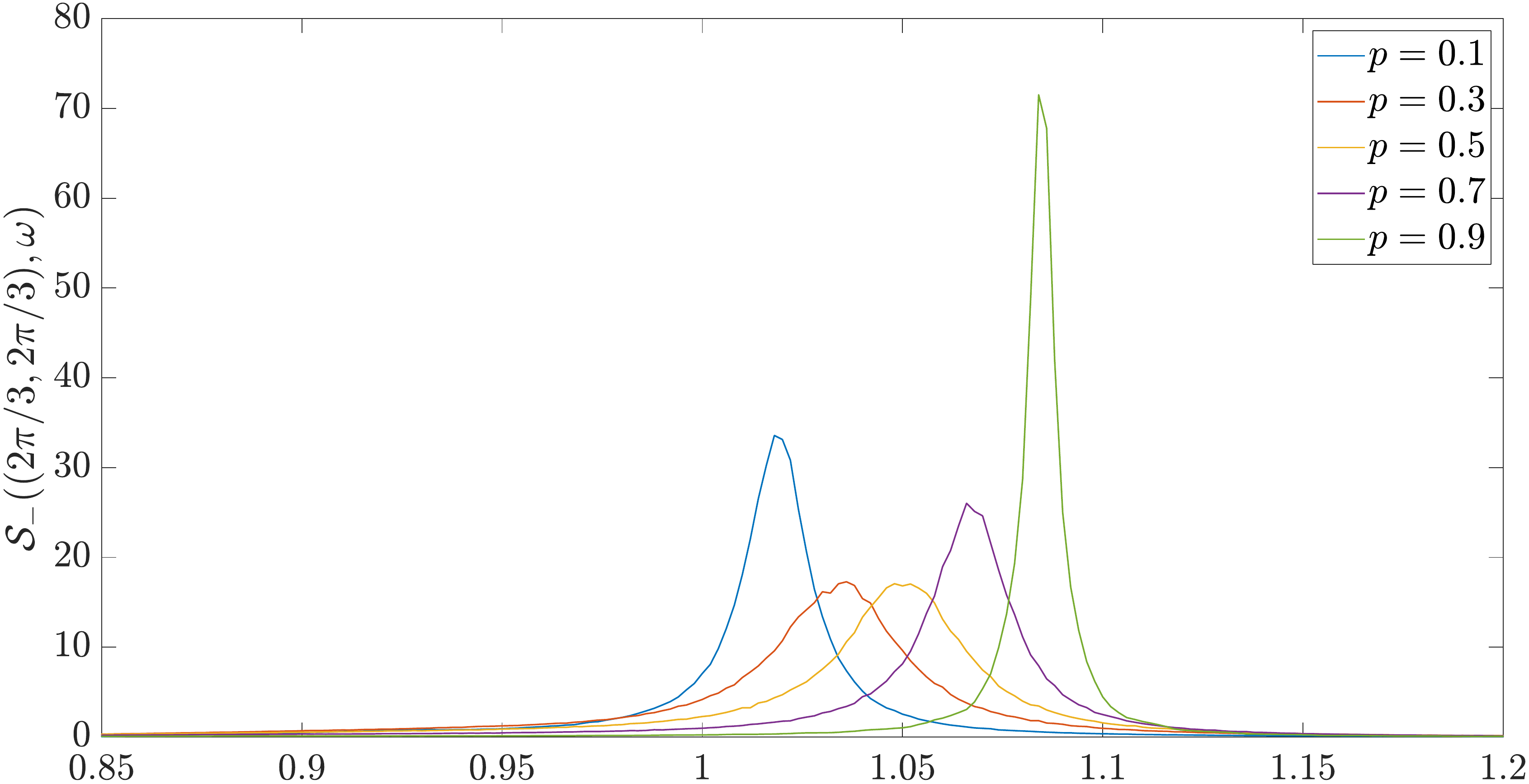}
		\label{fig:Ng2}
	\end{subfigure}
	\begin{subfigure}[b]{.9\columnwidth}
		\caption{}		
		\includegraphics[width=1\linewidth]{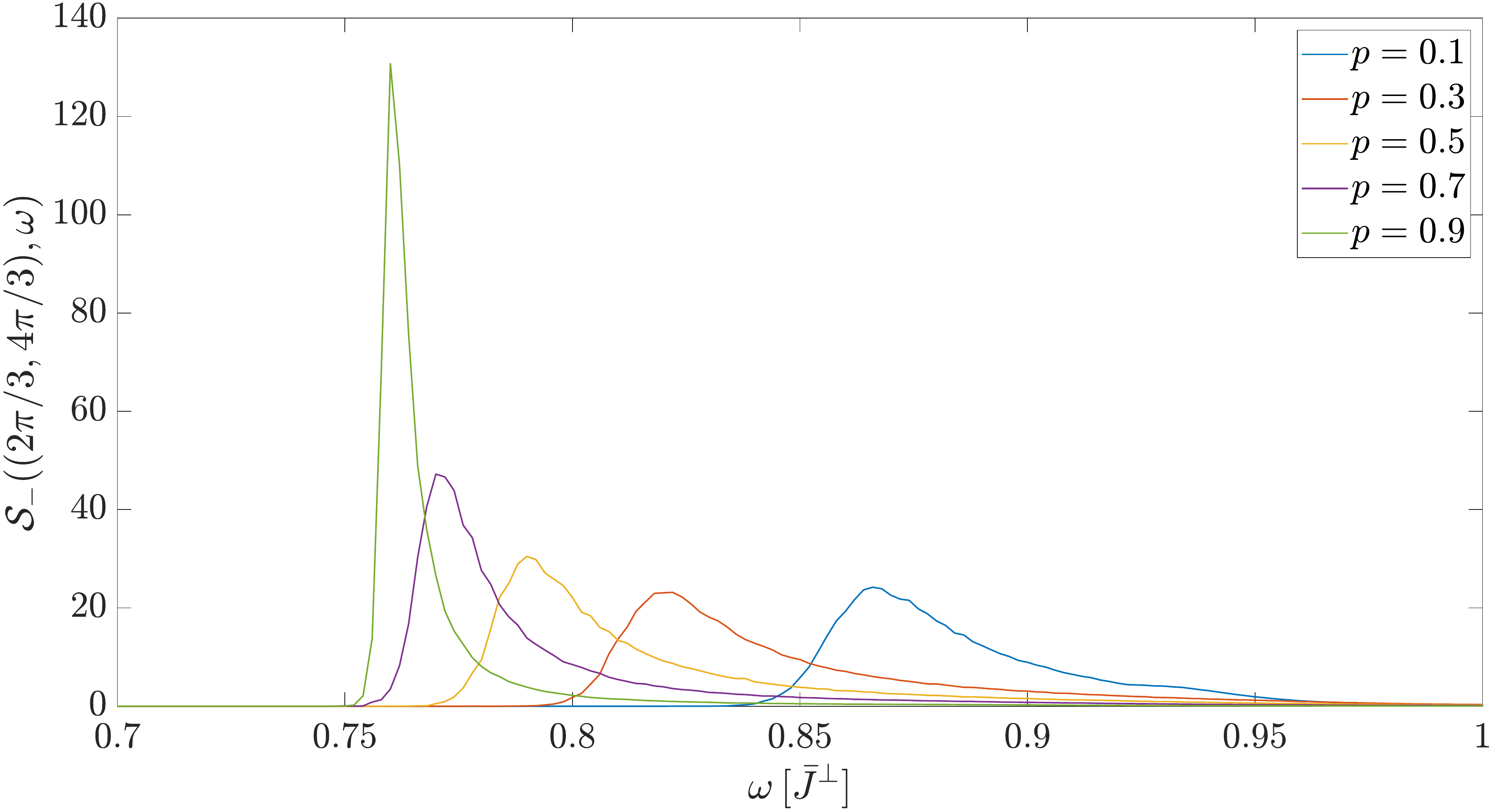}
		\label{fig:Ng2}
	\end{subfigure}
	\caption[]{
		The DSF $\mathcal{S}_-(k,\omega)$ is shown for bimodal inter-dimer disorder in the triangular lattice with $J^\perp=1$, $J^\parallel_1=1/15$ and $J^\parallel_2=1/5$ for 
		$p=0.5$ in (a) and for $p=0.1,0.3,0.5,0.7,0.9$ and momentas $\vec{k}=(0,0)$ in (b), $\vec{k}=(2\pi/3,2\pi/3)$ in (c) and $\vec{k}=(2\pi/3,4\pi/3)$ in (d).
	}
	\label{fig:Leg_Triangular}
\end{figure}

\subsubsection{Kagome lattice}

In the kagome lattice we considered a bimodal inter-dimer disorder of $J^\parallel_{1}=0.1$ and $J^\parallel_{1}=0.3$ and \mbox{$\mathcal{N}_{\rm d}=9747$} and  $\mathcal{N}_{\rm dc}=100$. We calculated the DSF for \mbox{$p=0.1,0.3,0.5,0.7,0.9$} and it is shown for $p=0.1,0.5,0.9$ in Fig.~\ref{fig:Leg_kagome}. In appendix C it is also depicted for $p=0.3,0.7$ in Fig.~\ref{fig:Rung_kagome_p0307}. The symmetric structure factor $\mathcal{S}_+(k,\omega)$ is depicted in Fig.~\ref{fig:Leg_Plus} in appendix C. Also for the kagome lattice the intensities in $\mathcal{S}_+(k,\omega)$ are very small compared to $\mathcal{S}_-(k,\omega)$.\par
\begin{figure}
	\centering
	\begin{subfigure}[b]{.9\columnwidth}
		\caption{}		
		\includegraphics[width=1\linewidth]{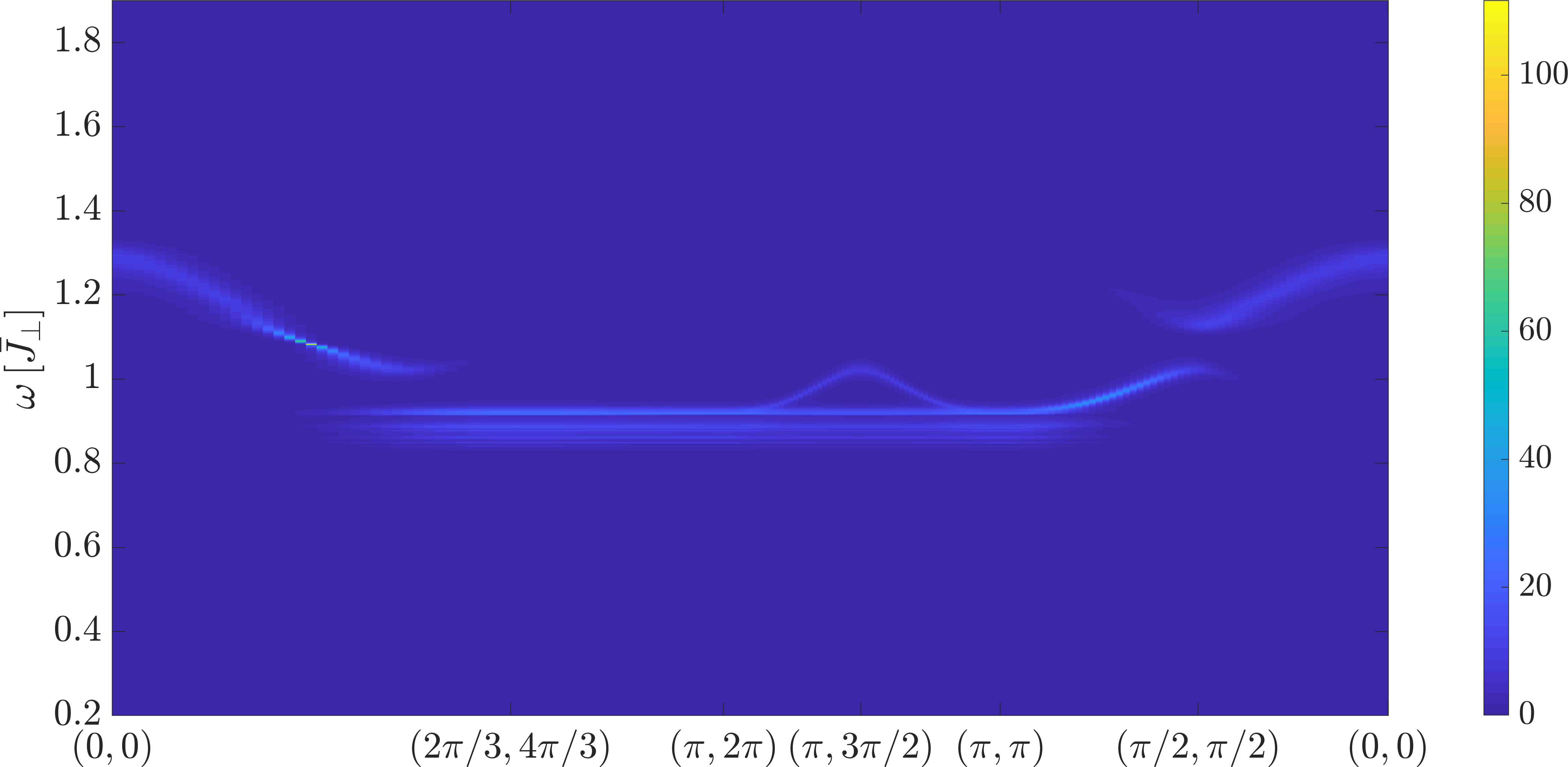}
		\label{fig:Ng1} 
	\end{subfigure}
	
	\begin{subfigure}[b]{.9\columnwidth}
		\caption{}		
		\includegraphics[width=1\linewidth]{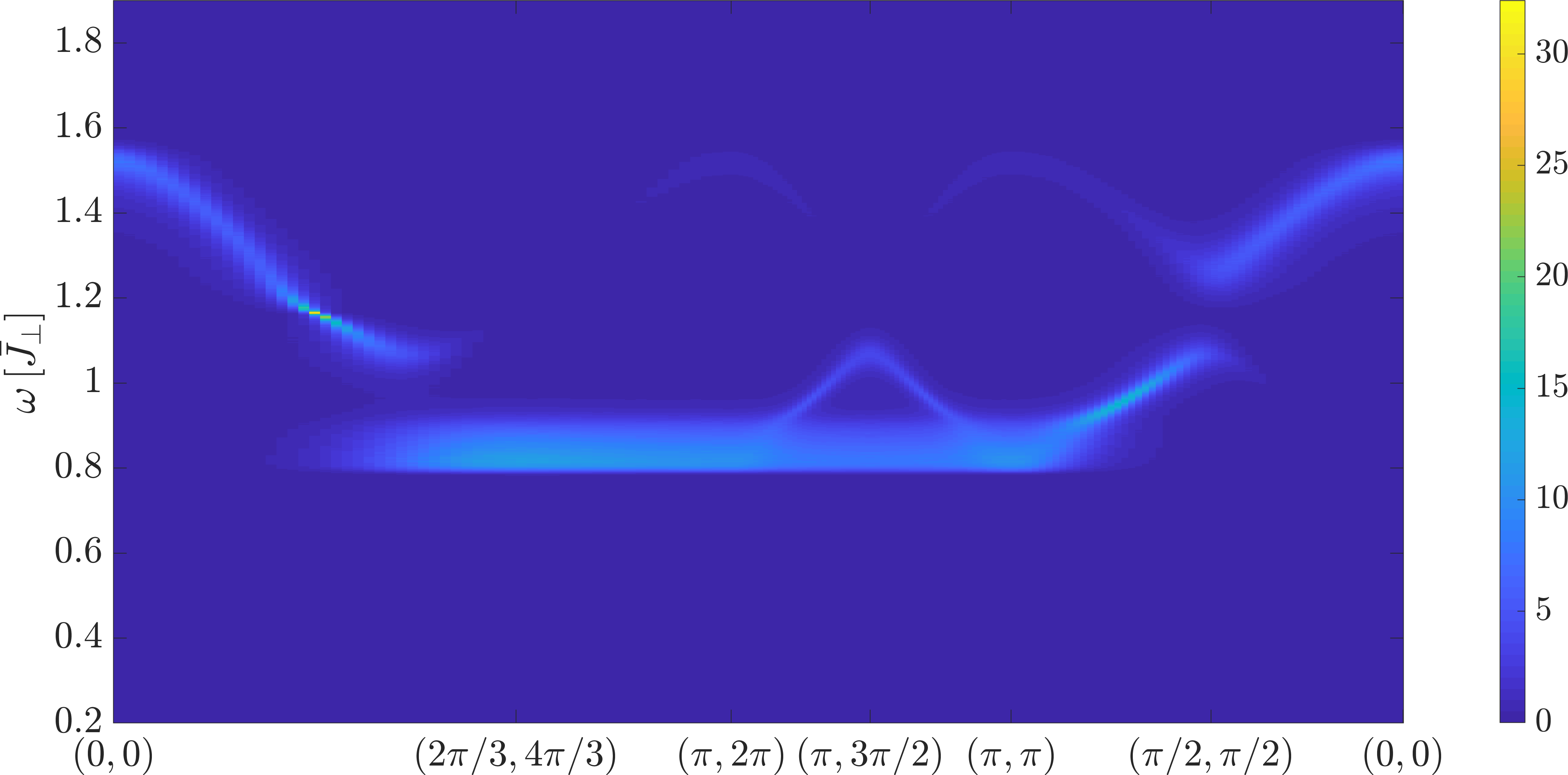}
		\label{fig:Ng2}
	\end{subfigure}
	
	\begin{subfigure}[b]{.9\columnwidth}
		\caption{}		
		\includegraphics[width=1\linewidth]{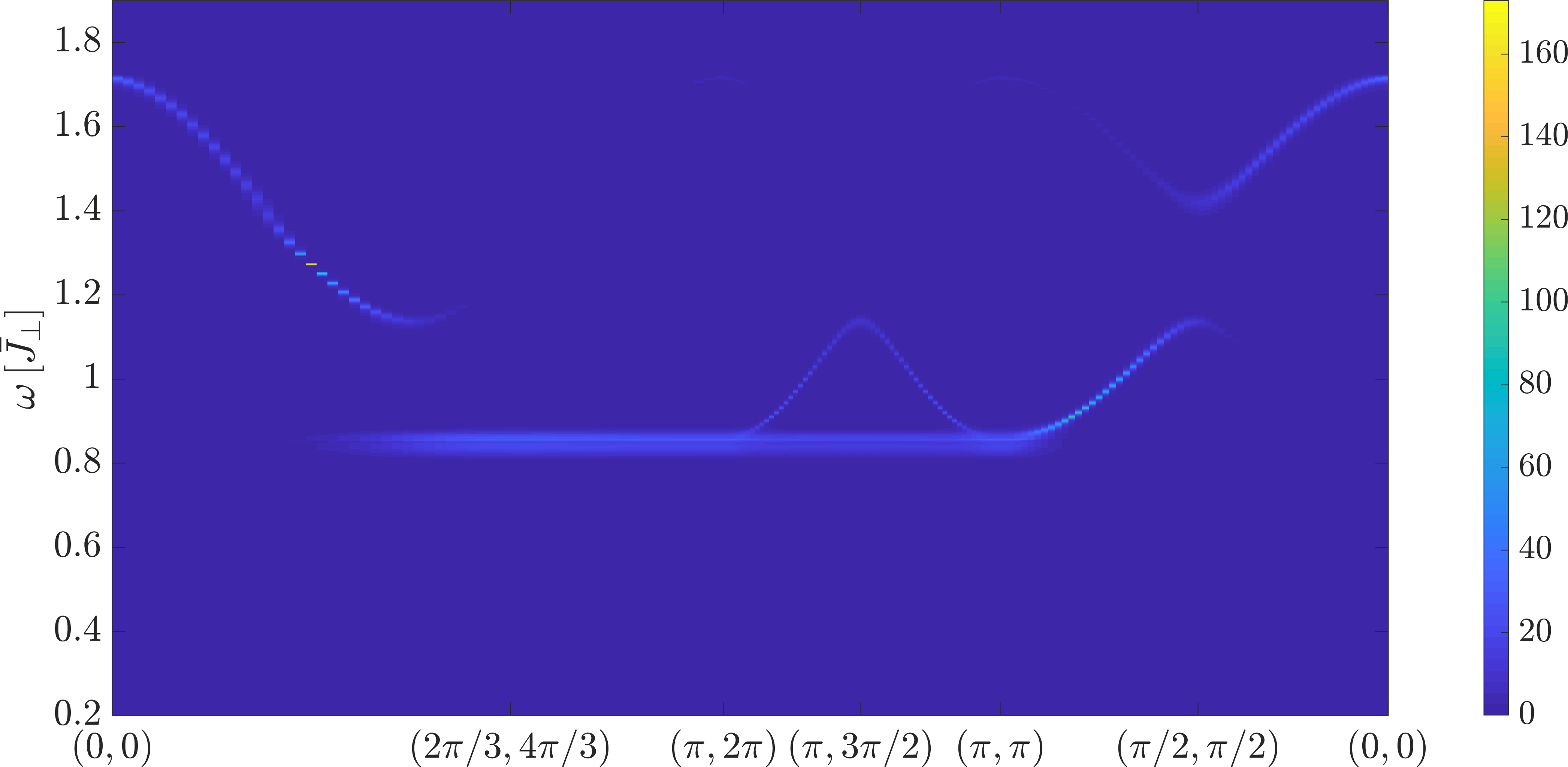}
		\label{fig:Ng2}
	\end{subfigure}	
	\caption[justification=justified]{
		The DSF $\mathcal{S}_-(k,\omega)$ is shown for bimodal inter-dimer disorder in the kagome lattice with $J^\perp=1$, \mbox{$J^\parallel_1=0.1$} and $J^\parallel_2=0.3$ for $p=0.1$ in (a), $p=0.5$ in (b) and $p=0.9$ in (c).
	}
	\label{fig:Leg_kagome}
\end{figure}
In contrast to the generic behavior that is also found for the square and the triangular lattice in the kagome lattice the minimum energy in the presence of disorder is lower than the bounds given by all exchanges equal $J^\parallel_{1}=0.1$ or $J^\parallel_{1}=0.3$. The crucial difference of the kagome lattice geometry is that its energetically lowest band is almost flat without disorder. The eigenstates in that flat band are localized on the hexagons of the kagome lattice up to order 7 in perturbation theory. This is possible because of destructive interference on the neighboring links to other hexagons. Inter-dimer disorder now breaks this destructive interference and the eigenstates get a larger localization length than in the clean case. These longer ranged correlated eigenstates can reach lower energies as was already seen for intra-dimer disorder in the kagome lattice. As a side note this rises the question if disorder can lead to new critical behavior of otherwise non-critical flat modes.

The effect of disorder on the two dispersive bands is weaker than the one on the flat band. This is related to the fact that without disorder the states in the dispersive bands are extended two-dimensional momentum states whereas in the flat band these are to a good approximation localized states in position space.
\begin{figure}[ht]
	\centering
	\begin{subfigure}[b]{.9\columnwidth}
		\caption{}		
		\includegraphics[width=1\linewidth]{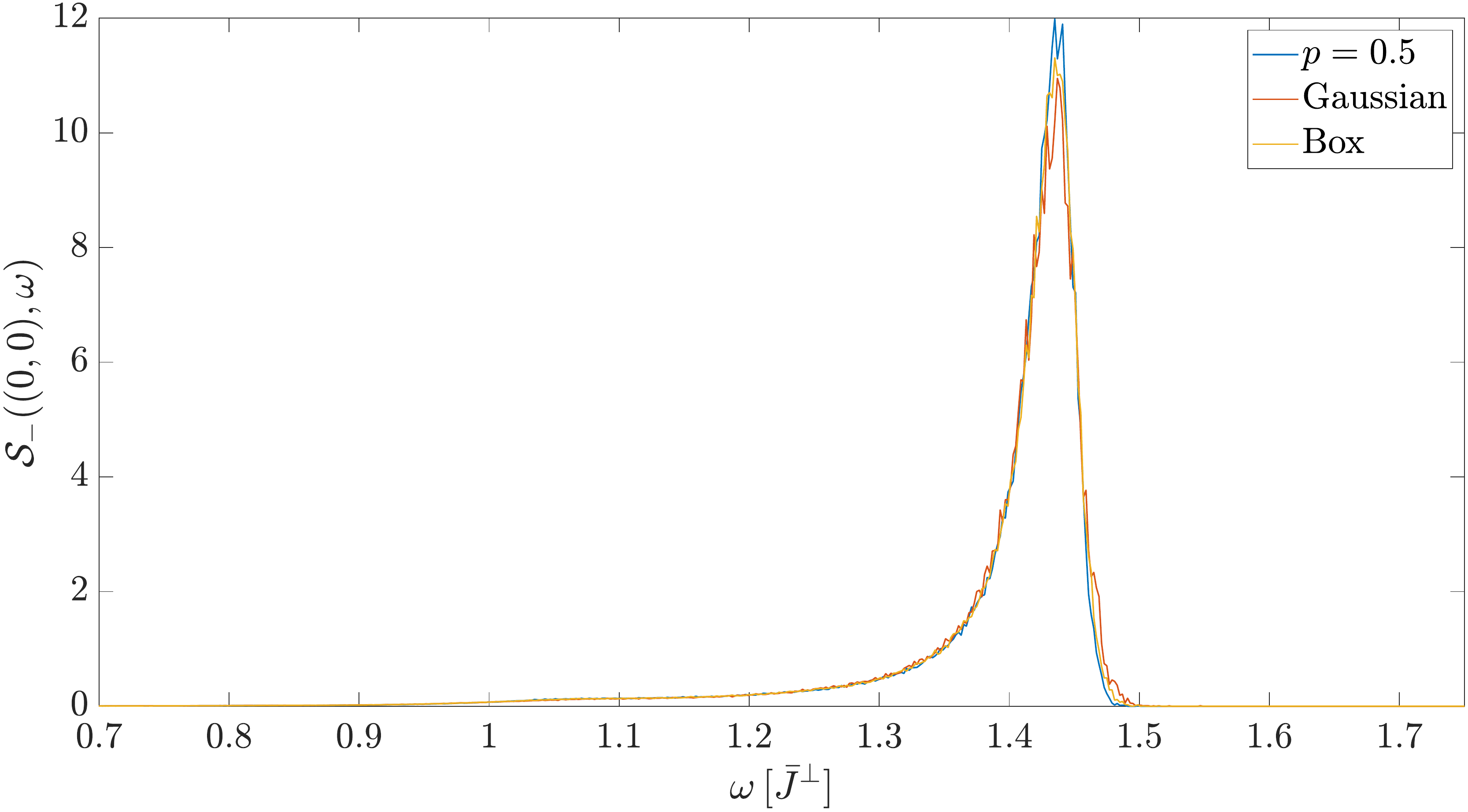}
		\label{fig:Ng1} 
	\end{subfigure}
	
	\begin{subfigure}[b]{.9\columnwidth}
		\caption{}		
		\includegraphics[width=1\linewidth]{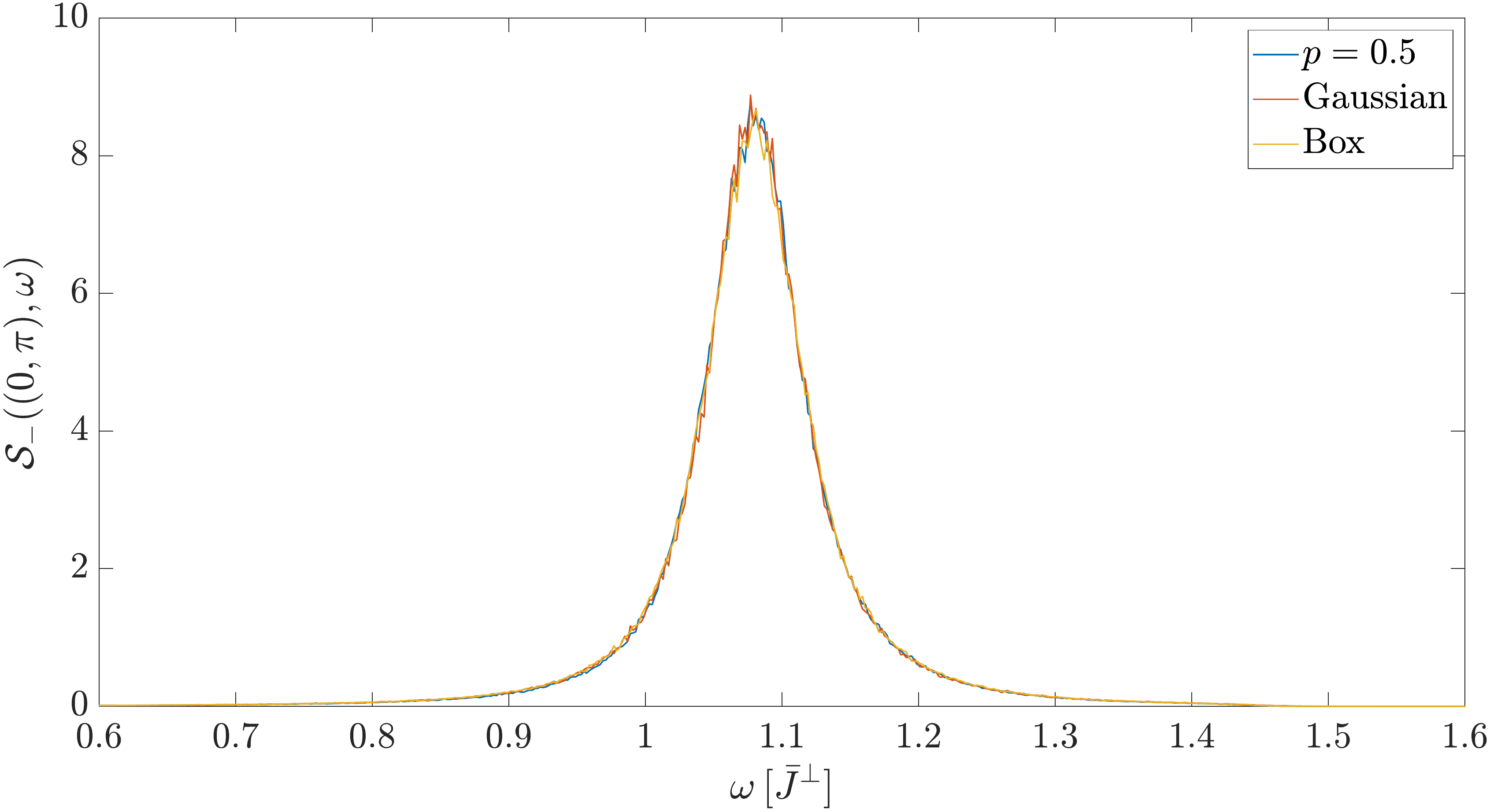}
		\label{fig:Ng2}
	\end{subfigure}
	
	\begin{subfigure}[b]{.9\columnwidth}
		\caption{}		
		\includegraphics[width=1\linewidth]{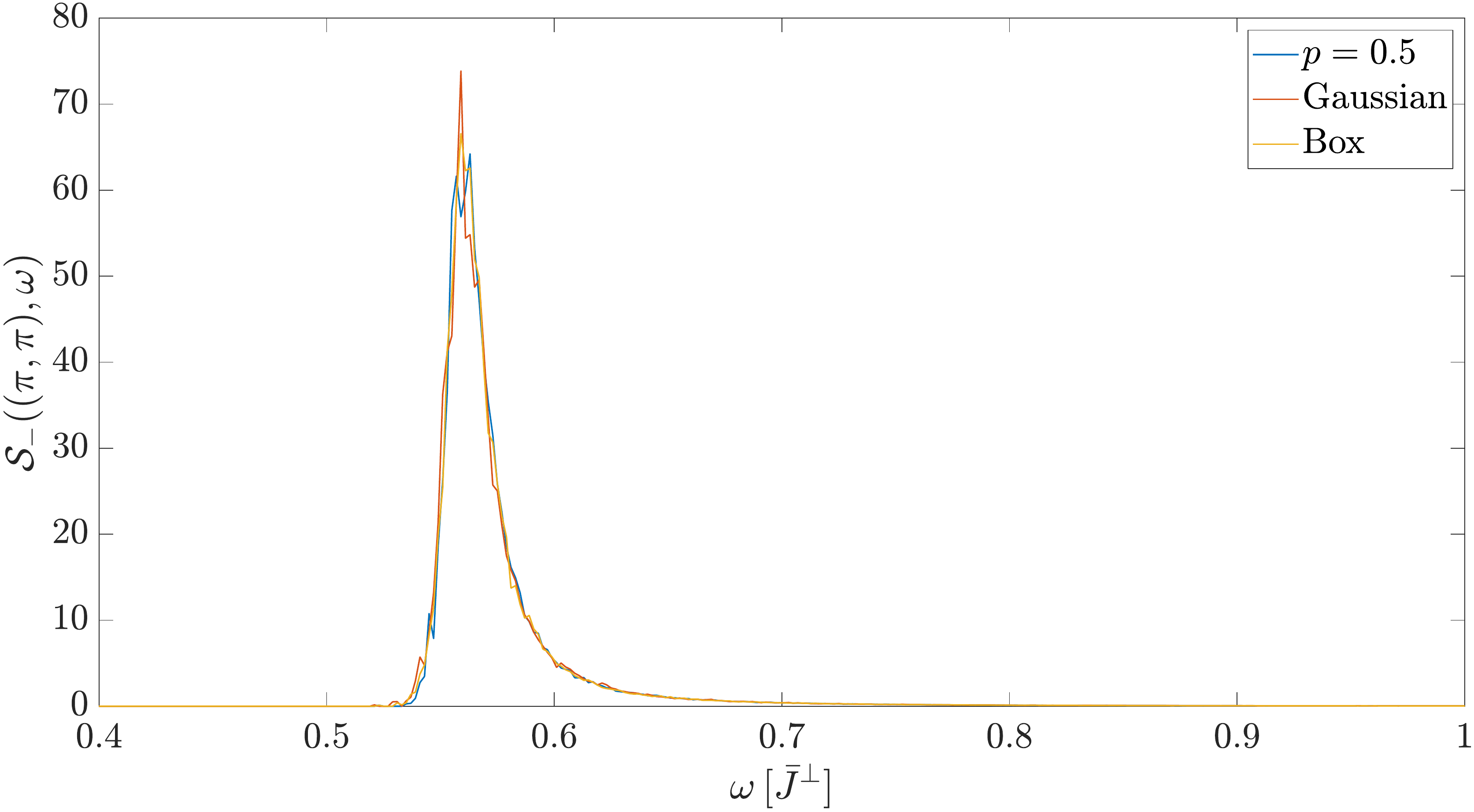}
		\label{fig:Ng2}
	\end{subfigure}
	\caption[]{
		The DSF $\mathcal{S}_-(k,\omega)$ is shown for bimodal inter-dimer disorder in the square lattice with $J^\perp=1$, $J^\parallel_1=0.1$ and $J^\parallel_1=0.3$ for 
		$p=0.5$ in blue. In red it is shown for a Gaussian distribution and in yellow for a box distribution with agreement in the first two moments with the bimodal distribution. In (a) $\vec{k}=(0,0)$, in (b) $\vec{k}=(0,\pi)$ and in (c) $\vec{k}=(\pi,\pi)$.
	}
	\label{fig:Leg_Continuous}
\end{figure}
Interestingly, already a small amount of disorder like $p=0.1,0.9$ has huge effects on the flat band. This might be of special interest for experiments in which it is not possible to reach larger $p$.

The kagome lattice has three bands in the clean case. Disorder leads to interactions between the states in these three bands. The dispersive bands can potentially interact with the almost flat band at low energies. We see that with disorder one can not find a gap anymore between the lowest band and the one in the middle. This rises the question how strong the states originating from the dispersive and the flat band interact and if the dispersive bands can get flatter or the flat band more dispersive.
\subsubsection{Continuous disorder}
In Fig.~\ref{fig:Leg_Continuous} the bimodal inter-dimer disorder on the square lattice with $p=0.5$ was compared with Gaussian and box disorder that coincides in the first two moments with the bimodal disorder. For the continuous disorder also $\mathcal{N}_{\rm d}=10000$ and  $\mathcal{N}_{\rm dc}=100$ was considered but the observable was only calculated up to order 7. One can see that the agreement is excellent. We conclude that for inter-dimer disorder on the square lattice agreement in the first two moments of the disorder is almost sufficient for a quantitative agreement. This is in strong contrast to the situation in one dimension where one can see significant differences between different kinds of leg disorder that only agree up to the first two moments.
\section{Conclusions}
\label{sec:conclusion}
The pCUT method together with a white-graph expansion is an efficient tool to investigate the fate of quasi-particles under quenched disorder in two dimensions. We showed this by calculating the one-triplon DSF of disordered Heisenberg bilayers on the square, triangular, and kagome lattice in the quantum paramagnetic phase. We found that disorder leads to a decrease of the minimum energy in the kagome lattice. Related to that one finds larger localization lengths in the flat band as long as local hexagon modes are still exact eigenstates. Frustration in the triangular lattice had the effect of stronger localization. In the square lattice stronger quantum correlations stabilized the mode at $\vec{k}=(\pi,\pi)$. For inter-dimer disorder agreement up to the second moment of the disorder distribution is almost sufficient for a quantitative agreement in the DSF. This is different for intra-dimer disorder where the bimodal character of the disorder plays a crucial role.\par

Quantitative calculations for the DSF in the presence of disorder are typically hard to obtain. We therefore believe that our approach will be of relevance for direct comparisons with experiments on disordered two-dimensional quantum magnets in the future.

\section{Acknowledgments}
MH and KPS acknowledge financial support from DFG Project No. SCHM 2511/10-1.


\begin{thebibliography}{36}%
\makeatletter
\providecommand \@ifxundefined [1]{%
 \@ifx{#1\undefined}
}%
\providecommand \@ifnum [1]{%
 \ifnum #1\expandafter \@firstoftwo
 \else \expandafter \@secondoftwo
 \fi
}%
\providecommand \@ifx [1]{%
 \ifx #1\expandafter \@firstoftwo
 \else \expandafter \@secondoftwo
 \fi
}%
\providecommand \natexlab [1]{#1}%
\providecommand \enquote  [1]{``#1''}%
\providecommand \bibnamefont  [1]{#1}%
\providecommand \bibfnamefont [1]{#1}%
\providecommand \citenamefont [1]{#1}%
\providecommand \href@noop [0]{\@secondoftwo}%
\providecommand \href [0]{\begingroup \@sanitize@url \@href}%
\providecommand \@href[1]{\@@startlink{#1}\@@href}%
\providecommand \@@href[1]{\endgroup#1\@@endlink}%
\providecommand \@sanitize@url [0]{\catcode `\\12\catcode `\$12\catcode
  `\&12\catcode `\#12\catcode `\^12\catcode `\_12\catcode `\%12\relax}%
\providecommand \@@startlink[1]{}%
\providecommand \@@endlink[0]{}%
\providecommand \url  [0]{\begingroup\@sanitize@url \@url }%
\providecommand \@url [1]{\endgroup\@href {#1}{\urlprefix }}%
\providecommand \urlprefix  [0]{URL }%
\providecommand \Eprint [0]{\href }%
\providecommand \doibase [0]{http://dx.doi.org/}%
\providecommand \selectlanguage [0]{\@gobble}%
\providecommand \bibinfo  [0]{\@secondoftwo}%
\providecommand \bibfield  [0]{\@secondoftwo}%
\providecommand \translation [1]{[#1]}%
\providecommand \BibitemOpen [0]{}%
\providecommand \bibitemStop [0]{}%
\providecommand \bibitemNoStop [0]{.\EOS\space}%
\providecommand \EOS [0]{\spacefactor3000\relax}%
\providecommand \BibitemShut  [1]{\csname bibitem#1\endcsname}%
\let\auto@bib@innerbib\@empty
\bibitem [{\citenamefont {Anderson}(1958)}]{anderson1958absence}%
  \BibitemOpen
  \bibfield  {author} {\bibinfo {author} {\bibfnamefont {P.~W.}\ \bibnamefont
  {Anderson}},\ }\href@noop {} {\bibfield  {journal} {\bibinfo  {journal}
  {Physical Review}\ }\textbf {\bibinfo {volume} {109}},\ \bibinfo {pages}
  {1492} (\bibinfo {year} {1958})}\BibitemShut {NoStop}%
\bibitem [{\citenamefont {Andrade}\ \emph {et~al.}(2018)\citenamefont
  {Andrade}, \citenamefont {Hoyos}, \citenamefont {Rachel},\ and\ \citenamefont
  {Vojta}}]{andrade2018cluster}%
  \BibitemOpen
  \bibfield  {author} {\bibinfo {author} {\bibfnamefont {E.~C.}\ \bibnamefont
  {Andrade}}, \bibinfo {author} {\bibfnamefont {J.~A.}\ \bibnamefont {Hoyos}},
  \bibinfo {author} {\bibfnamefont {S.}~\bibnamefont {Rachel}}, \ and\ \bibinfo
  {author} {\bibfnamefont {M.}~\bibnamefont {Vojta}},\ }\href@noop {}
  {\bibfield  {journal} {\bibinfo  {journal} {Physical Review Letters}\
  }\textbf {\bibinfo {volume} {120}},\ \bibinfo {pages} {097204} (\bibinfo
  {year} {2018})}\BibitemShut {NoStop}%
\bibitem [{\citenamefont {M{\'e}lin}\ \emph {et~al.}(2002)\citenamefont
  {M{\'e}lin}, \citenamefont {Lin}, \citenamefont {Lajk{\'o}}, \citenamefont
  {Rieger},\ and\ \citenamefont {Igl{\'o}i}}]{melin2002strongly}%
  \BibitemOpen
  \bibfield  {author} {\bibinfo {author} {\bibfnamefont {R.}~\bibnamefont
  {M{\'e}lin}}, \bibinfo {author} {\bibfnamefont {Y.-C.}\ \bibnamefont {Lin}},
  \bibinfo {author} {\bibfnamefont {P.}~\bibnamefont {Lajk{\'o}}}, \bibinfo
  {author} {\bibfnamefont {H.}~\bibnamefont {Rieger}}, \ and\ \bibinfo {author}
  {\bibfnamefont {F.}~\bibnamefont {Igl{\'o}i}},\ }\href@noop {} {\bibfield
  {journal} {\bibinfo  {journal} {Physical Review B}\ }\textbf {\bibinfo
  {volume} {65}},\ \bibinfo {pages} {104415} (\bibinfo {year}
  {2002})}\BibitemShut {NoStop}%
\bibitem [{\citenamefont {Harris}(1974)}]{harris74}%
  \BibitemOpen
  \bibfield  {author} {\bibinfo {author} {\bibfnamefont {A.}~\bibnamefont
  {Harris}},\ }\href@noop {} {\bibfield  {journal} {\bibinfo  {journal} {J.
  Phys. C: Solid State Phys.}\ }\textbf {\bibinfo {volume} {7}},\ \bibinfo
  {pages} {1671} (\bibinfo {year} {1974})}\BibitemShut {NoStop}%
\bibitem [{\citenamefont {Vojta}(2006)}]{vojta2006rare}%
  \BibitemOpen
  \bibfield  {author} {\bibinfo {author} {\bibfnamefont {T.}~\bibnamefont
  {Vojta}},\ }\href@noop {} {\bibfield  {journal} {\bibinfo  {journal} {Journal
  of Physics A: Mathematical and General}\ }\textbf {\bibinfo {volume} {39}},\
  \bibinfo {pages} {R143} (\bibinfo {year} {2006})}\BibitemShut {NoStop}%
\bibitem [{\citenamefont {Povarov}\ \emph {et~al.}(2015)\citenamefont
  {Povarov}, \citenamefont {Wulf}, \citenamefont {H{\"u}vonen}, \citenamefont
  {Ollivier}, \citenamefont {Paduan-Filho},\ and\ \citenamefont
  {Zheludev}}]{povarov2015dynamics}%
  \BibitemOpen
  \bibfield  {author} {\bibinfo {author} {\bibfnamefont {K.~Y.}\ \bibnamefont
  {Povarov}}, \bibinfo {author} {\bibfnamefont {E.}~\bibnamefont {Wulf}},
  \bibinfo {author} {\bibfnamefont {D.}~\bibnamefont {H{\"u}vonen}}, \bibinfo
  {author} {\bibfnamefont {J.}~\bibnamefont {Ollivier}}, \bibinfo {author}
  {\bibfnamefont {A.}~\bibnamefont {Paduan-Filho}}, \ and\ \bibinfo {author}
  {\bibfnamefont {A.}~\bibnamefont {Zheludev}},\ }\href@noop {} {\bibfield
  {journal} {\bibinfo  {journal} {Physical Review B}\ }\textbf {\bibinfo
  {volume} {92}},\ \bibinfo {pages} {024429} (\bibinfo {year}
  {2015})}\BibitemShut {NoStop}%
\bibitem [{\citenamefont {Smirnov}\ \emph {et~al.}(2017)\citenamefont
  {Smirnov}, \citenamefont {Soldatov}, \citenamefont {Petrenko}, \citenamefont
  {Takata}, \citenamefont {Kida}, \citenamefont {Hagiwara}, \citenamefont
  {Shapiro},\ and\ \citenamefont {Zhitomirsky}}]{smirnov2017order}%
  \BibitemOpen
  \bibfield  {author} {\bibinfo {author} {\bibfnamefont {A.}~\bibnamefont
  {Smirnov}}, \bibinfo {author} {\bibfnamefont {T.}~\bibnamefont {Soldatov}},
  \bibinfo {author} {\bibfnamefont {O.}~\bibnamefont {Petrenko}}, \bibinfo
  {author} {\bibfnamefont {A.}~\bibnamefont {Takata}}, \bibinfo {author}
  {\bibfnamefont {T.}~\bibnamefont {Kida}}, \bibinfo {author} {\bibfnamefont
  {M.}~\bibnamefont {Hagiwara}}, \bibinfo {author} {\bibfnamefont {A.~Y.}\
  \bibnamefont {Shapiro}}, \ and\ \bibinfo {author} {\bibfnamefont
  {M.}~\bibnamefont {Zhitomirsky}},\ }\href@noop {} {\bibfield  {journal}
  {\bibinfo  {journal} {Physical Review Letters}\ }\textbf {\bibinfo {volume}
  {119}},\ \bibinfo {pages} {047204} (\bibinfo {year} {2017})}\BibitemShut
  {NoStop}%
\bibitem [{\citenamefont {Mannig}\ \emph {et~al.}(2018)\citenamefont {Mannig},
  \citenamefont {Povarov}, \citenamefont {Ollivier},\ and\ \citenamefont
  {Zheludev}}]{mannig2018spin}%
  \BibitemOpen
  \bibfield  {author} {\bibinfo {author} {\bibfnamefont {A.}~\bibnamefont
  {Mannig}}, \bibinfo {author} {\bibfnamefont {K.~Y.}\ \bibnamefont {Povarov}},
  \bibinfo {author} {\bibfnamefont {J.}~\bibnamefont {Ollivier}}, \ and\
  \bibinfo {author} {\bibfnamefont {A.}~\bibnamefont {Zheludev}},\ }\href@noop
  {} {\bibfield  {journal} {\bibinfo  {journal} {Physical Review B}\ }\textbf
  {\bibinfo {volume} {98}},\ \bibinfo {pages} {214419} (\bibinfo {year}
  {2018})}\BibitemShut {NoStop}%
\bibitem [{\citenamefont {Igl{\'o}i}\ and\ \citenamefont
  {Monthus}(2018)}]{igloi2018strong}%
  \BibitemOpen
  \bibfield  {author} {\bibinfo {author} {\bibfnamefont {F.}~\bibnamefont
  {Igl{\'o}i}}\ and\ \bibinfo {author} {\bibfnamefont {C.}~\bibnamefont
  {Monthus}},\ }\href@noop {} {\bibfield  {journal} {\bibinfo  {journal} {The
  European Physical Journal B}\ }\textbf {\bibinfo {volume} {91}},\ \bibinfo
  {pages} {290} (\bibinfo {year} {2018})}\BibitemShut {NoStop}%
\bibitem [{\citenamefont {Lin}(2001)}]{lin2001strongly}%
  \BibitemOpen
  \bibfield  {author} {\bibinfo {author} {\bibfnamefont {Y.-C.}\ \bibnamefont
  {Lin}},\ }\href@noop {} {\bibfield  {journal} {\bibinfo  {journal} {PhD
  thesis, Universit{\"a}t zu K{\"o}ln}\ } (\bibinfo {year} {2001})}\BibitemShut
  {NoStop}%
\bibitem [{\citenamefont {Singh}\ and\ \citenamefont
  {Young}(2017)}]{singh2017critical}%
  \BibitemOpen
  \bibfield  {author} {\bibinfo {author} {\bibfnamefont {R.}~\bibnamefont
  {Singh}}\ and\ \bibinfo {author} {\bibfnamefont {A.}~\bibnamefont {Young}},\
  }\href@noop {} {\bibfield  {journal} {\bibinfo  {journal} {Physical Review
  E}\ }\textbf {\bibinfo {volume} {96}},\ \bibinfo {pages} {022139} (\bibinfo
  {year} {2017})}\BibitemShut {NoStop}%
\bibitem [{\citenamefont {Thompson}\ and\ \citenamefont
  {Singh}(2019)}]{thompson2019griffiths}%
  \BibitemOpen
  \bibfield  {author} {\bibinfo {author} {\bibfnamefont {F.}~\bibnamefont
  {Thompson}}\ and\ \bibinfo {author} {\bibfnamefont {R.~R.}\ \bibnamefont
  {Singh}},\ }\href@noop {} {\bibfield  {journal} {\bibinfo  {journal}
  {Physical Review E}\ }\textbf {\bibinfo {volume} {99}},\ \bibinfo {pages}
  {032129} (\bibinfo {year} {2019})}\BibitemShut {NoStop}%
\bibitem [{\citenamefont {Vojta}(2013)}]{vojta2013excitation}%
  \BibitemOpen
  \bibfield  {author} {\bibinfo {author} {\bibfnamefont {M.}~\bibnamefont
  {Vojta}},\ }\href@noop {} {\bibfield  {journal} {\bibinfo  {journal}
  {Physical Review Letters}\ }\textbf {\bibinfo {volume} {111}},\ \bibinfo
  {pages} {097202} (\bibinfo {year} {2013})}\BibitemShut {NoStop}%
\bibitem [{\citenamefont {H{\"o}rmann}\ \emph {et~al.}(2018)\citenamefont
  {H{\"o}rmann}, \citenamefont {Wunderlich},\ and\ \citenamefont
  {Schmidt}}]{hormann2018dynamic}%
  \BibitemOpen
  \bibfield  {author} {\bibinfo {author} {\bibfnamefont {M.}~\bibnamefont
  {H{\"o}rmann}}, \bibinfo {author} {\bibfnamefont {P.}~\bibnamefont
  {Wunderlich}}, \ and\ \bibinfo {author} {\bibfnamefont {K.~P.}\ \bibnamefont
  {Schmidt}},\ }\href@noop {} {\bibfield  {journal} {\bibinfo  {journal}
  {Physical Review Letters}\ }\textbf {\bibinfo {volume} {121}},\ \bibinfo
  {pages} {167201} (\bibinfo {year} {2018})}\BibitemShut {NoStop}%
\bibitem [{\citenamefont {Schmidt}\ and\ \citenamefont
  {Uhrig}(2003)}]{schmidt03}%
  \BibitemOpen
  \bibfield  {author} {\bibinfo {author} {\bibfnamefont {K.~P.}\ \bibnamefont
  {Schmidt}}\ and\ \bibinfo {author} {\bibfnamefont {G.~S.}\ \bibnamefont
  {Uhrig}},\ }\href@noop {} {\bibfield  {journal} {\bibinfo  {journal} {Phys.
  Rev. Lett.}\ }\textbf {\bibinfo {volume} {90}},\ \bibinfo {pages} {227204}
  (\bibinfo {year} {2003})}\BibitemShut {NoStop}%
\bibitem [{\citenamefont {Weihong}(1997)}]{weihong1997various}%
  \BibitemOpen
  \bibfield  {author} {\bibinfo {author} {\bibfnamefont {Z.}~\bibnamefont
  {Weihong}},\ }\href@noop {} {\bibfield  {journal} {\bibinfo  {journal}
  {Physical Review B}\ }\textbf {\bibinfo {volume} {55}},\ \bibinfo {pages}
  {12267} (\bibinfo {year} {1997})}\BibitemShut {NoStop}%
\bibitem [{\citenamefont {Coester}\ \emph {et~al.}(2016)\citenamefont
  {Coester}, \citenamefont {Joshi}, \citenamefont {Vojta},\ and\ \citenamefont
  {Schmidt}}]{coester2016linked}%
  \BibitemOpen
  \bibfield  {author} {\bibinfo {author} {\bibfnamefont {K.}~\bibnamefont
  {Coester}}, \bibinfo {author} {\bibfnamefont {D.~G.}\ \bibnamefont {Joshi}},
  \bibinfo {author} {\bibfnamefont {M.}~\bibnamefont {Vojta}}, \ and\ \bibinfo
  {author} {\bibfnamefont {K.~P.}\ \bibnamefont {Schmidt}},\ }\href@noop {}
  {\bibfield  {journal} {\bibinfo  {journal} {Physical Review B}\ }\textbf
  {\bibinfo {volume} {94}},\ \bibinfo {pages} {125109} (\bibinfo {year}
  {2016})}\BibitemShut {NoStop}%
\bibitem [{\citenamefont {Wenzel}\ \emph {et~al.}(2008)\citenamefont {Wenzel},
  \citenamefont {Bogacz},\ and\ \citenamefont {Janke}}]{wenzel08}%
  \BibitemOpen
  \bibfield  {author} {\bibinfo {author} {\bibfnamefont {S.}~\bibnamefont
  {Wenzel}}, \bibinfo {author} {\bibfnamefont {L.}~\bibnamefont {Bogacz}}, \
  and\ \bibinfo {author} {\bibfnamefont {W.}~\bibnamefont {Janke}},\
  }\href@noop {} {\bibfield  {journal} {\bibinfo  {journal} {Phys. Rev. Lett.}\
  }\textbf {\bibinfo {volume} {101}},\ \bibinfo {pages} {127202} (\bibinfo
  {year} {2008})}\BibitemShut {NoStop}%
\bibitem [{\citenamefont {Sknepnek}\ \emph {et~al.}(2004)\citenamefont
  {Sknepnek}, \citenamefont {Vojta},\ and\ \citenamefont
  {Vojta}}]{sknepnek2004exotic}%
  \BibitemOpen
  \bibfield  {author} {\bibinfo {author} {\bibfnamefont {R.}~\bibnamefont
  {Sknepnek}}, \bibinfo {author} {\bibfnamefont {T.}~\bibnamefont {Vojta}}, \
  and\ \bibinfo {author} {\bibfnamefont {M.}~\bibnamefont {Vojta}},\
  }\href@noop {} {\bibfield  {journal} {\bibinfo  {journal} {Physical Review
  Letters}\ }\textbf {\bibinfo {volume} {93}},\ \bibinfo {pages} {097201}
  (\bibinfo {year} {2004})}\BibitemShut {NoStop}%
\bibitem [{\citenamefont {Sandvik}(2006)}]{sandvik2006quantum}%
  \BibitemOpen
  \bibfield  {author} {\bibinfo {author} {\bibfnamefont {A.~W.}\ \bibnamefont
  {Sandvik}},\ }\href@noop {} {\bibfield  {journal} {\bibinfo  {journal}
  {Physical Review Letters}\ }\textbf {\bibinfo {volume} {96}},\ \bibinfo
  {pages} {207201} (\bibinfo {year} {2006})}\BibitemShut {NoStop}%
\bibitem [{\citenamefont {Vajk}\ and\ \citenamefont
  {Greven}(2002)}]{vajk2002quantum}%
  \BibitemOpen
  \bibfield  {author} {\bibinfo {author} {\bibfnamefont {O.}~\bibnamefont
  {Vajk}}\ and\ \bibinfo {author} {\bibfnamefont {M.}~\bibnamefont {Greven}},\
  }\href@noop {} {\bibfield  {journal} {\bibinfo  {journal} {Physical Review
  Letters}\ }\textbf {\bibinfo {volume} {89}},\ \bibinfo {pages} {177202}
  (\bibinfo {year} {2002})}\BibitemShut {NoStop}%
\bibitem [{\citenamefont {Puschmann}\ \emph {et~al.}(2019)\citenamefont
  {Puschmann}, \citenamefont {Crewse}, \citenamefont {Hoyos},\ and\
  \citenamefont {Vojta}}]{puschmann2019collective}%
  \BibitemOpen
  \bibfield  {author} {\bibinfo {author} {\bibfnamefont {M.}~\bibnamefont
  {Puschmann}}, \bibinfo {author} {\bibfnamefont {J.}~\bibnamefont {Crewse}},
  \bibinfo {author} {\bibfnamefont {J.~A.}\ \bibnamefont {Hoyos}}, \ and\
  \bibinfo {author} {\bibfnamefont {T.}~\bibnamefont {Vojta}},\ }\href@noop {}
  {\bibfield  {journal} {\bibinfo  {journal} {arXiv preprint arXiv:1911.04452}\
  } (\bibinfo {year} {2019})}\BibitemShut {NoStop}%
\bibitem [{\citenamefont {Singh}\ and\ \citenamefont
  {Elstner}(1998)}]{singh1998quantum}%
  \BibitemOpen
  \bibfield  {author} {\bibinfo {author} {\bibfnamefont {R.~R.}\ \bibnamefont
  {Singh}}\ and\ \bibinfo {author} {\bibfnamefont {N.}~\bibnamefont
  {Elstner}},\ }\href@noop {} {\bibfield  {journal} {\bibinfo  {journal}
  {Physical Review Letters}\ }\textbf {\bibinfo {volume} {81}},\ \bibinfo
  {pages} {4732} (\bibinfo {year} {1998})}\BibitemShut {NoStop}%
\bibitem [{\citenamefont {Kawamura}(1992)}]{Kawamura_1992}%
  \BibitemOpen
  \bibfield  {author} {\bibinfo {author} {\bibfnamefont {H.}~\bibnamefont
  {Kawamura}},\ }\href@noop {} {\bibfield  {journal} {\bibinfo  {journal}
  {Journal of the Physical Society of Japan}\ }\textbf {\bibinfo {volume}
  {61}},\ \bibinfo {pages} {1299} (\bibinfo {year} {1992})}\BibitemShut
  {NoStop}%
\bibitem [{\citenamefont {Singh}\ and\ \citenamefont
  {Huse}(2007)}]{singh2007ground}%
  \BibitemOpen
  \bibfield  {author} {\bibinfo {author} {\bibfnamefont {R.~R.}\ \bibnamefont
  {Singh}}\ and\ \bibinfo {author} {\bibfnamefont {D.~A.}\ \bibnamefont
  {Huse}},\ }\href@noop {} {\bibfield  {journal} {\bibinfo  {journal} {Physical
  Review B}\ }\textbf {\bibinfo {volume} {76}},\ \bibinfo {pages} {180407}
  (\bibinfo {year} {2007})}\BibitemShut {NoStop}%
\bibitem [{\citenamefont {Evenbly}\ and\ \citenamefont
  {Vidal}(2010)}]{evenbly2010frustrated}%
  \BibitemOpen
  \bibfield  {author} {\bibinfo {author} {\bibfnamefont {G.}~\bibnamefont
  {Evenbly}}\ and\ \bibinfo {author} {\bibfnamefont {G.}~\bibnamefont
  {Vidal}},\ }\href@noop {} {\bibfield  {journal} {\bibinfo  {journal}
  {Physical Review Letters}\ }\textbf {\bibinfo {volume} {104}},\ \bibinfo
  {pages} {187203} (\bibinfo {year} {2010})}\BibitemShut {NoStop}%
\bibitem [{\citenamefont {Depenbrock}\ \emph {et~al.}(2012)\citenamefont
  {Depenbrock}, \citenamefont {McCulloch},\ and\ \citenamefont
  {Schollw{\"o}ck}}]{depenbrock2012nature}%
  \BibitemOpen
  \bibfield  {author} {\bibinfo {author} {\bibfnamefont {S.}~\bibnamefont
  {Depenbrock}}, \bibinfo {author} {\bibfnamefont {I.~P.}\ \bibnamefont
  {McCulloch}}, \ and\ \bibinfo {author} {\bibfnamefont {U.}~\bibnamefont
  {Schollw{\"o}ck}},\ }\href@noop {} {\bibfield  {journal} {\bibinfo  {journal}
  {Physical Review Letters}\ }\textbf {\bibinfo {volume} {109}},\ \bibinfo
  {pages} {067201} (\bibinfo {year} {2012})}\BibitemShut {NoStop}%
\bibitem [{\citenamefont {Knetter}\ and\ \citenamefont
  {Uhrig}(2000)}]{knetter2000perturbation}%
  \BibitemOpen
  \bibfield  {author} {\bibinfo {author} {\bibfnamefont {C.}~\bibnamefont
  {Knetter}}\ and\ \bibinfo {author} {\bibfnamefont {G.~S.}\ \bibnamefont
  {Uhrig}},\ }\href@noop {} {\bibfield  {journal} {\bibinfo  {journal} {The
  European Physical Journal B-Condensed Matter and Complex Systems}\ }\textbf
  {\bibinfo {volume} {13}},\ \bibinfo {pages} {209} (\bibinfo {year}
  {2000})}\BibitemShut {NoStop}%
\bibitem [{\citenamefont {Knetter}\ \emph {et~al.}(2003)\citenamefont
  {Knetter}, \citenamefont {Schmidt},\ and\ \citenamefont
  {Uhrig}}]{knetter2003structure}%
  \BibitemOpen
  \bibfield  {author} {\bibinfo {author} {\bibfnamefont {C.}~\bibnamefont
  {Knetter}}, \bibinfo {author} {\bibfnamefont {K.~P.}\ \bibnamefont
  {Schmidt}}, \ and\ \bibinfo {author} {\bibfnamefont {G.~S.}\ \bibnamefont
  {Uhrig}},\ }\href@noop {} {\bibfield  {journal} {\bibinfo  {journal} {Journal
  of Physics A: Mathematical and General}\ }\textbf {\bibinfo {volume} {36}},\
  \bibinfo {pages} {7889} (\bibinfo {year} {2003})}\BibitemShut {NoStop}%
\bibitem [{\citenamefont {Coester}\ and\ \citenamefont
  {Schmidt}(2015)}]{coester2015optimizing}%
  \BibitemOpen
  \bibfield  {author} {\bibinfo {author} {\bibfnamefont {K.}~\bibnamefont
  {Coester}}\ and\ \bibinfo {author} {\bibfnamefont {K.}~\bibnamefont
  {Schmidt}},\ }\href@noop {} {\bibfield  {journal} {\bibinfo  {journal}
  {Physical Review E}\ }\textbf {\bibinfo {volume} {92}},\ \bibinfo {pages}
  {022118} (\bibinfo {year} {2015})}\BibitemShut {NoStop}%
\bibitem [{\citenamefont {Siek}\ \emph {et~al.}(2000)\citenamefont {Siek},
  \citenamefont {Lee},\ and\ \citenamefont {Lumsdaine}}]{boost_graph}%
  \BibitemOpen
  \bibfield  {author} {\bibinfo {author} {\bibfnamefont {J.}~\bibnamefont
  {Siek}}, \bibinfo {author} {\bibfnamefont {L.-Q.}\ \bibnamefont {Lee}}, \
  and\ \bibinfo {author} {\bibfnamefont {A.}~\bibnamefont {Lumsdaine}},\
  }\href@noop {} {\enquote {\bibinfo {title} {Boost graph library},}\ }\bibinfo
  {howpublished} {http://www.boost.org/libs/graph/} (\bibinfo {year}
  {2000})\BibitemShut {NoStop}%
\bibitem [{\citenamefont {Eilmes}\ \emph {et~al.}(1998)\citenamefont {Eilmes},
  \citenamefont {R{\"o}mer},\ and\ \citenamefont {Schreiber}}]{eilmes1998two}%
  \BibitemOpen
  \bibfield  {author} {\bibinfo {author} {\bibfnamefont {A.}~\bibnamefont
  {Eilmes}}, \bibinfo {author} {\bibfnamefont {R.~A.}\ \bibnamefont
  {R{\"o}mer}}, \ and\ \bibinfo {author} {\bibfnamefont {M.}~\bibnamefont
  {Schreiber}},\ }\href@noop {} {\bibfield  {journal} {\bibinfo  {journal} {The
  European Physical Journal B-Condensed Matter and Complex Systems}\ }\textbf
  {\bibinfo {volume} {1}},\ \bibinfo {pages} {29} (\bibinfo {year}
  {1998})}\BibitemShut {NoStop}%
\bibitem [{\citenamefont {Theodorou}\ and\ \citenamefont
  {Cohen}(1976)}]{theodorou1976extended}%
  \BibitemOpen
  \bibfield  {author} {\bibinfo {author} {\bibfnamefont {G.}~\bibnamefont
  {Theodorou}}\ and\ \bibinfo {author} {\bibfnamefont {M.~H.}\ \bibnamefont
  {Cohen}},\ }\href@noop {} {\bibfield  {journal} {\bibinfo  {journal}
  {Physical Review B}\ }\textbf {\bibinfo {volume} {13}},\ \bibinfo {pages}
  {4597} (\bibinfo {year} {1976})}\BibitemShut {NoStop}%
\bibitem [{\citenamefont {Derrida}\ and\ \citenamefont
  {Stauffer}(1985)}]{derrida1985corrections}%
  \BibitemOpen
  \bibfield  {author} {\bibinfo {author} {\bibfnamefont {B.}~\bibnamefont
  {Derrida}}\ and\ \bibinfo {author} {\bibfnamefont {D.}~\bibnamefont
  {Stauffer}},\ }\href@noop {} {\bibfield  {journal} {\bibinfo  {journal}
  {Journal de Physique}\ }\textbf {\bibinfo {volume} {46}},\ \bibinfo {pages}
  {1623} (\bibinfo {year} {1985})}\BibitemShut {NoStop}%
\bibitem [{\citenamefont {Powalski}\ \emph {et~al.}(2013)\citenamefont
  {Powalski}, \citenamefont {Coester}, \citenamefont {Moessner},\ and\
  \citenamefont {Schmidt}}]{powalski2013disorder}%
  \BibitemOpen
  \bibfield  {author} {\bibinfo {author} {\bibfnamefont {M.}~\bibnamefont
  {Powalski}}, \bibinfo {author} {\bibfnamefont {K.}~\bibnamefont {Coester}},
  \bibinfo {author} {\bibfnamefont {R.}~\bibnamefont {Moessner}}, \ and\
  \bibinfo {author} {\bibfnamefont {K.}~\bibnamefont {Schmidt}},\ }\href@noop
  {} {\bibfield  {journal} {\bibinfo  {journal} {Physical Review B}\ }\textbf
  {\bibinfo {volume} {87}},\ \bibinfo {pages} {054404} (\bibinfo {year}
  {2013})}\BibitemShut {NoStop}%
\bibitem [{\citenamefont {Sykes}\ and\ \citenamefont
  {Essam}(1964)}]{sykes1964exact}%
  \BibitemOpen
  \bibfield  {author} {\bibinfo {author} {\bibfnamefont {M.~F.}\ \bibnamefont
  {Sykes}}\ and\ \bibinfo {author} {\bibfnamefont {J.~W.}\ \bibnamefont
  {Essam}},\ }\href@noop {} {\bibfield  {journal} {\bibinfo  {journal} {Journal
  of Mathematical Physics}\ }\textbf {\bibinfo {volume} {5}},\ \bibinfo {pages}
  {1117} (\bibinfo {year} {1964})}\BibitemShut {NoStop}%
\end{thebibliography}

%

\clearpage

\begin{appendix}

\section*{Appendix}
In Appendix A the densities of states for the sorts of disorder considered in the main body of the paper are given. In Appendix B results for the inverse participation ration (IPR) and the finite-size scaling of the DSF are discussed. Additional figures for the symmetric and anti-symmetric DSF not contained in the main text are depicted in Appendix C.

\section{Density of states}
The non-normalized density of states is defined as

\begin{equation}
\mathrm{DOS}_{nn}(\omega) = \sum_{n,r} \delta(\omega-\omega_n),
\end{equation} 
where $n$ runs over the energies of one sample and $r$ over the number of samples. With this the density of states (DOS) is defined as 
\begin{equation}
\mathrm{DOS}(\omega) = \frac{1}{\int d\omega\,\mathrm{DOS}_{nn}(\omega)}\,\mathrm{DOS}_{nn}(\omega).
\end{equation}
\par
\begin{figure}
	\centering
	\begin{subfigure}[b]{.9\columnwidth}
		\caption{}		
		\includegraphics[width=1\linewidth]{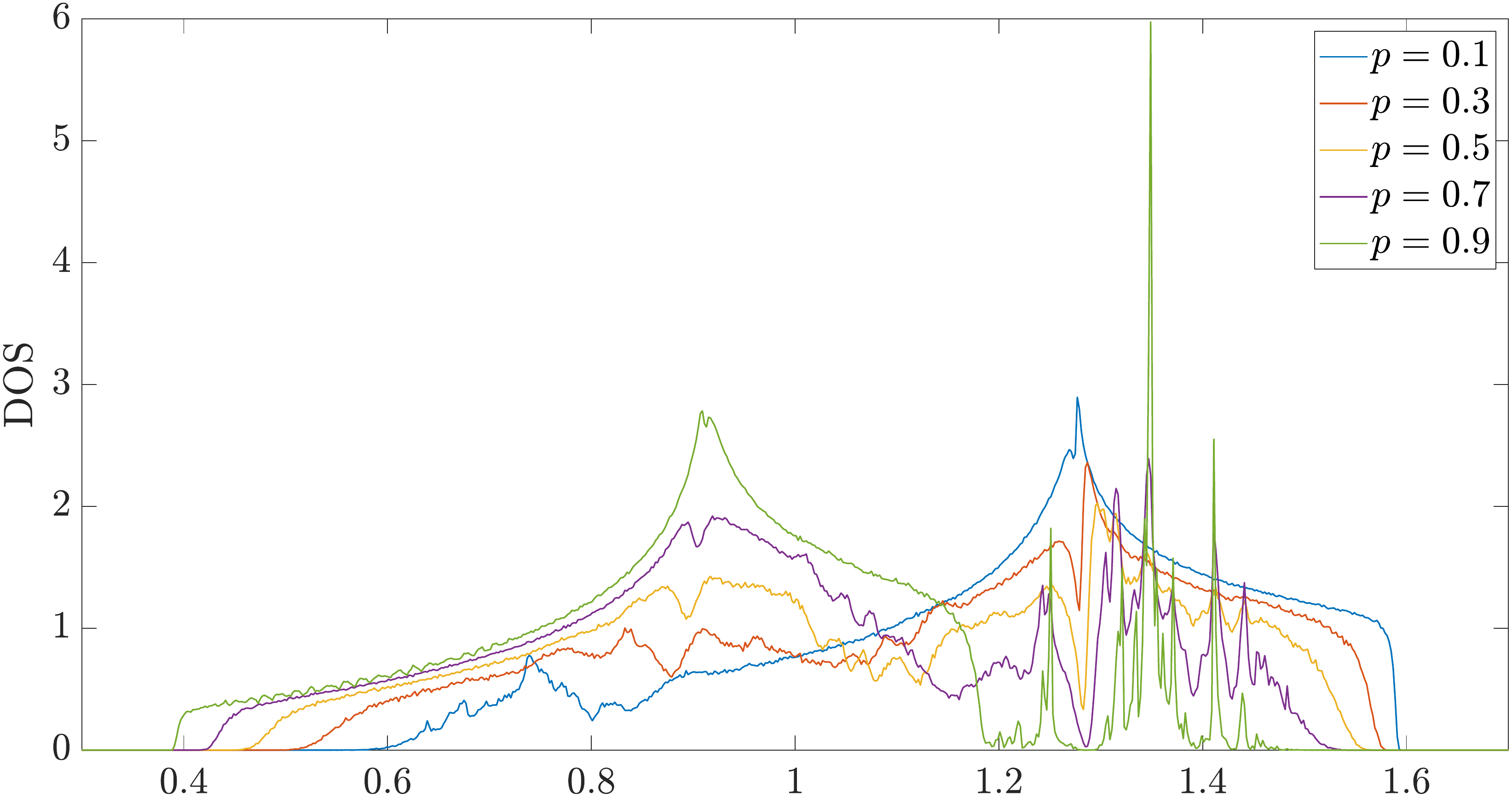}
		\label{fig:DOS_Square_Rung} 
	\end{subfigure}
	
	\begin{subfigure}[b]{.9\columnwidth}
		\caption{}		
		\includegraphics[width=1\linewidth]{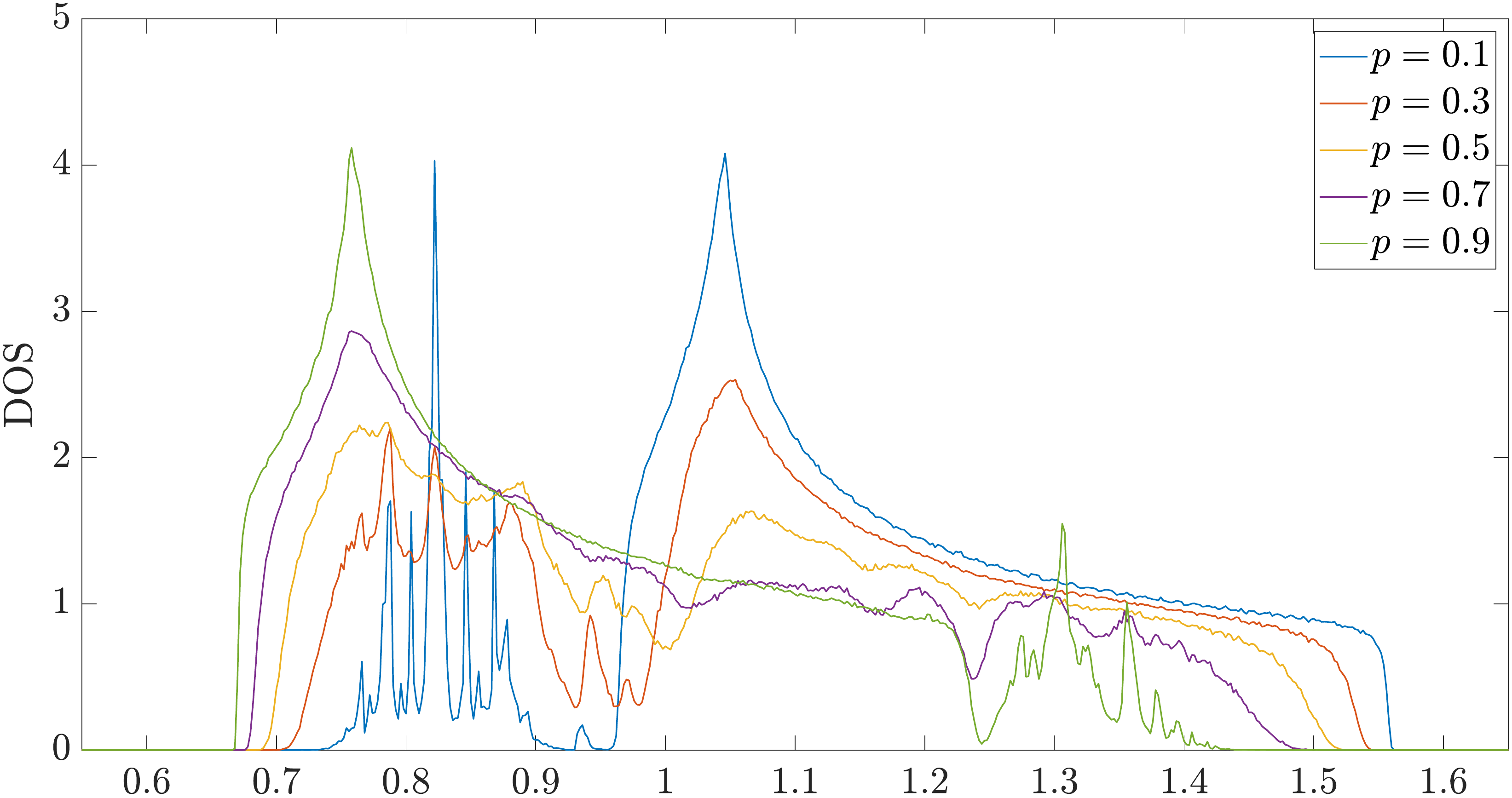}
		\label{fig:DOS_Triangular_Rung}
	\end{subfigure}
	
	\begin{subfigure}[b]{.9\columnwidth}
		\caption{}		
		\includegraphics[width=1\linewidth]{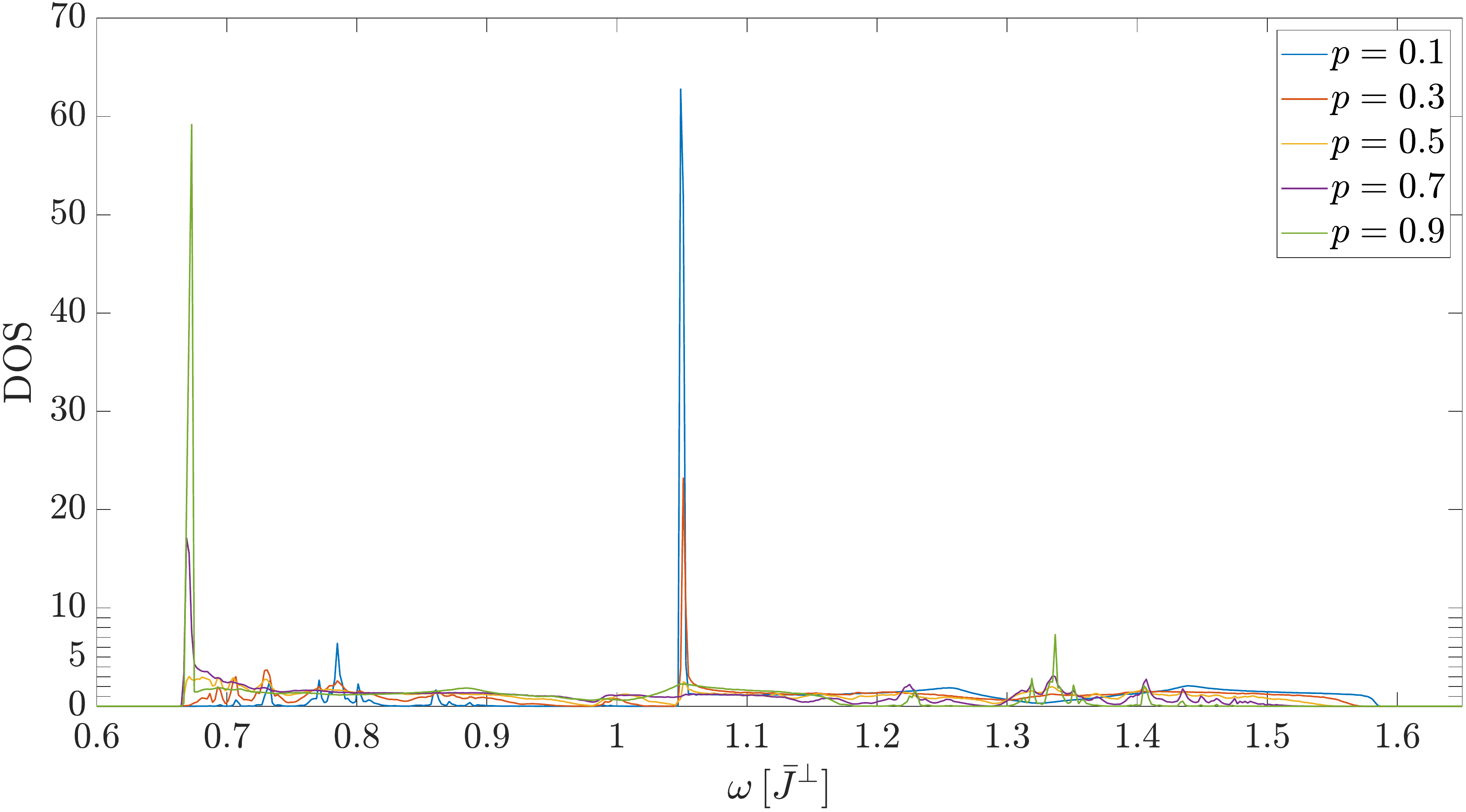}
		\label{fig:DOS_kagome_Rung}
	\end{subfigure}

	\caption[]{
		The DOS is shown for bimodal intra-dimer disorder, $p=0.1,0.3,0.5,0.7,0.9$ and the corresponding disorder configurations of the main body of the paper in (a) for the square, in (b) for the triangular and  in (c) for the kagome lattice.
	}
	\label{fig:Rung_DOS}
\end{figure}
\begin{figure}
	\centering
	\begin{subfigure}[b]{.9\columnwidth}
		\caption{}		
		\includegraphics[width=1\linewidth]{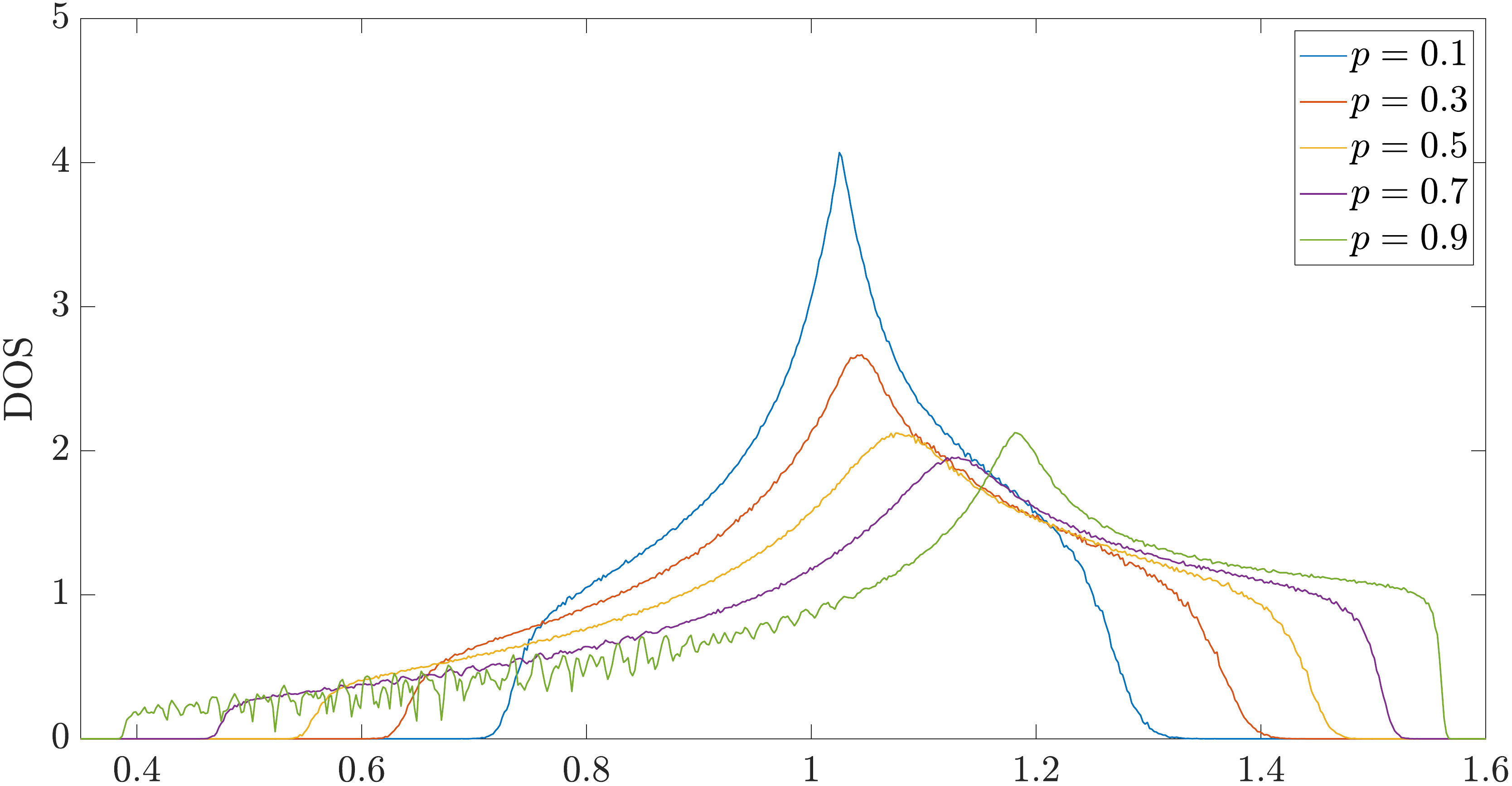}
		\label{fig:DOS_Square_Leg} 
	\end{subfigure}
	
	\begin{subfigure}[b]{.9\columnwidth}
		\caption{}		
		\includegraphics[width=1\linewidth]{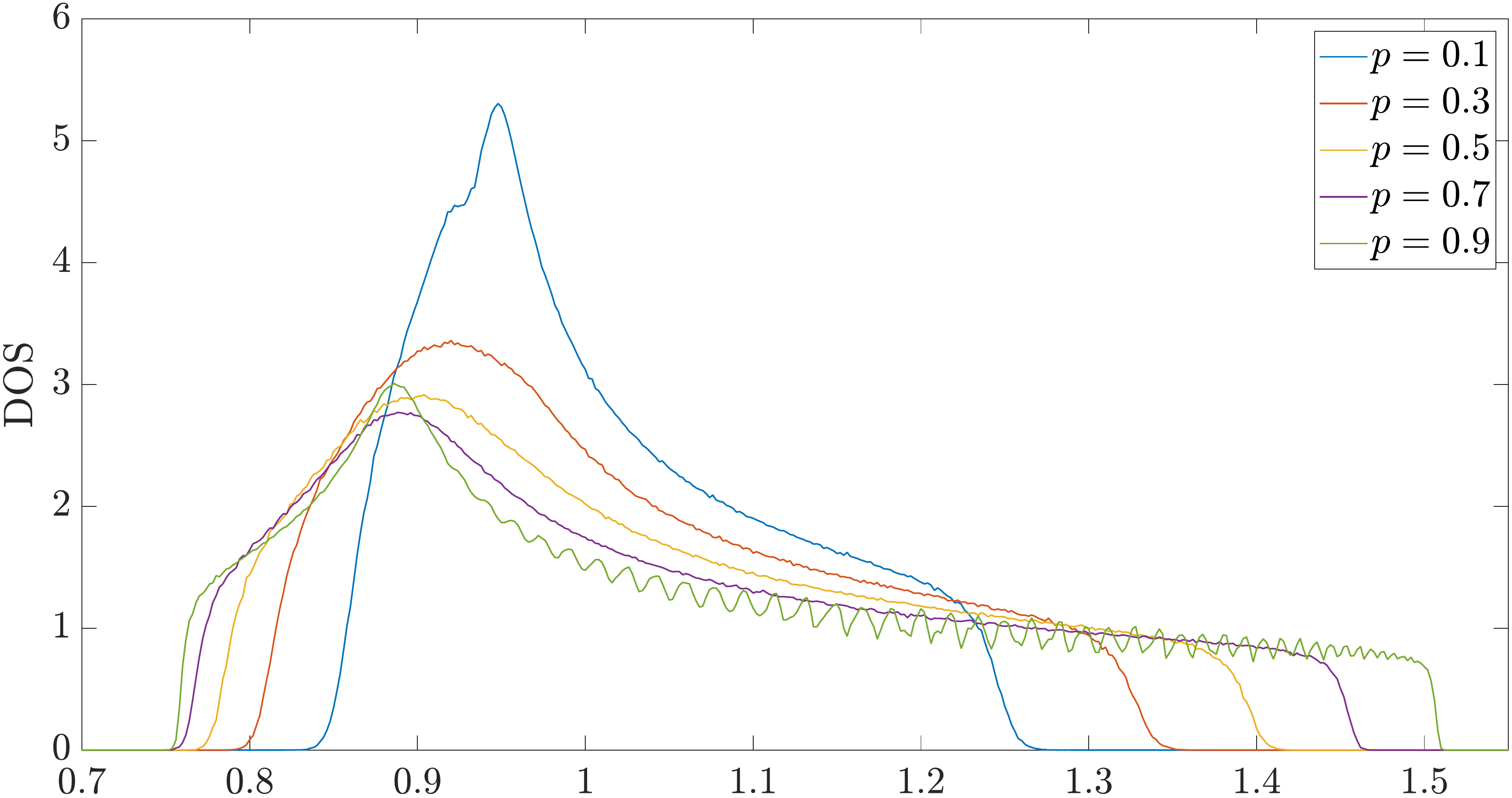}
		\label{fig:DOS_Triangular_Leg}
	\end{subfigure}
	
	\begin{subfigure}[b]{.9\columnwidth}
		\caption{}		
		\includegraphics[width=1\linewidth]{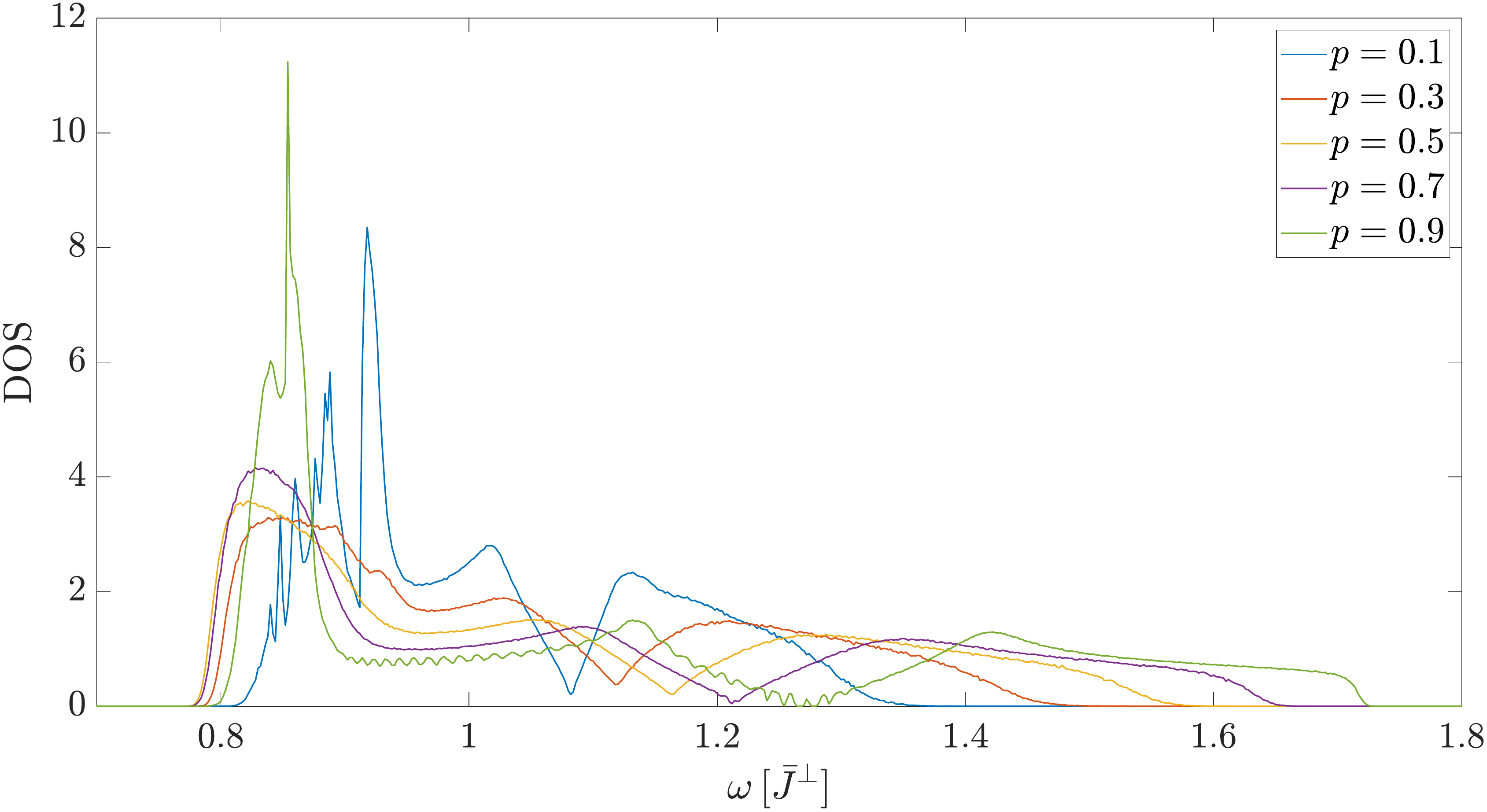}
		\label{fig:DOS_kagome_Leg}
	\end{subfigure}

	\caption[]{
		The DOS is shown for bimodal inter-dimer disorder, $p=0.1,0.3,0.5,0.7,0.9$ and the corresponding disorder configurations of the main body of the paper in (a) for the square, in (b) for the triangular and  in (c) for the kagome lattice.
	}
	\label{fig:Leg_DOS}
\end{figure}

In Fig.~\ref{fig:Rung_DOS} it is shown for the intra-dimer disorder and in Fig.~\ref{fig:Leg_DOS} for the inter-dimer disorder of the main body of the paper. As already mentioned in the main body of the paper one can see roughly peaked structures for intra-dimer disorder in the energy regions of the intra-dimer value that is taken with a small enough probability compared to the percolation threshold of the lattice. In the kagome lattice one furthermore finds peaks in the DOS that belong to the states originating from the flat band.\par
One also finds these peaks for inter-dimer disorder. Apart from this and in contrast to intra-dimer disorder, the DOS is smooth for inter-dimer disorder.

\section{Inverse participation ratio and finite-size scaling}
The inverse participation ratio ($\mathrm{IPR}$)
\begin{equation}
\mathrm{IPR} = 1/\sum_{\nu} \left\vert \Braket{n\vert\nu}\right\vert^4,
\end{equation}
with $\Ket{\nu}$ denoting position states, is a simple and intuitive measure for the localization length of a normalized eigenfunction $\ket{n}$. Suppose $\ket{n}$ is a plane wave eigenstate of a periodic one-dimensional chain with one atom in the unit cell. Then the $\mathrm{IPR}$ will be $\mathcal{N}_{d}$. For all other extended states the $\mathrm{IPR}$ will be smaller but always remain $\propto \mathcal{N}_{d}$. For a perfectly localized eigenstate with $\Ket{n}=\Ket{\nu}$ the value of the $\mathrm{IPR}$ is one. An exponentially localized state has a finite $\mathrm{IPR}$ for sufficiently large system sizes.\par
\begin{figure}
	\centering
	\begin{subfigure}[b]{.9\columnwidth}
		\caption{}		
		\includegraphics[width=1\linewidth]{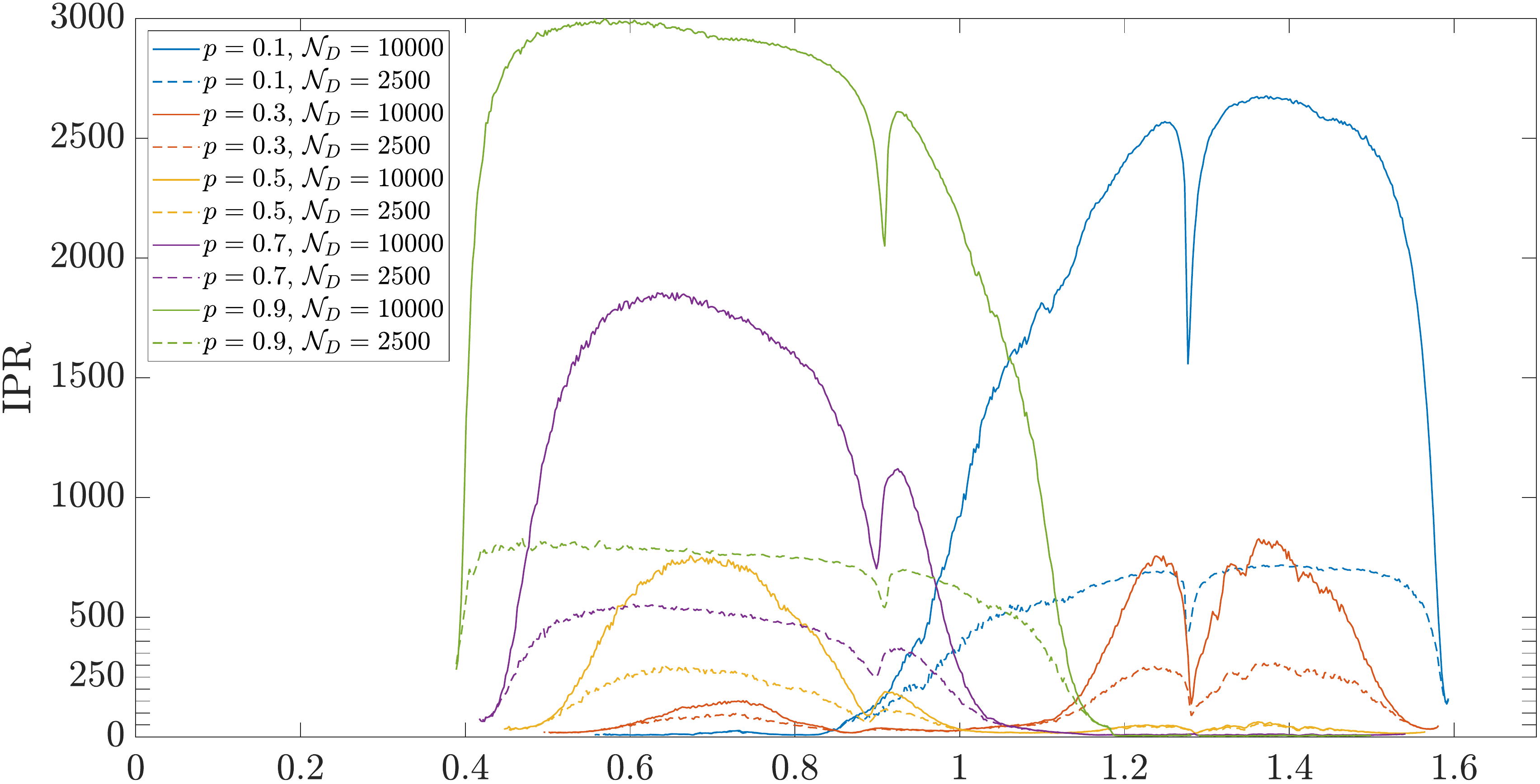}
		\label{fig:IPR_Square_Rung} 
	\end{subfigure}
	
	\begin{subfigure}[b]{.9\columnwidth}
		\caption{}		
		\includegraphics[width=1\linewidth]{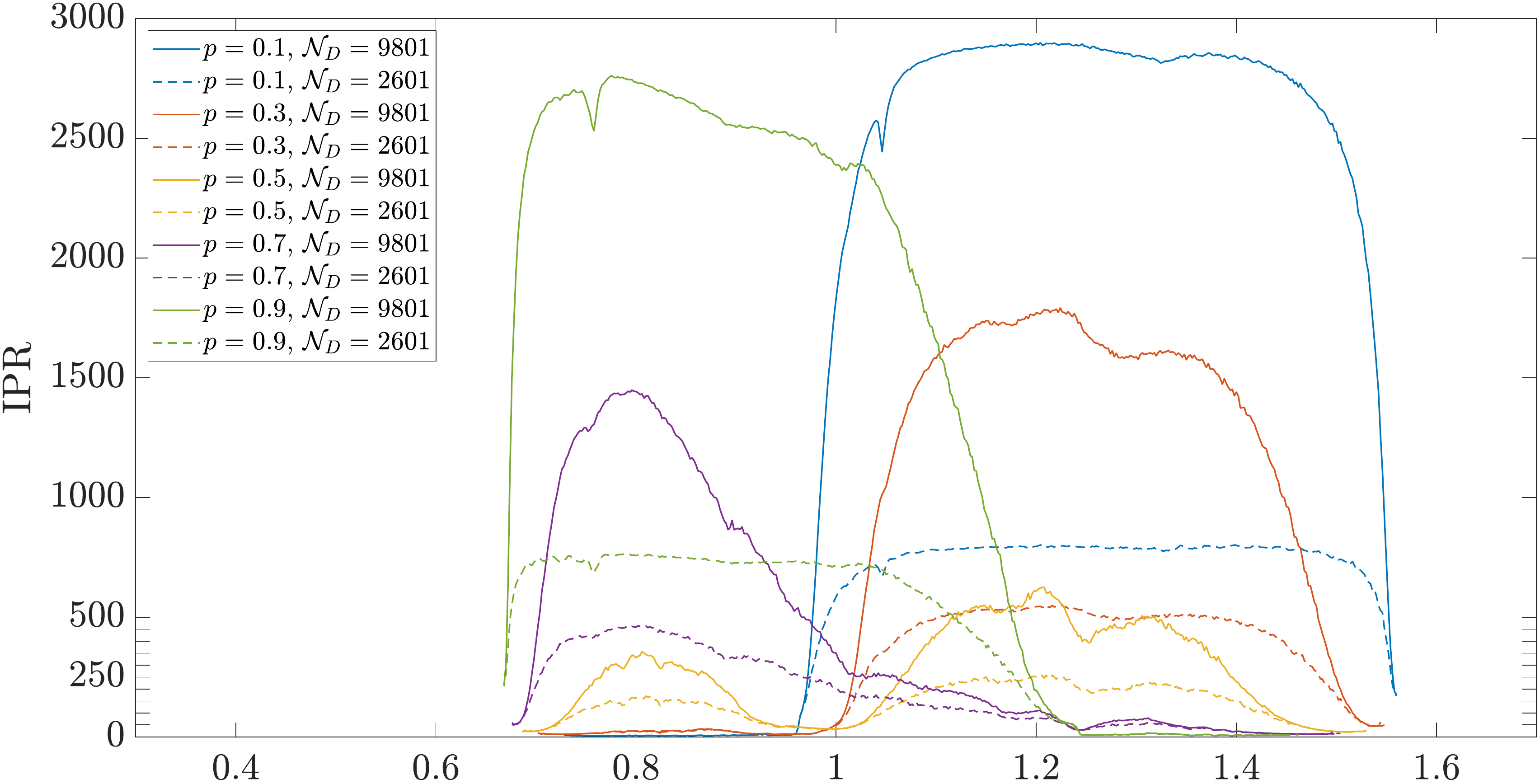}
		\label{fig:IPR_Triangular_Rung}
	\end{subfigure}
	
	\begin{subfigure}[b]{.9\columnwidth}
		\caption{}		
		\includegraphics[width=1\linewidth]{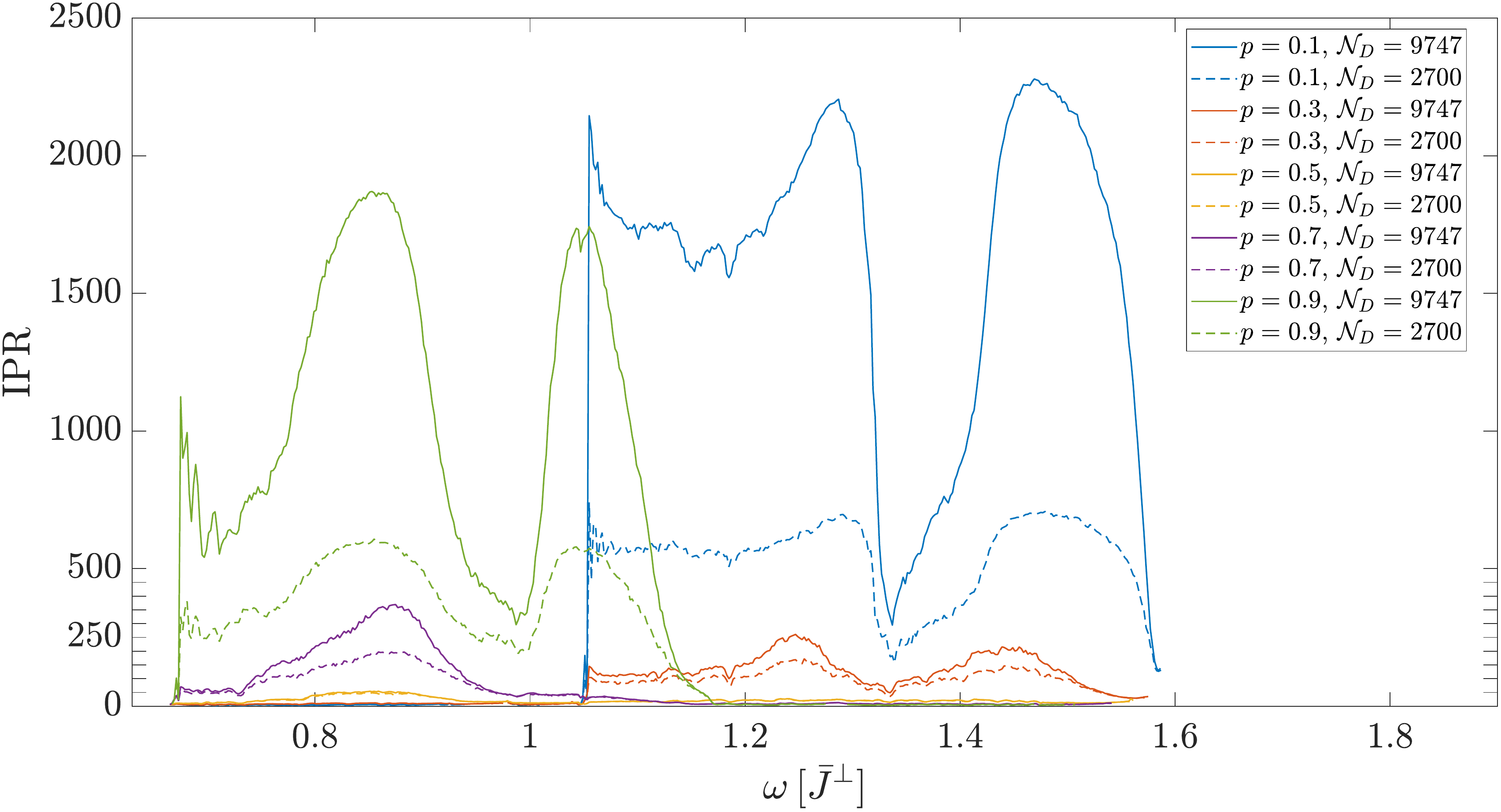}
		\label{fig:IPR_kagome_Rung}
	\end{subfigure}

	\caption[]{
		The IPR is shown for bimodal intra-dimer disorder, $p=0.1,0.3,0.5,0.7,0.9$ and the corresponding disorder configurations of the main body of the paper in (a) for the square, in (b) for the triangular and  in (c) for the kagome lattice.
	}
	\label{fig:Rung_IPR}
\end{figure}
\begin{figure}
	\centering
	\begin{subfigure}[b]{.9\columnwidth}
		\caption{}		
		\includegraphics[width=1\linewidth]{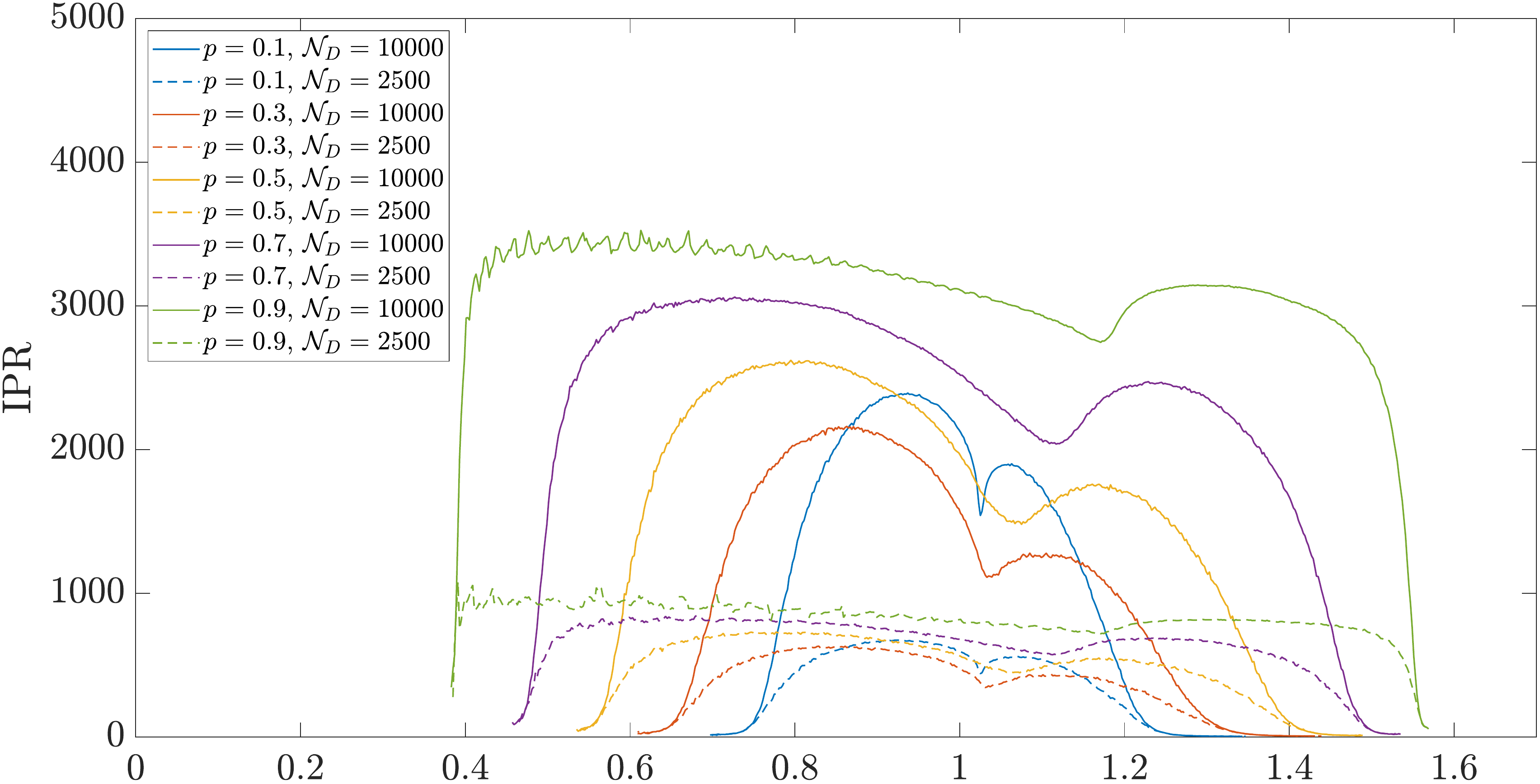}
		\label{fig:IPR_Square_Leg} 
	\end{subfigure}
	
	\begin{subfigure}[b]{.9\columnwidth}
		\caption{}		
		\includegraphics[width=1\linewidth]{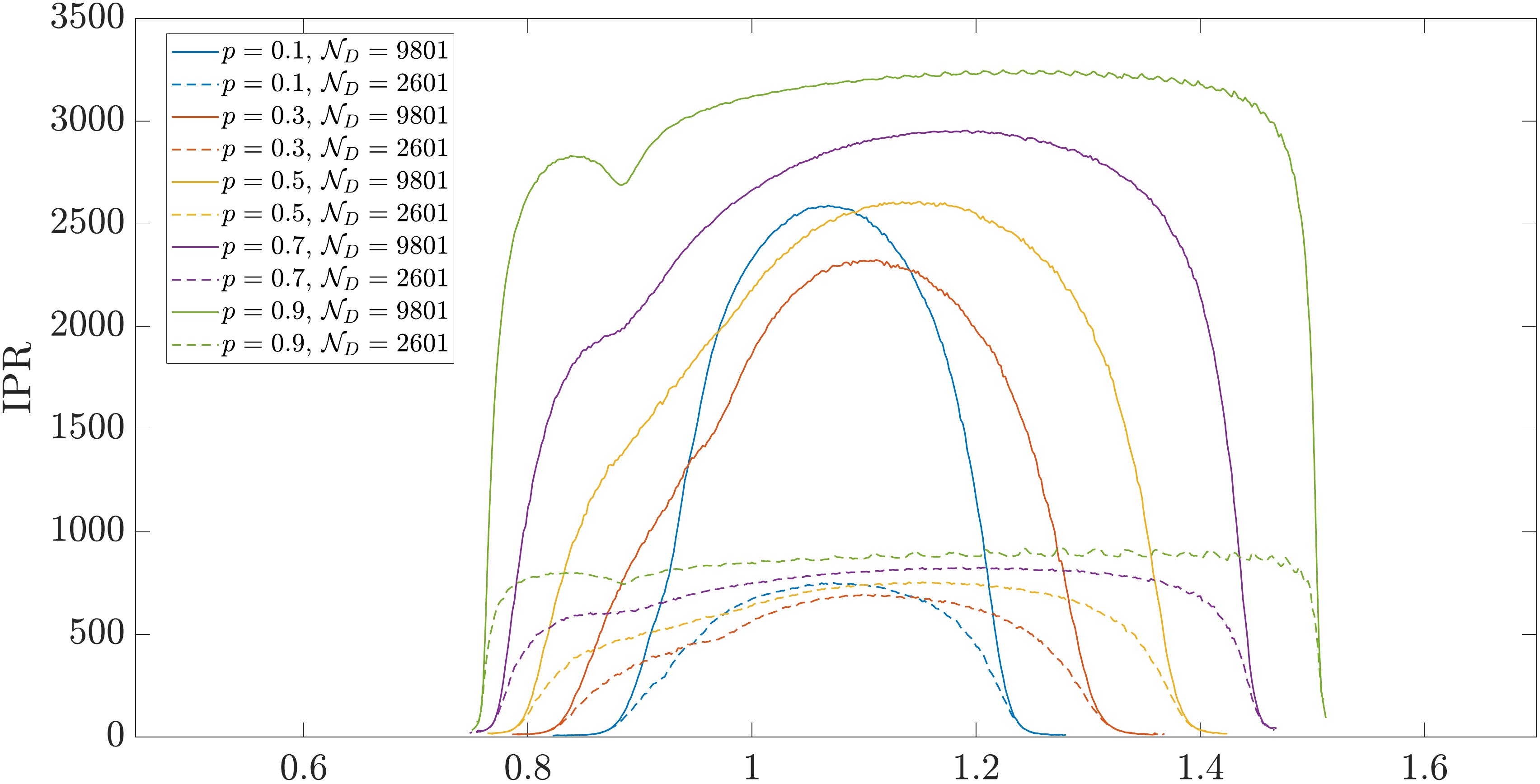}
		\label{fig:IPR_Triangular_Leg}
	\end{subfigure}
	
	\begin{subfigure}[b]{.9\columnwidth}
		\caption{}		
		\includegraphics[width=1\linewidth]{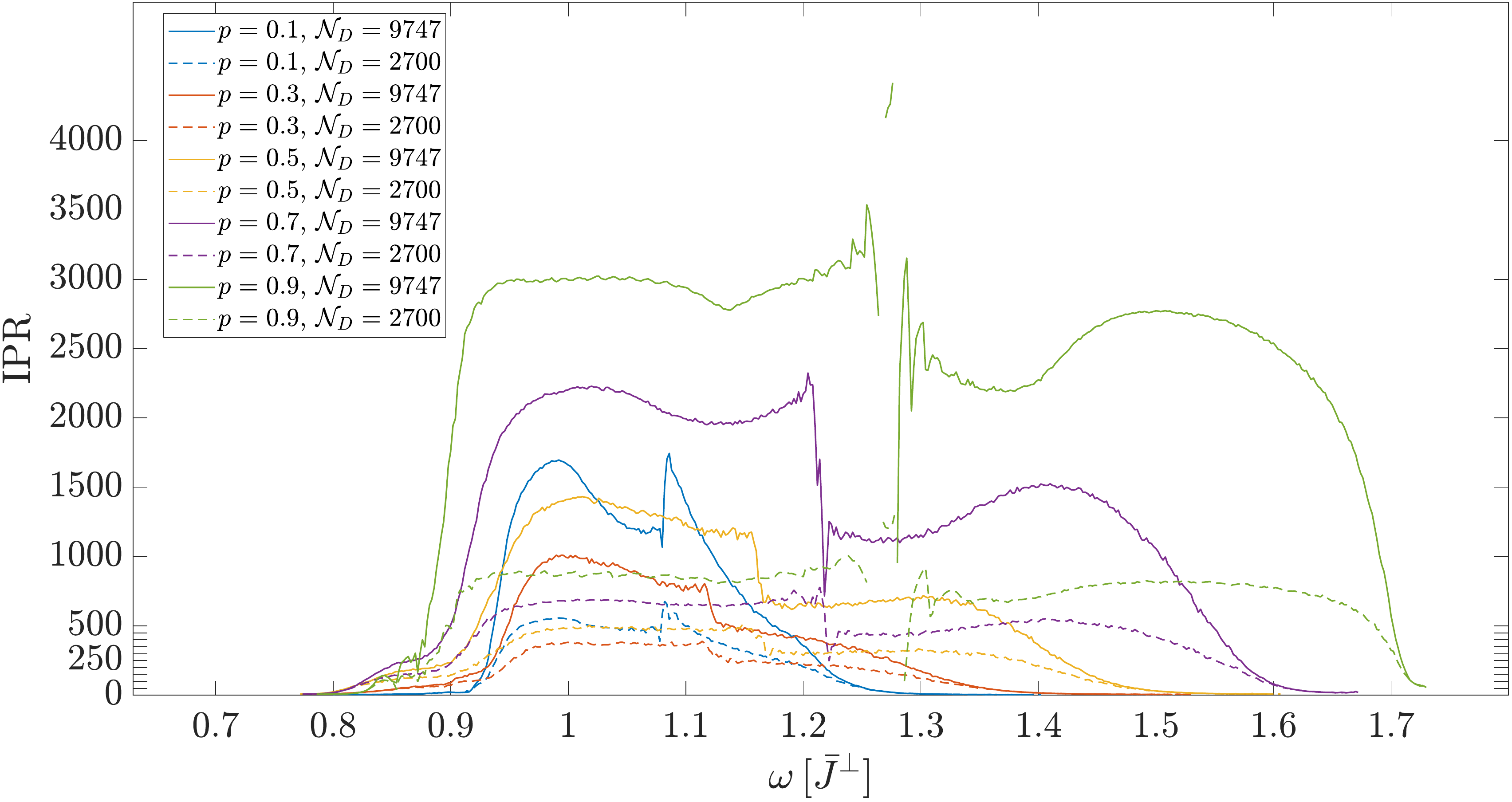}
		\label{fig:IPR_kagome_Leg}
	\end{subfigure}
	\caption[]{
		The IPR is shown for bimodal inter-dimer disorder, $p=0.1,0.3,0.5,0.7,0.9$ and the corresponding disorder configurations of the main body of the paper in (a) for the square, in (b) for the triangular and  in (c) for the kagome lattice.
	}
	\label{fig:Leg_IPR}
\end{figure}

In Fig.~\ref{fig:Rung_IPR} the IPR is shown for the intra-dimer disorder and in Fig.~\ref{fig:Leg_IPR} for the inter-dimer disorder of the main body of the paper. To analyze the scaling behavior two different system sizes where compared. In the square lattice these were $\mathcal{N}_{\rm d}=10000$ and $\mathcal{N}_{\rm d}=2500$, in the triangular lattice $\mathcal{N}_{\rm d}=9801$ and $\mathcal{N}_{\rm d}=2601$ and in the kagome lattice $\mathcal{N}_{\rm d}=9747$ and $\mathcal{N}_{\rm d}=2700$.\par
For intra-dimer disorder the IPR shows smaller values in the non-percolating regions and for $p=0.5$ in the region with the smaller effective ratio between inter- and intra-dimer coupling. In the other regions the IPR still shows scaling of extended stats. The reason for that is that the localization length in these regions is not yet small compared to the system size. The IPR for highest and smallest energies always has a comparatively low value because the energy eigenstates of these energies have a lot of weight on either consecutive $J^\perp_{1}$- or $J^\perp_{2}$-dimers. This means for larger and larger systems the IPR should actually start to rise again for the extremal energies. Because our system size is too small to cover this we only see minima of the IPR at these energies yet. In the kagome lattice it is remarkable how small the IPR is for $p=0.5$ compared with the other values of $p$. In the energy regions of the flat band also much smaller values of the IPR are seen compared to those of the dispersive bands.\par
For inter-dimer disorder in the square and triangular lattice the IPR shows scaling of extended states except for the extremal energies. The reason for that is again that the system size is not yet large enough compared to the localization lenghts. In the kagome lattice the situation is similar except for the low-energy regions linked to the flat band. Towards this region we see that the difference in the IPR of the smaller and larger system gets smaller and smaller. \par
\begin{figure}
	\centering
	\begin{subfigure}[b]{.9\columnwidth}
		\caption{}		
		\includegraphics[width=1\linewidth]{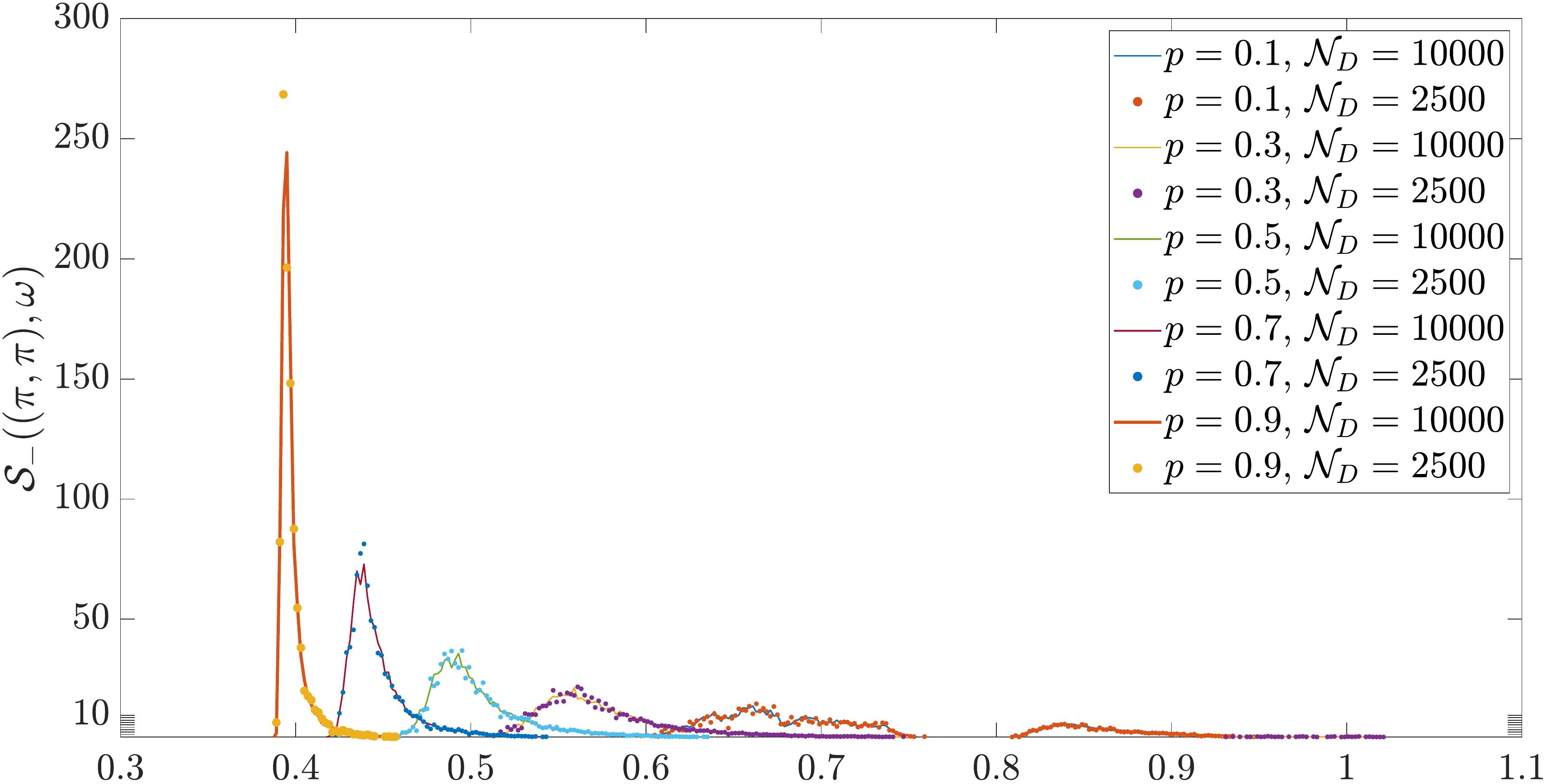}
		\label{fig:Ng1} 
	\end{subfigure}
	
	\begin{subfigure}[b]{.9\columnwidth}
		\caption{}		
		\includegraphics[width=1\linewidth]{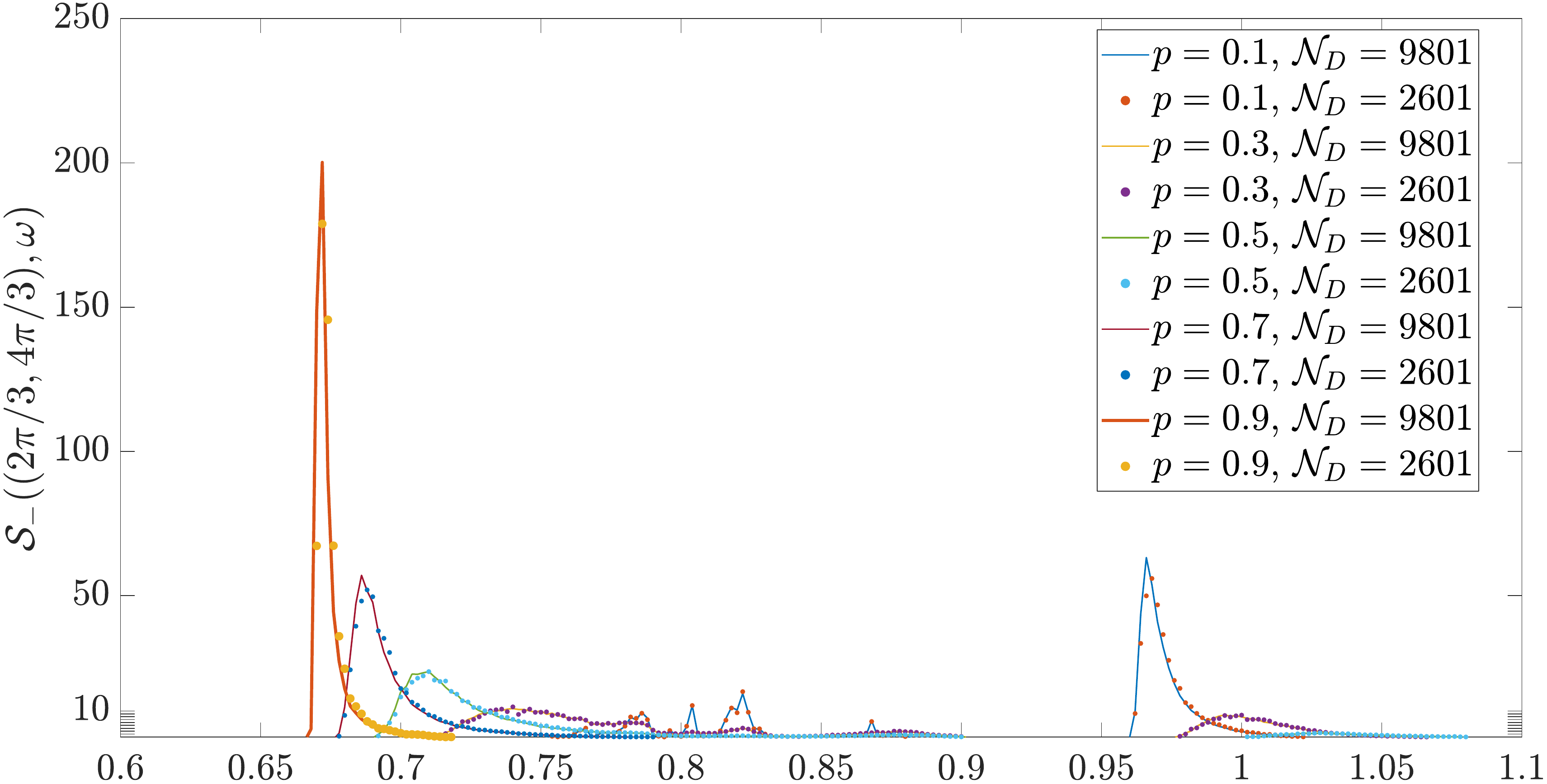}
		\label{fig:Ng2}
	\end{subfigure}
	
	\begin{subfigure}[b]{.9\columnwidth}
		\caption{}		
		\includegraphics[width=1\linewidth]{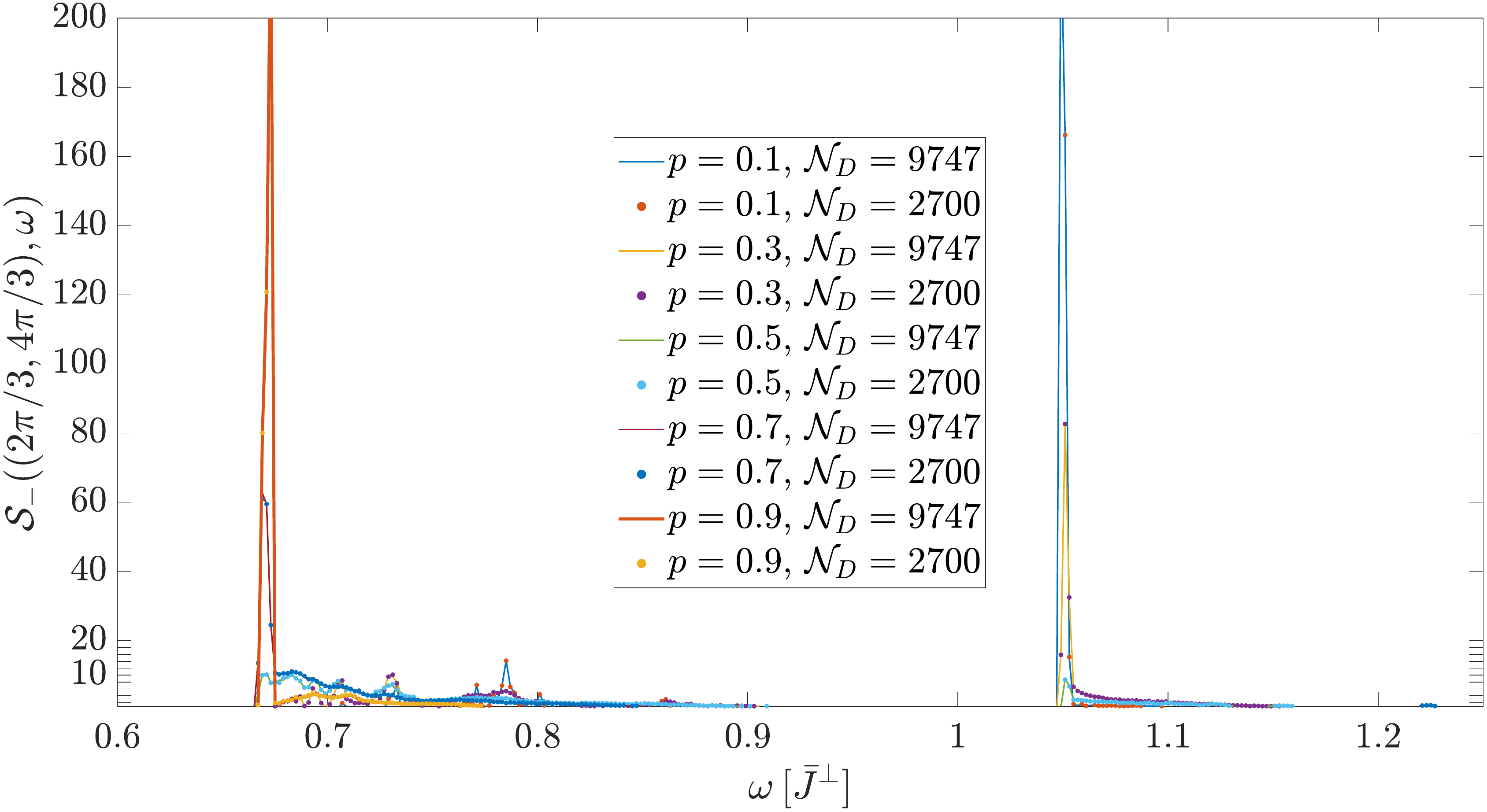}
		\label{fig:Ng2}
	\end{subfigure}
	
	\caption[]{
		The DSF $\mathcal{S}_-(k,\omega)$ and its finite-size scaling behaviour is shown for bimodal intra-dimer disorder, $p=0.1,0.3,0.5,0.7,0.9$ and the corresponding disorder configurations of the main body of the paper in (a) for the square, in (b) for the triangular lattice and in (c) for the kagome lattice. The momentum $\vec{k}$ is the gap momentum in the corresponding lattices, i.e. $\vec{k}=(\pi,\pi)$ for the square, $\vec{k}=(2\pi/3,4\pi/3)$ for the triangular and $\vec{k}=(2\pi/3,4\pi/3)$ for the kagome lattice.}
	\label{fig:Rung_FiniteSize}
\end{figure}
\begin{figure}
	\centering
	\begin{subfigure}[b]{.9\columnwidth}
		\caption{}		
		\includegraphics[width=1\linewidth]{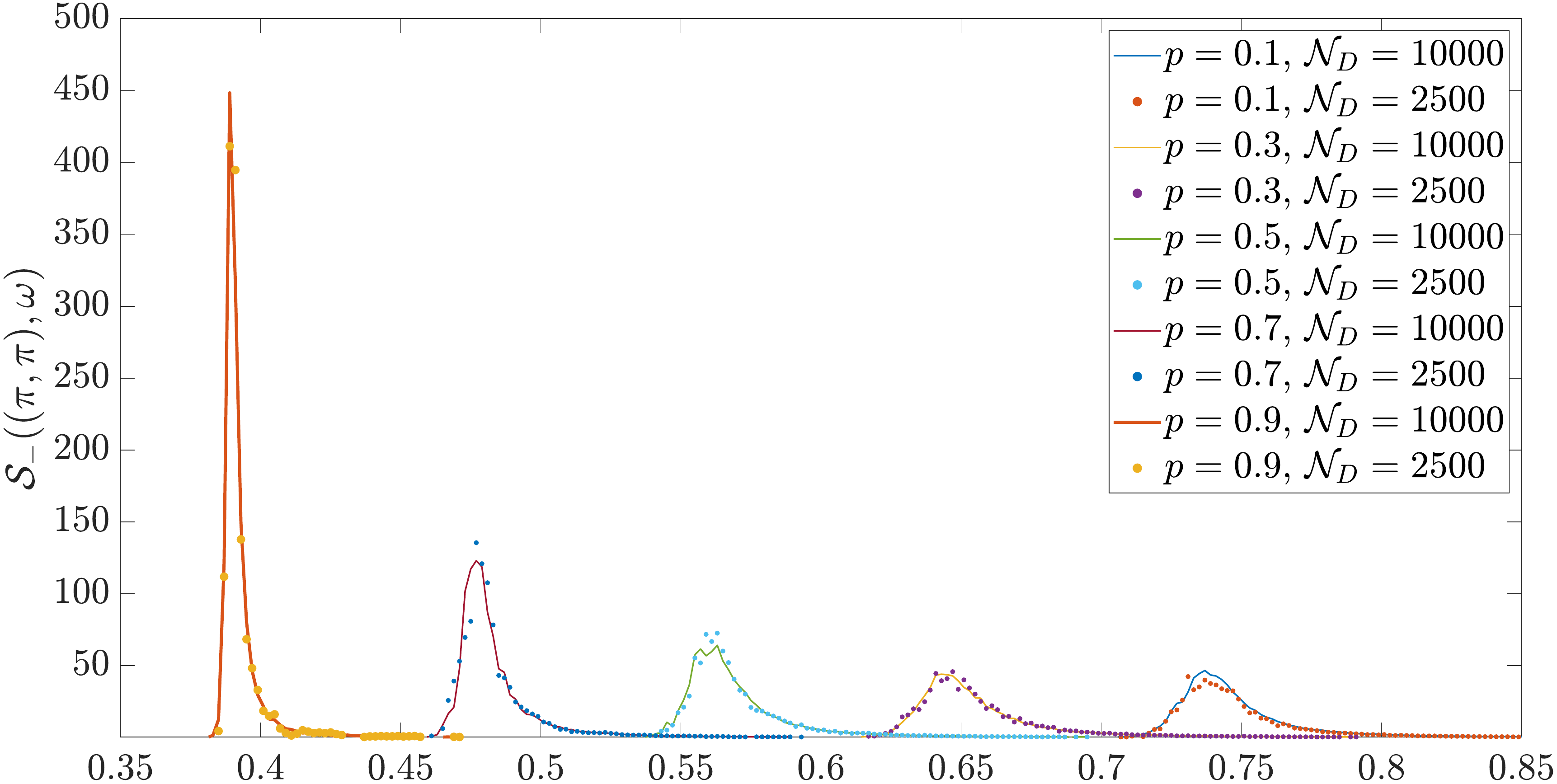}
		\label{fig:Ng1} 
	\end{subfigure}
	
	\begin{subfigure}[b]{.9\columnwidth}
		\caption{}		
		\includegraphics[width=1\linewidth]{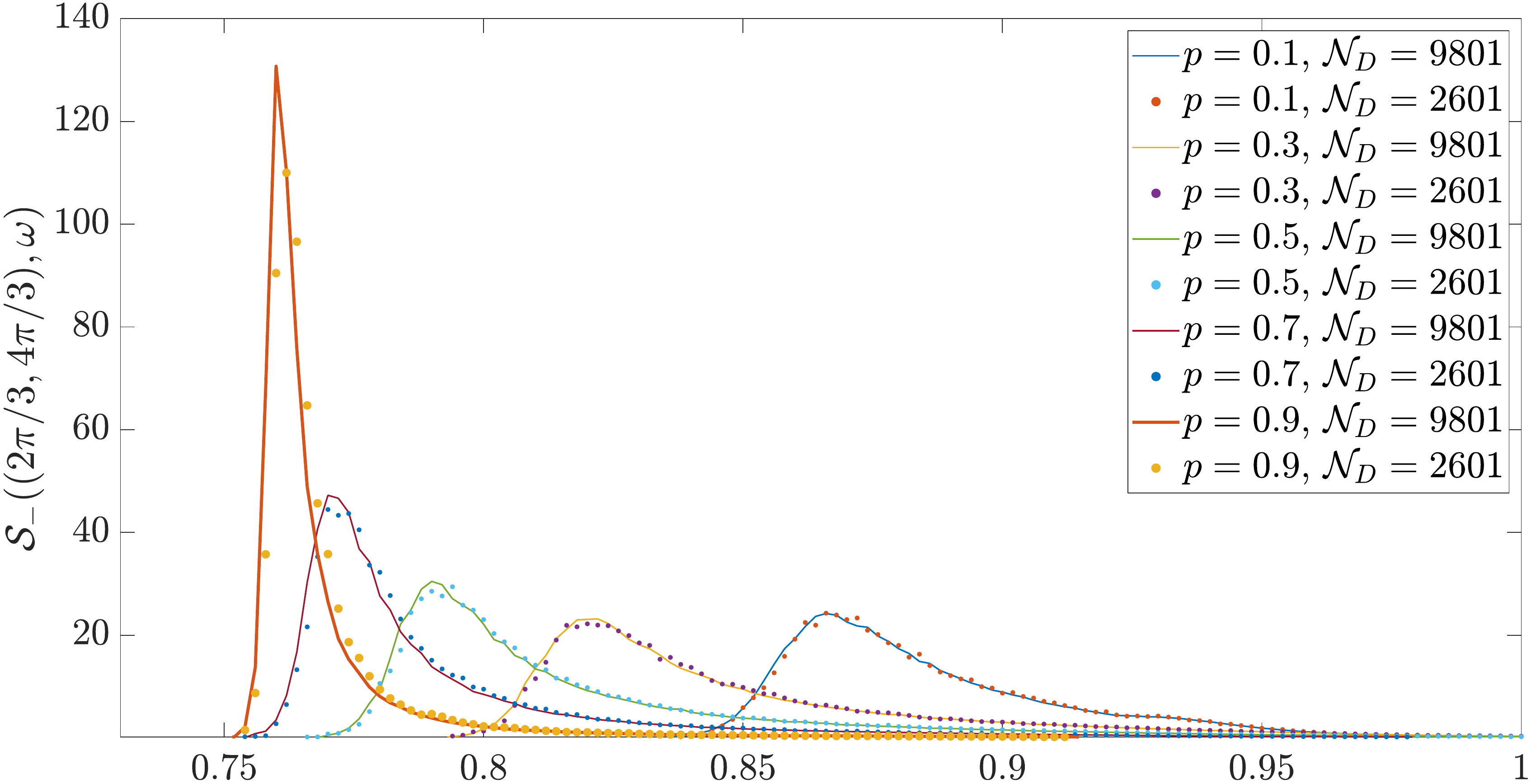}
		\label{fig:Ng2}
	\end{subfigure}
	
	\begin{subfigure}[b]{.9\columnwidth}
		\caption{}		
		\includegraphics[width=1\linewidth]{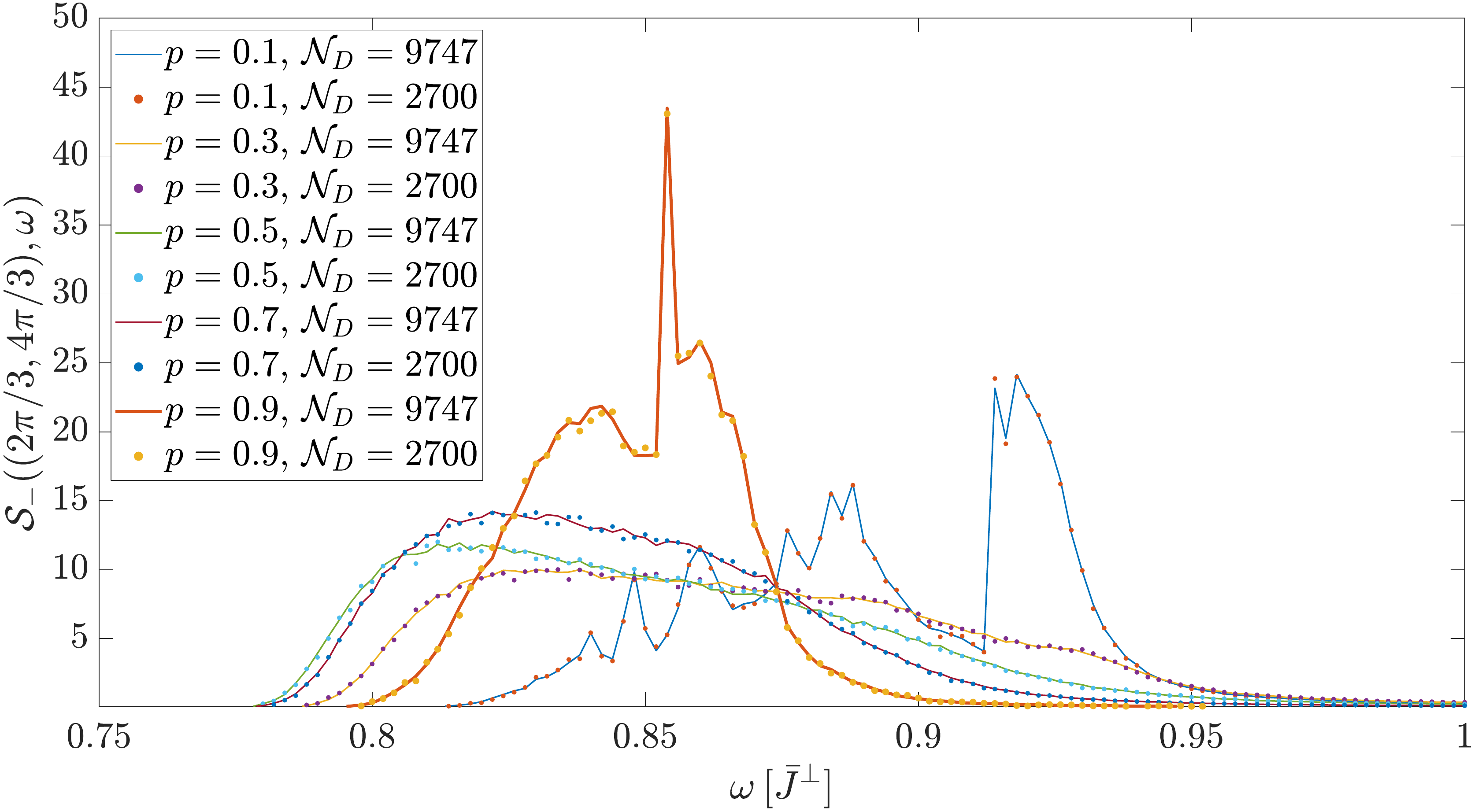}
		\label{fig:Ng2}
	\end{subfigure}
	
	\caption[]{
		The DSF $\mathcal{S}_-(k,\omega)$ and its finite-size scaling behaviour is shown for bimodal inter-dimer disorder, $p=0.1,0.3,0.5,0.7,0.9$ and the corresponding disorder configurations of the main body of the paper in (a) for the square, in (b) for the triangular lattice and in (c) for the kagome lattice. The momentum $\vec{k}$ is the gap momentum in the corresponding lattices, i.e. $\vec{k}=(\pi,\pi)$ for the square, $\vec{k}=(2\pi/3,4\pi/3)$ for the triangular and $\vec{k}=(2\pi/3,4\pi/3)$ for the kagome lattice.}
	\label{fig:LegMinus_FiniteSize}
\end{figure}
\begin{figure}
	\centering
	\begin{subfigure}[b]{.9\columnwidth}
		\caption{}		
		\includegraphics[width=1\linewidth]{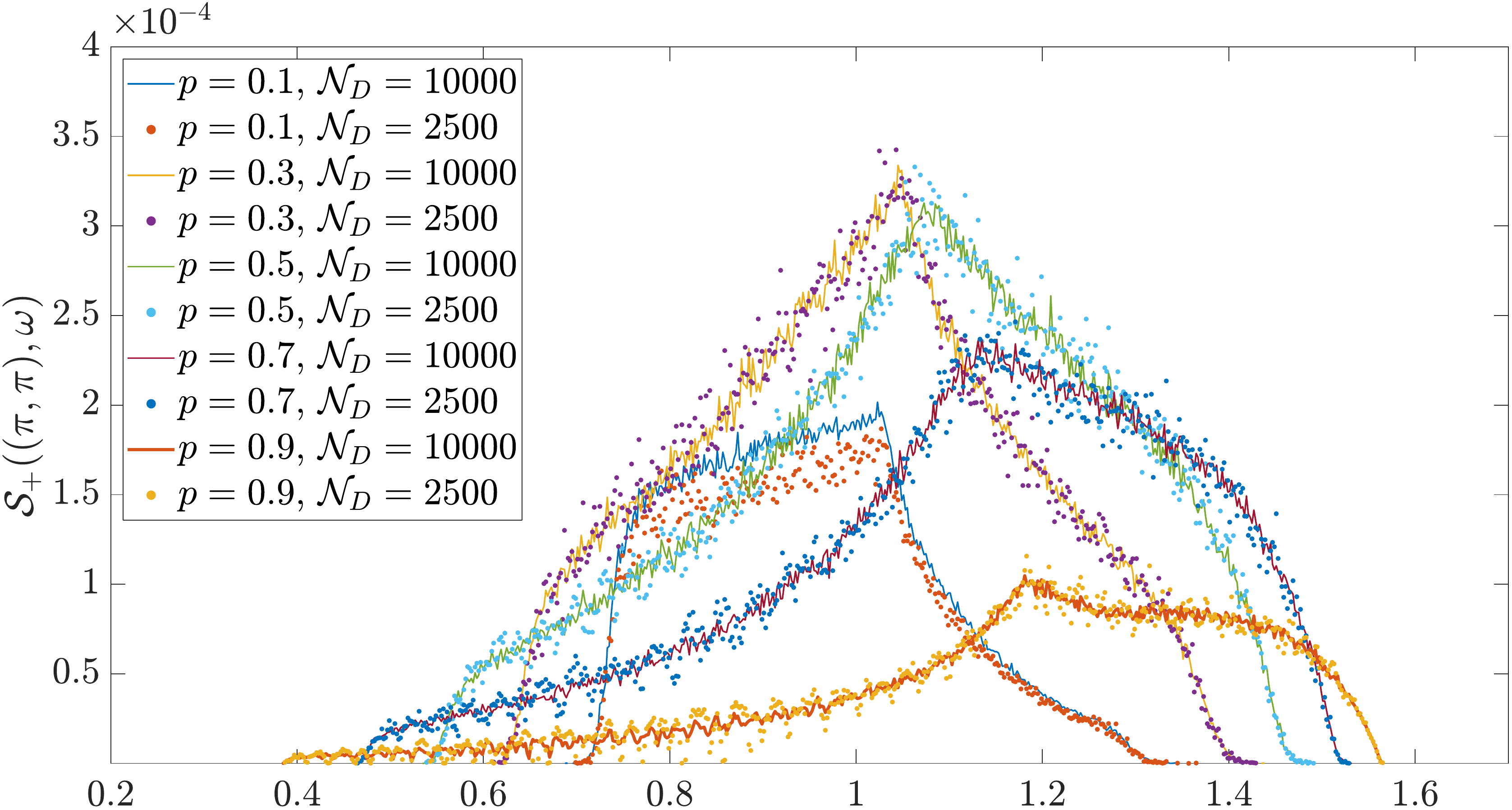}
		\label{fig:Ng1} 
	\end{subfigure}
	
	\begin{subfigure}[b]{.9\columnwidth}
		\caption{}		
		\includegraphics[width=1\linewidth]{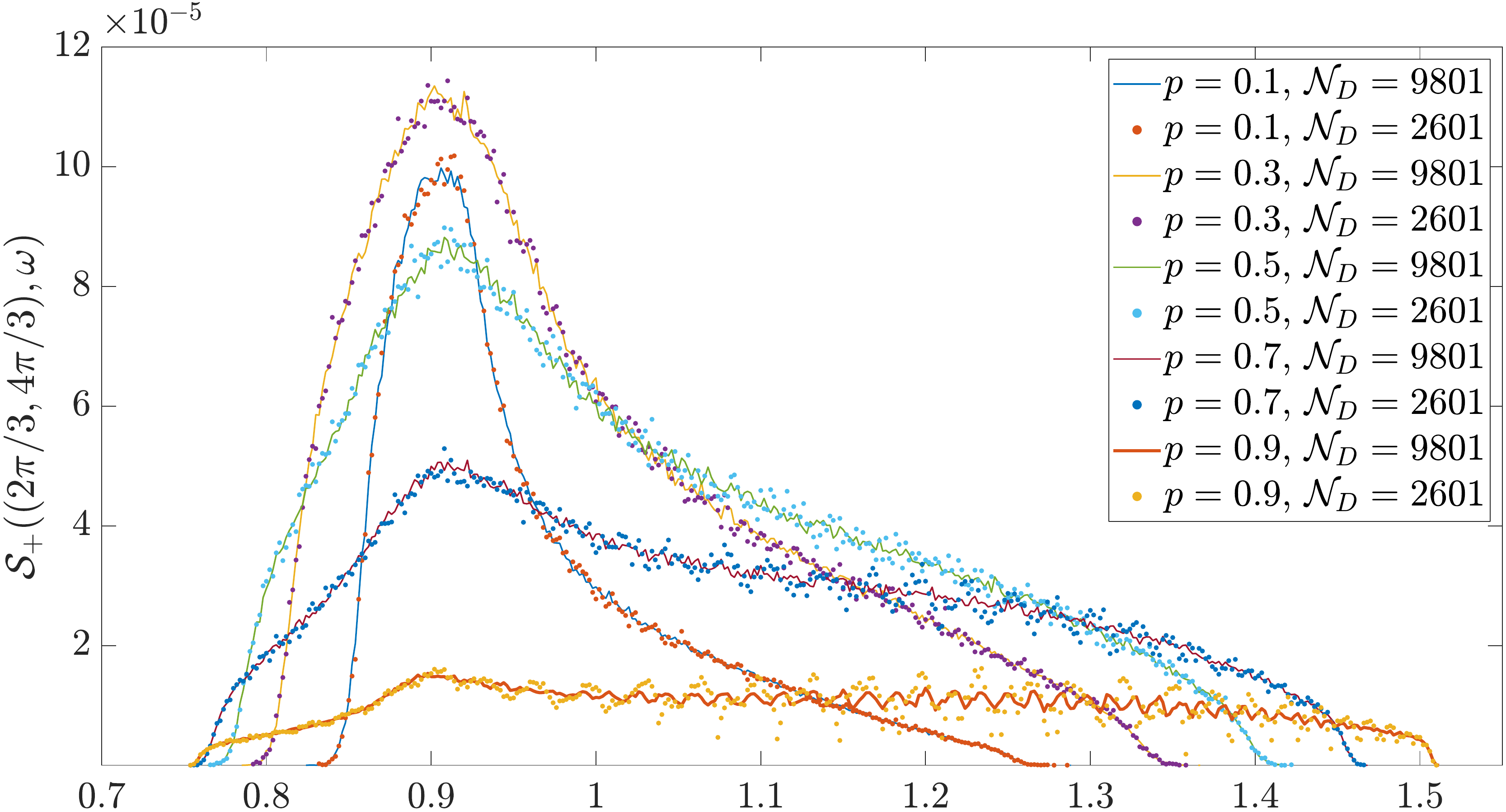}
		\label{fig:Ng2}
	\end{subfigure}
	
	\begin{subfigure}[b]{.9\columnwidth}
		\caption{}		
		\includegraphics[width=1\linewidth]{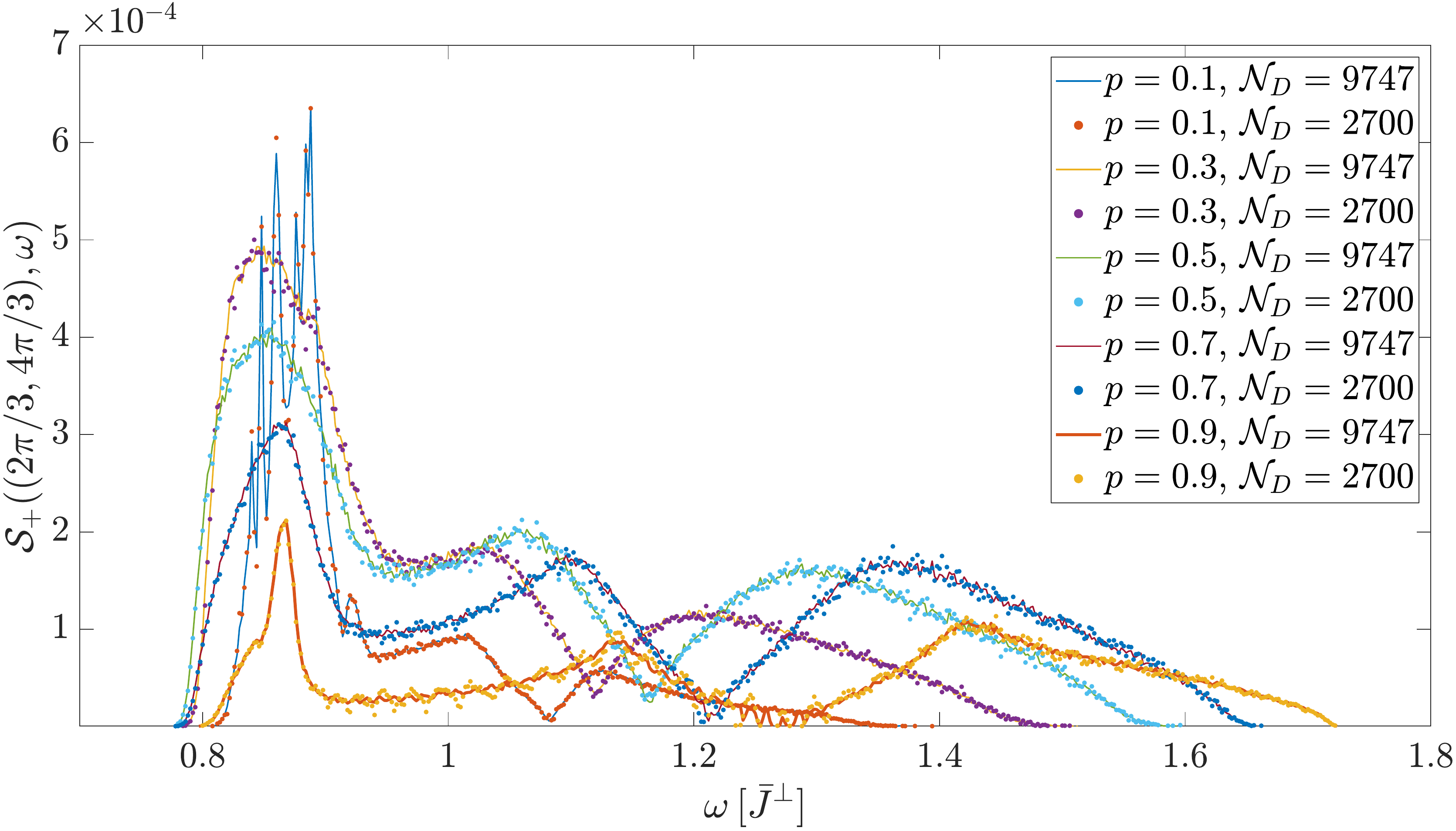}
		\label{fig:Ng2}
	\end{subfigure}
	
	\caption[]{
		The DSF $\mathcal{S}_+(k,\omega)$ and its finite-size scaling behaviour is shown for bimodal inter-dimer disorder, $p=0.1,0.3,0.5,0.7,0.9$ and the corresponding disorder configurations of the main body of the paper in (a) for the square, in (b) for the triangular lattice and in (c) for the kagome lattice. The momentum $\vec{k}$ is the gap momentum in the corresponding lattices, i.e. $\vec{k}=(\pi,\pi)$ for the square, $\vec{k}=(2\pi/3,4\pi/3)$ for the triangular and $\vec{k}=(2\pi/3,4\pi/3)$ for the kagome lattice.}
	\label{fig:LegPlus_FiniteSize}
\end{figure}

For the same system sizes as for the IPR also $\mathcal{S}_-(\vec{k},\omega)$ was compared for intra-dimer disorder in Fig.~\ref{fig:Rung_FiniteSize} and for inter-dimer disorder in Fig.~\ref{fig:LegMinus_FiniteSize}. $\mathcal{S}_+ (\vec{k},\omega)$ was compared in Fig.~\ref{fig:LegPlus_FiniteSize}. One sees that even though for many energies the IPR indicated that the localization length is larger than the system size the agreement of the DSF is still quite good.

\section{DSF}
This last appendix contains further results for the symmetric and anti-symmetric DSF, which are not contained in the main body of the article.

\begin{figure}[H]
	\centering
	\begin{subfigure}[b]{.9\columnwidth}
		\caption{}		
		\includegraphics[width=1\linewidth]{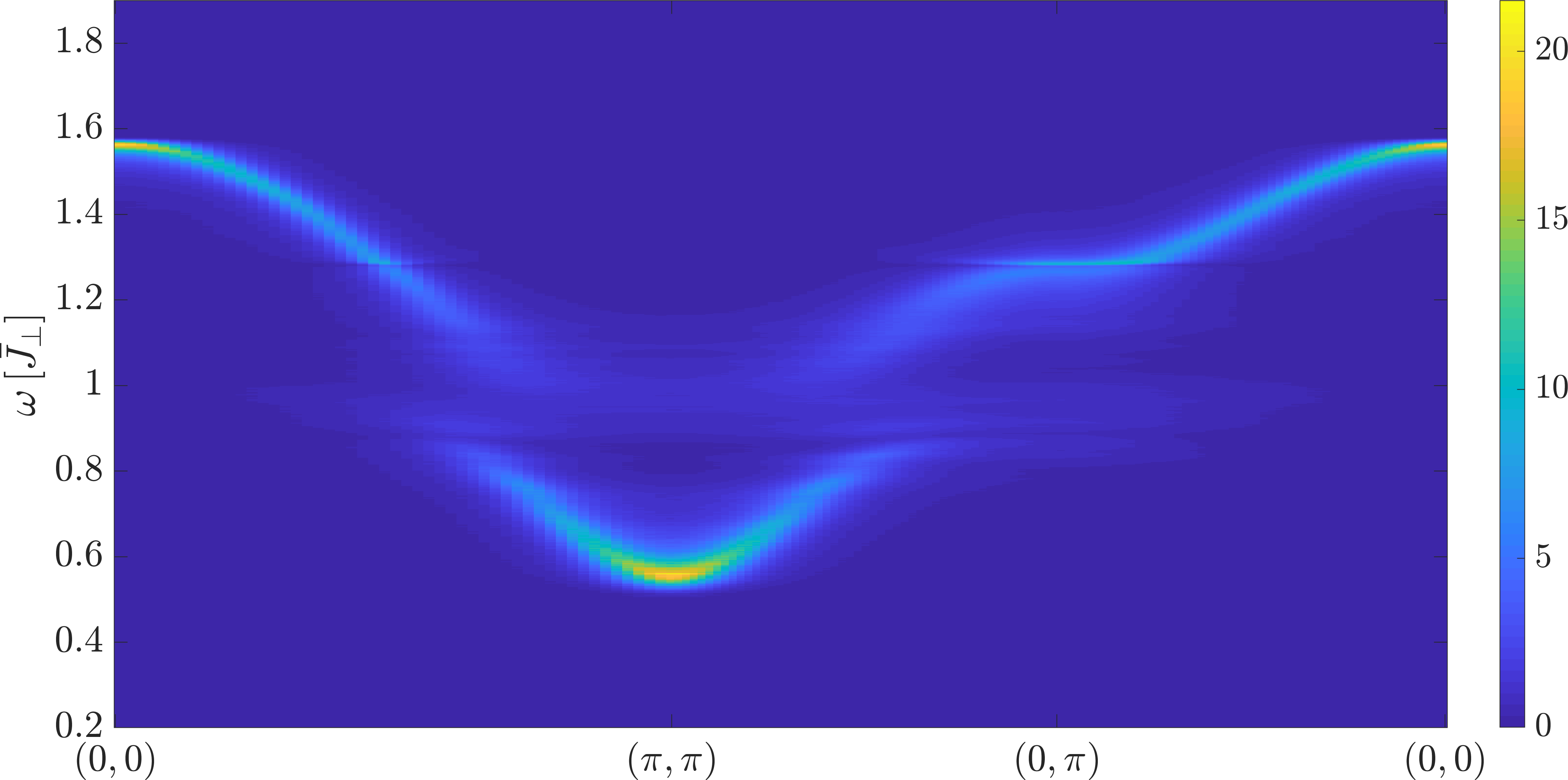}
		\label{fig:Rung_Square_Minus_03} 
	\end{subfigure}
	
	\begin{subfigure}[b]{.9\columnwidth}
		\caption{}		
		\includegraphics[width=1\linewidth]{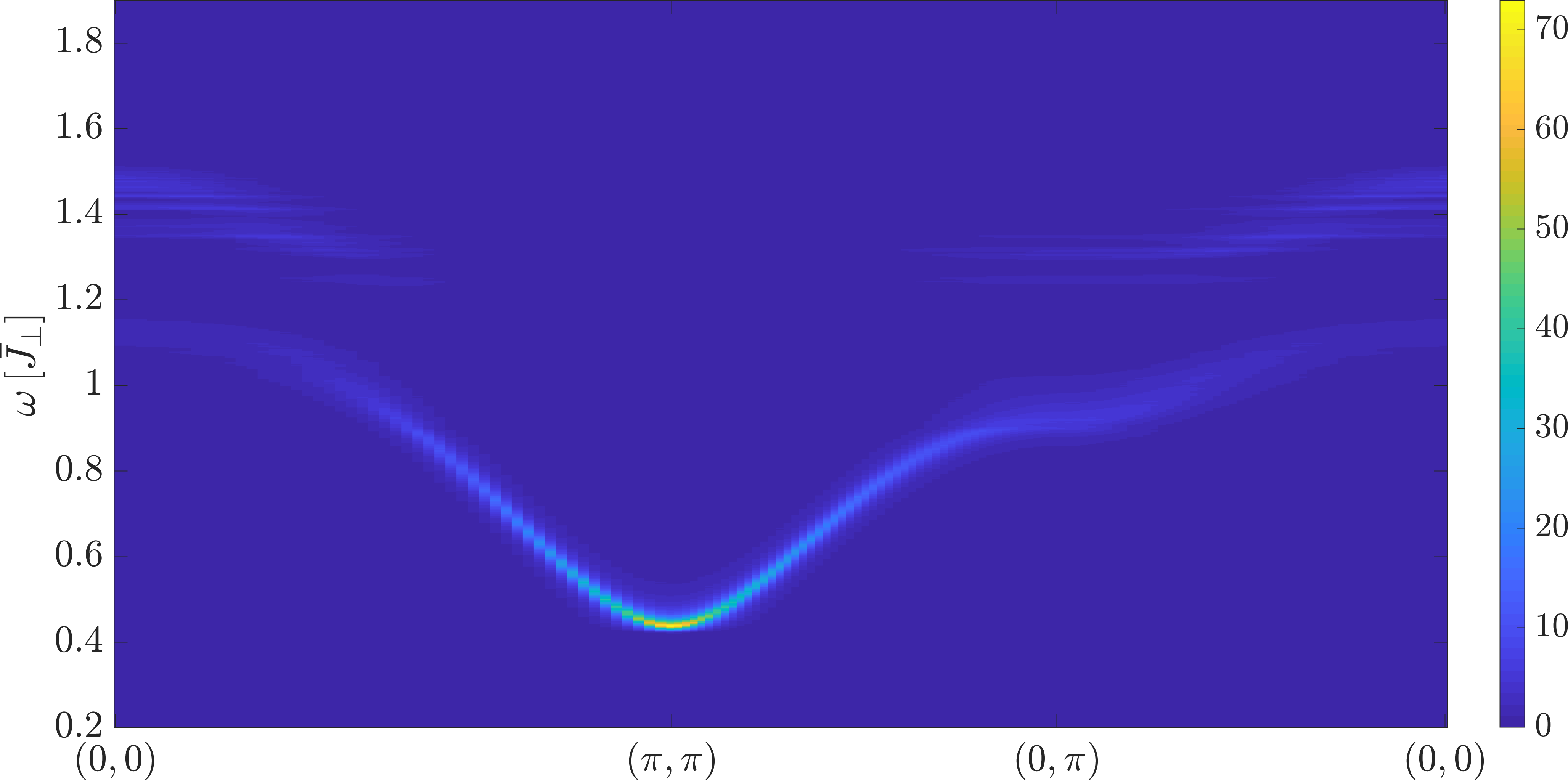}
		\label{fig:Rung_Square_Minus_07}
	\end{subfigure}

\begin{subfigure}[b]{.9\columnwidth}
	\caption{}		
	\includegraphics[width=1\linewidth]{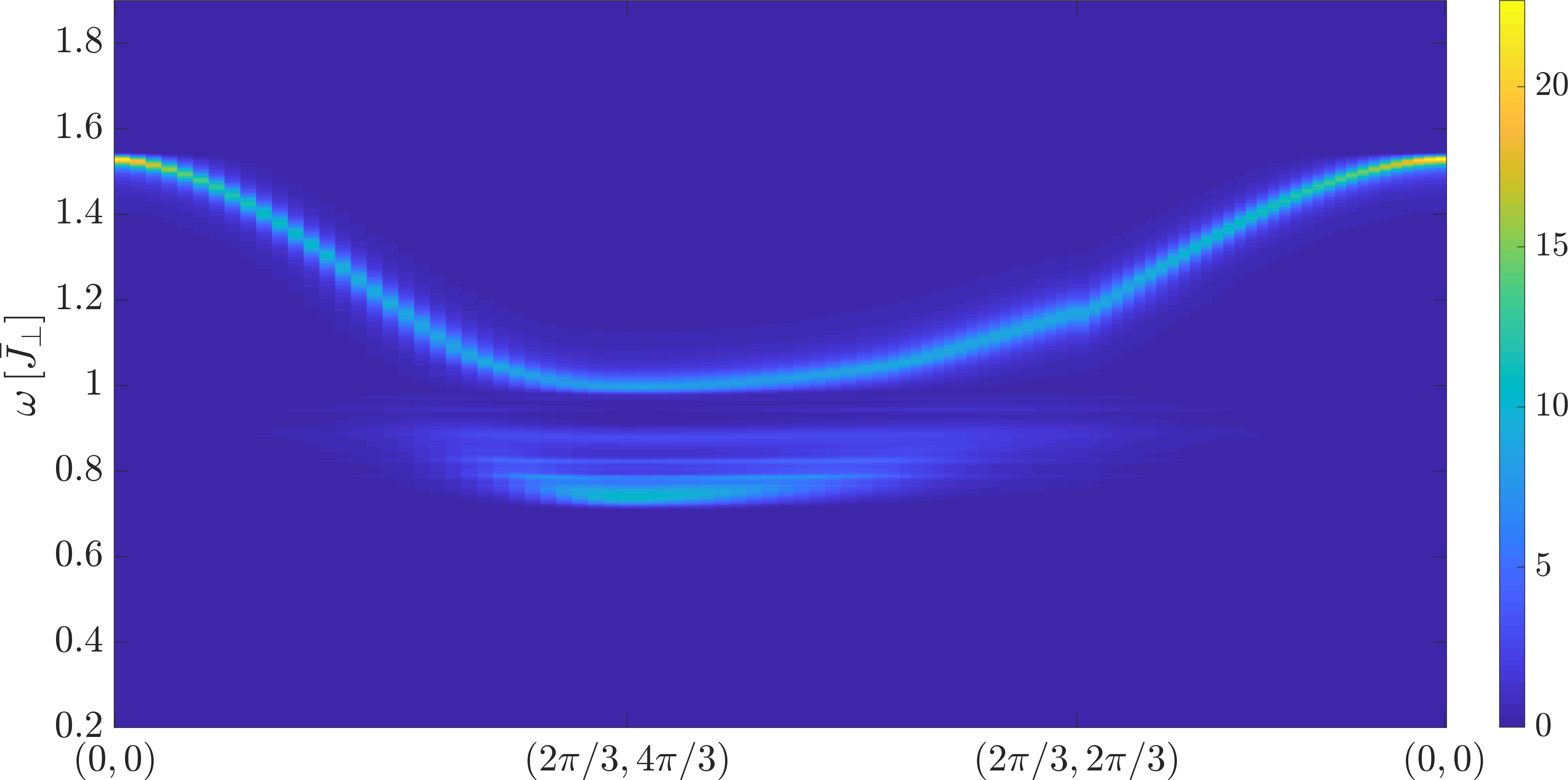}
	\label{fig:Rung_Triangular_Minus_03} 
\end{subfigure}

\begin{subfigure}[b]{.9\columnwidth}
	\caption{}		
	\includegraphics[width=1\linewidth]{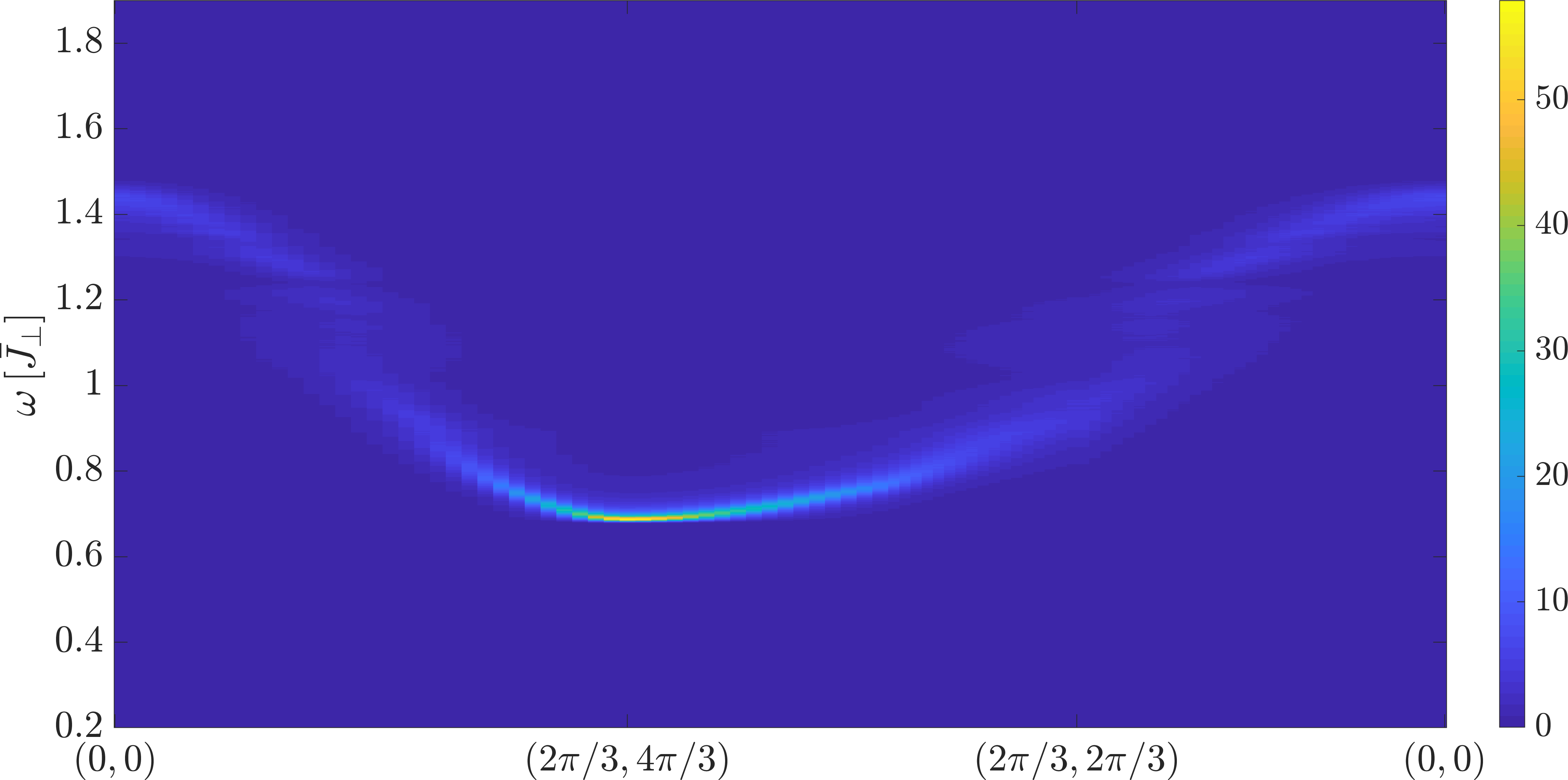}
	\label{fig:Rung_Triangular_Minus_07}
\end{subfigure}

	\caption[]{
		(a) and (b): The DSF $\mathcal{S}_-(k,\omega)$ is shown for bimodal intra-dimer disorder in the square lattice. The disorder is identical to the one in the main body of the paper. In (a) $p=0.3$ and in (b) $p=0.7$.\\
		(c) and (d): The DSF $\mathcal{S}_-(k,\omega)$ is shown for bimodal intra-dimer disorder in the triangular lattice. The disorder is identical to the one in the main body of the paper. In (c) $p=0.3$ and in (d) $p=0.7$.
	}
	\label{fig:Rung_Square_p0307}
\end{figure}

\begin{figure}[H]
	\centering
	\begin{subfigure}[b]{.9\columnwidth}
		\caption{}		
		\includegraphics[width=1\linewidth]{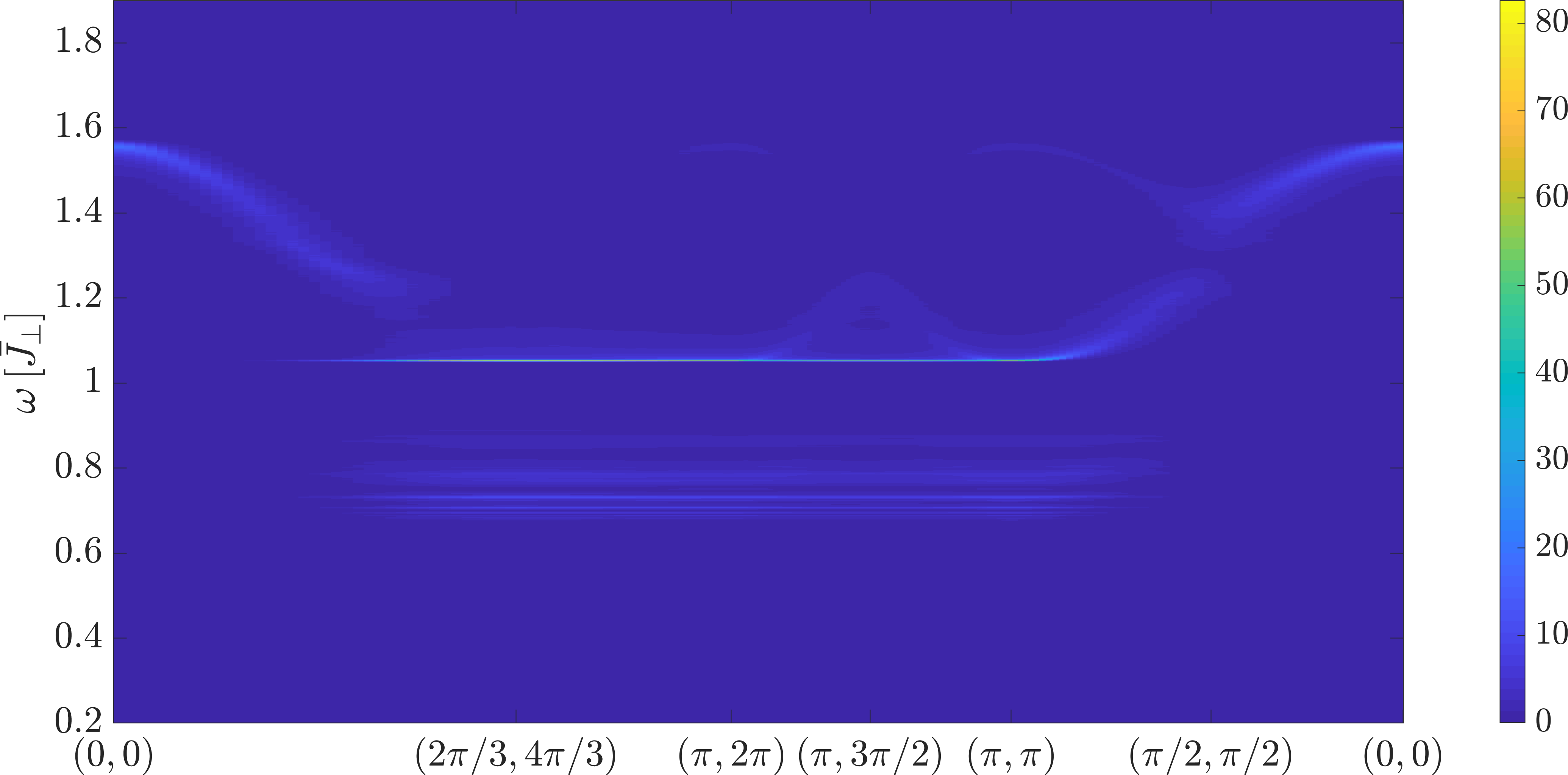}
		\label{fig:Rung_kagome_Minus_03} 
	\end{subfigure}
	
	\begin{subfigure}[b]{.9\columnwidth}
		\caption{}		
		\includegraphics[width=1\linewidth]{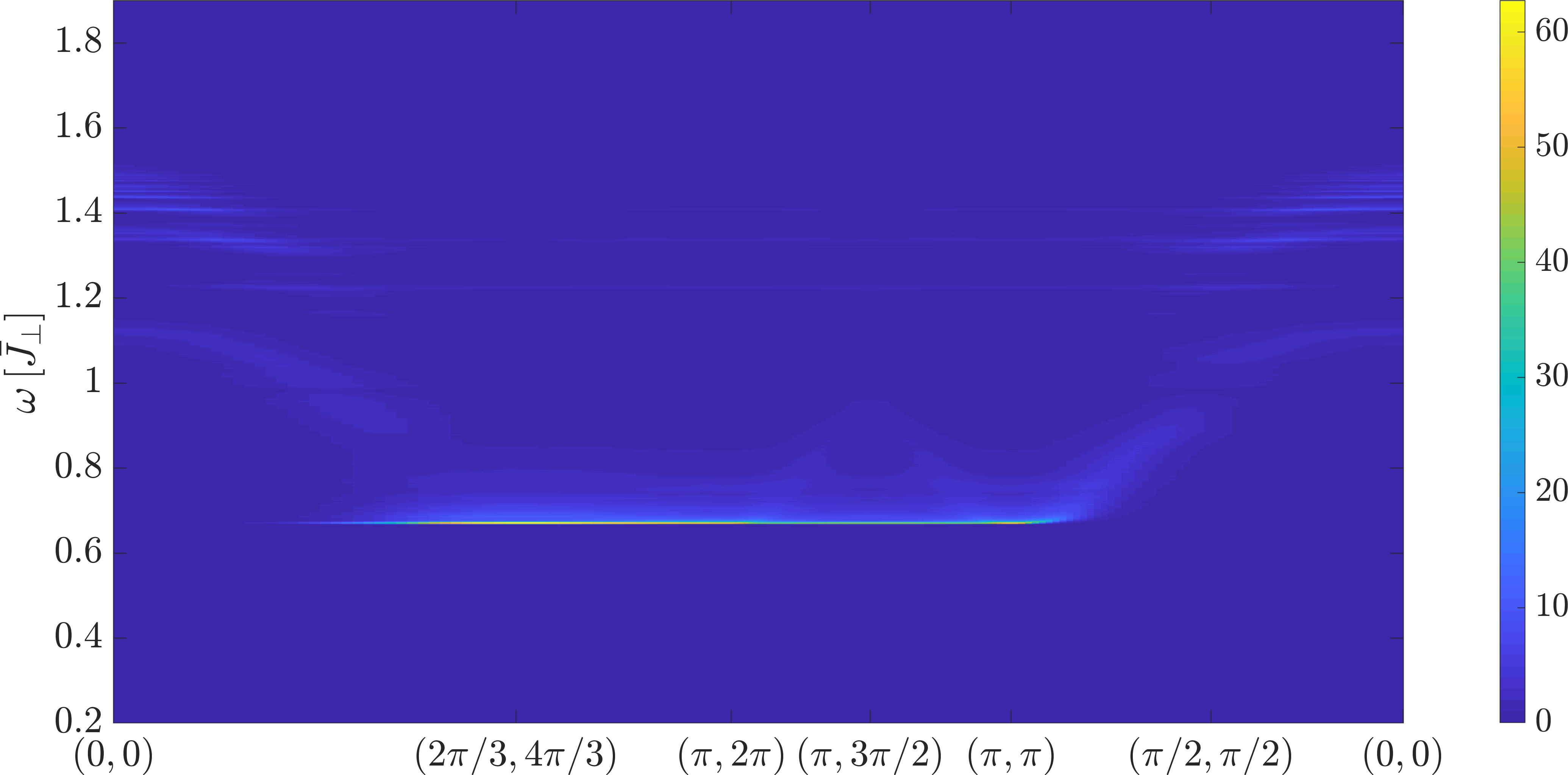}
		\label{fig:Rung_kagome_Minus_07}
	\end{subfigure}

\begin{subfigure}[b]{.9\columnwidth}
	\caption{}		
	\includegraphics[width=1\linewidth]{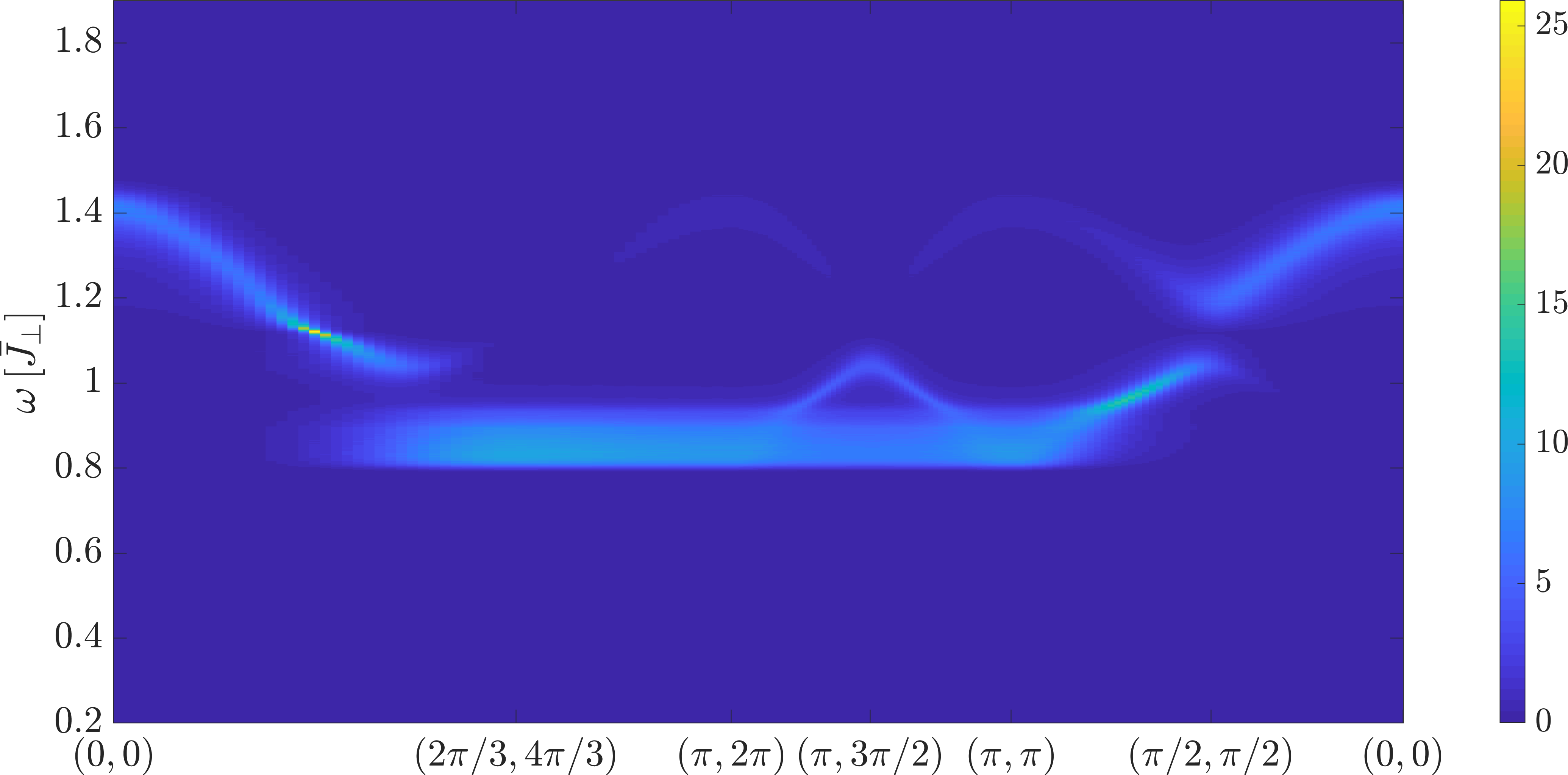}
	\label{fig:Leg_kagome_Minus_03} 
\end{subfigure}

\begin{subfigure}[b]{.9\columnwidth}
	\caption{}		
	\includegraphics[width=1\linewidth]{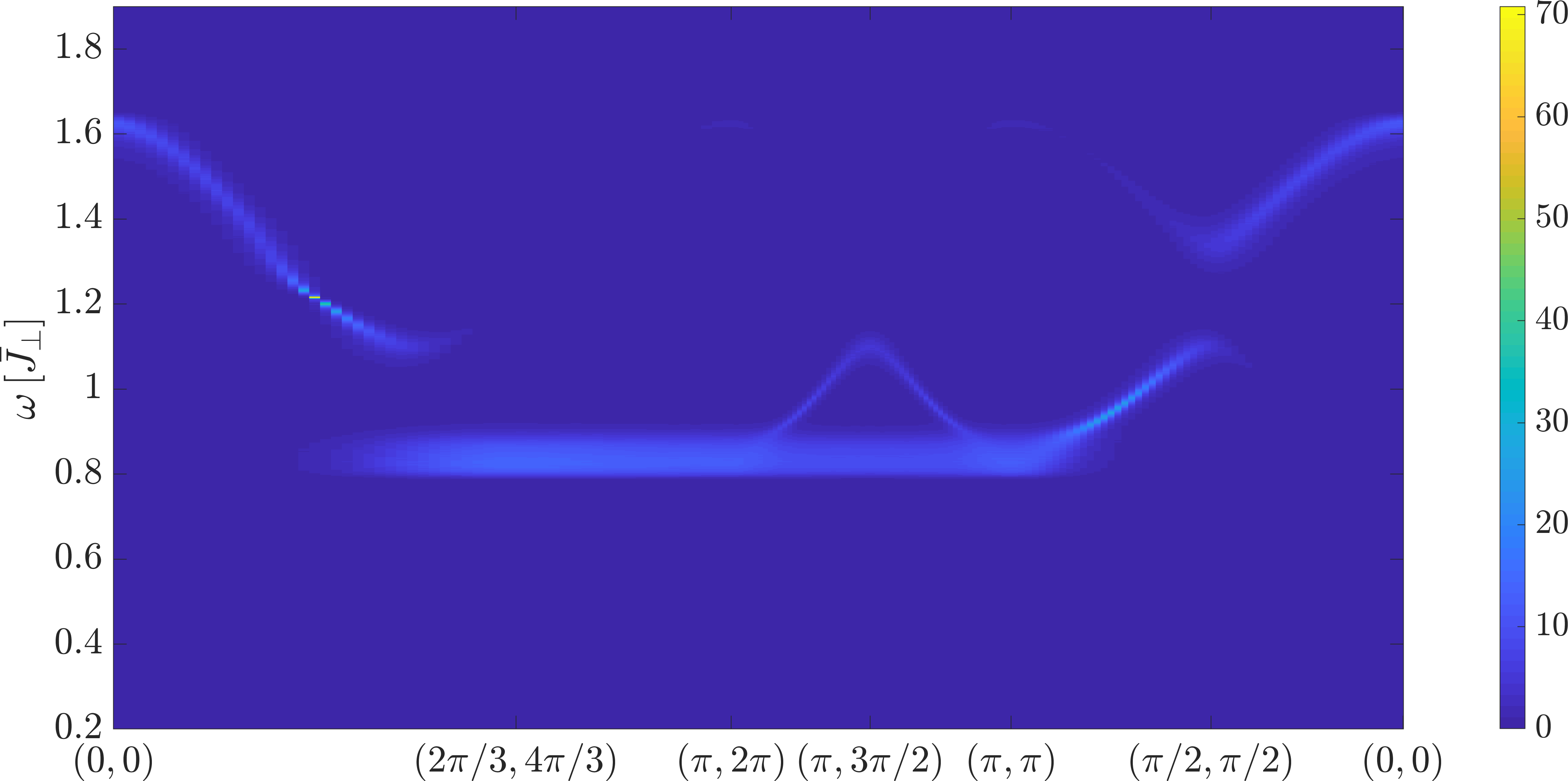}
	\label{fig:Leg_kagome_Minus_07}
\end{subfigure}

	\caption[]{
		(a) and (b): The DSF $\mathcal{S}_-(k,\omega)$ is shown for bimodal intra-dimer disorder in the kagome lattice. The disorder is identical to the one in the main body of the paper. In (a) $p=0.3$ and in (b) $p=0.7$.\\
		(c) and (d): The DSF $\mathcal{S}_-(k,\omega)$ is shown for bimodal inter-dimer disorder in the kagome lattice. The disorder is identical to the one in the main body of the paper. In (c) $p=0.3$ and in (d) $p=0.7$.
	}
	\label{fig:Rung_kagome_p0307}
\end{figure}

\begin{figure}[H]
	\centering
	\begin{subfigure}[b]{.9\columnwidth}
		\caption{}		
		\includegraphics[width=1\linewidth]{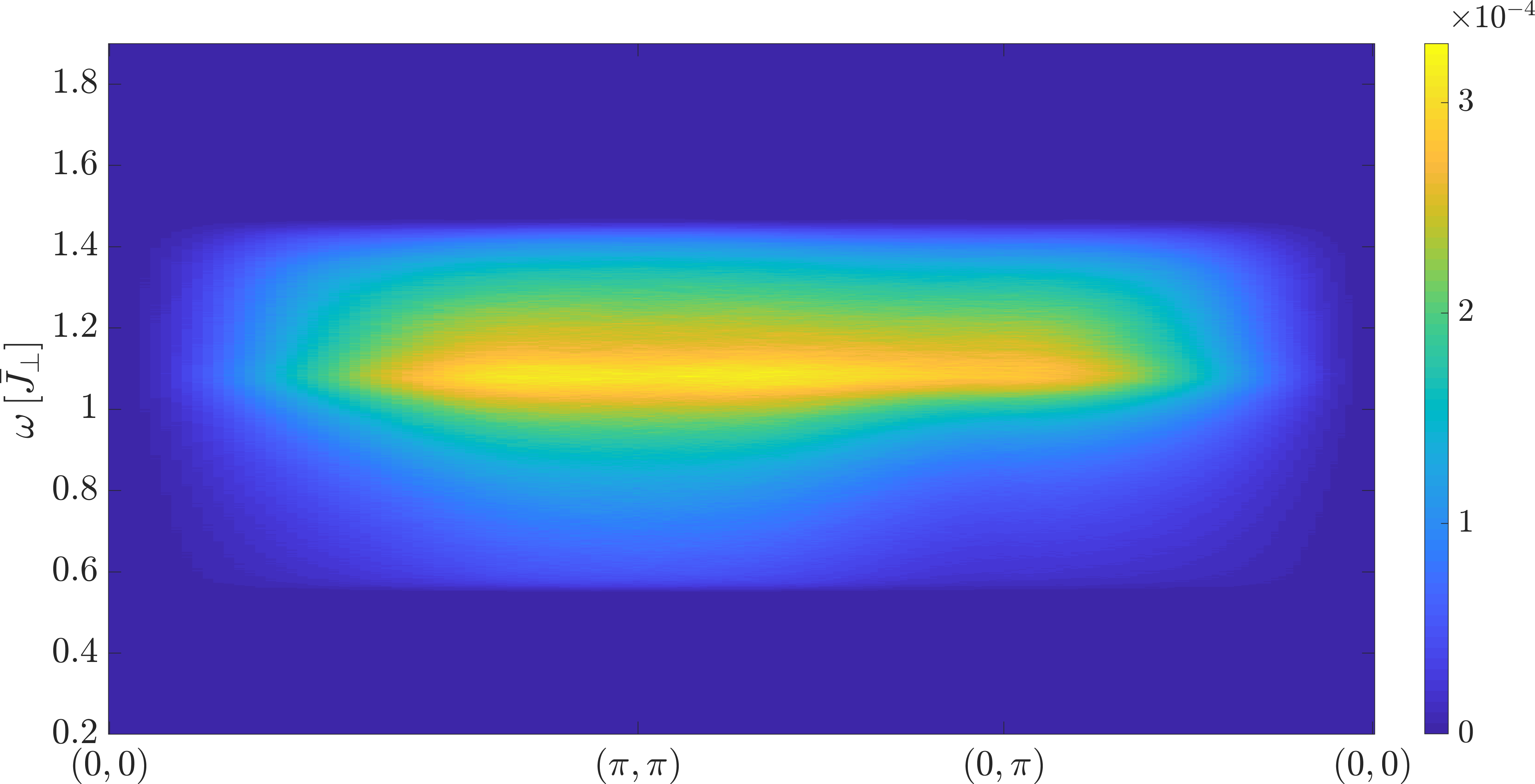}
		\label{fig:Ng1} 
	\end{subfigure}
	
	\begin{subfigure}[b]{.9\columnwidth}
		\caption{}		
		\includegraphics[width=1\linewidth]{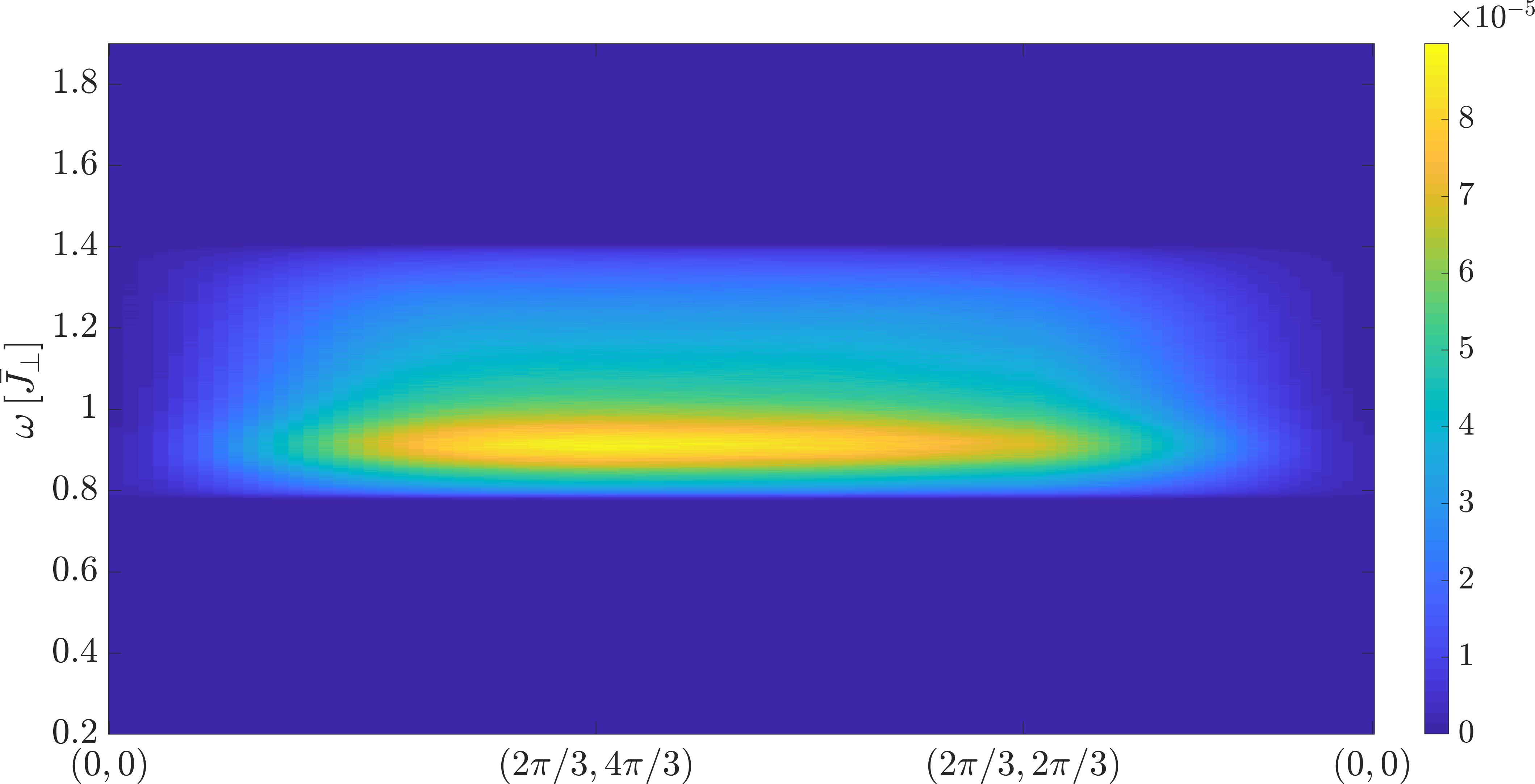}
		\label{fig:Ng2}
	\end{subfigure}
	
	\begin{subfigure}[b]{.9\columnwidth}
		\caption{}		
		\includegraphics[width=1\linewidth]{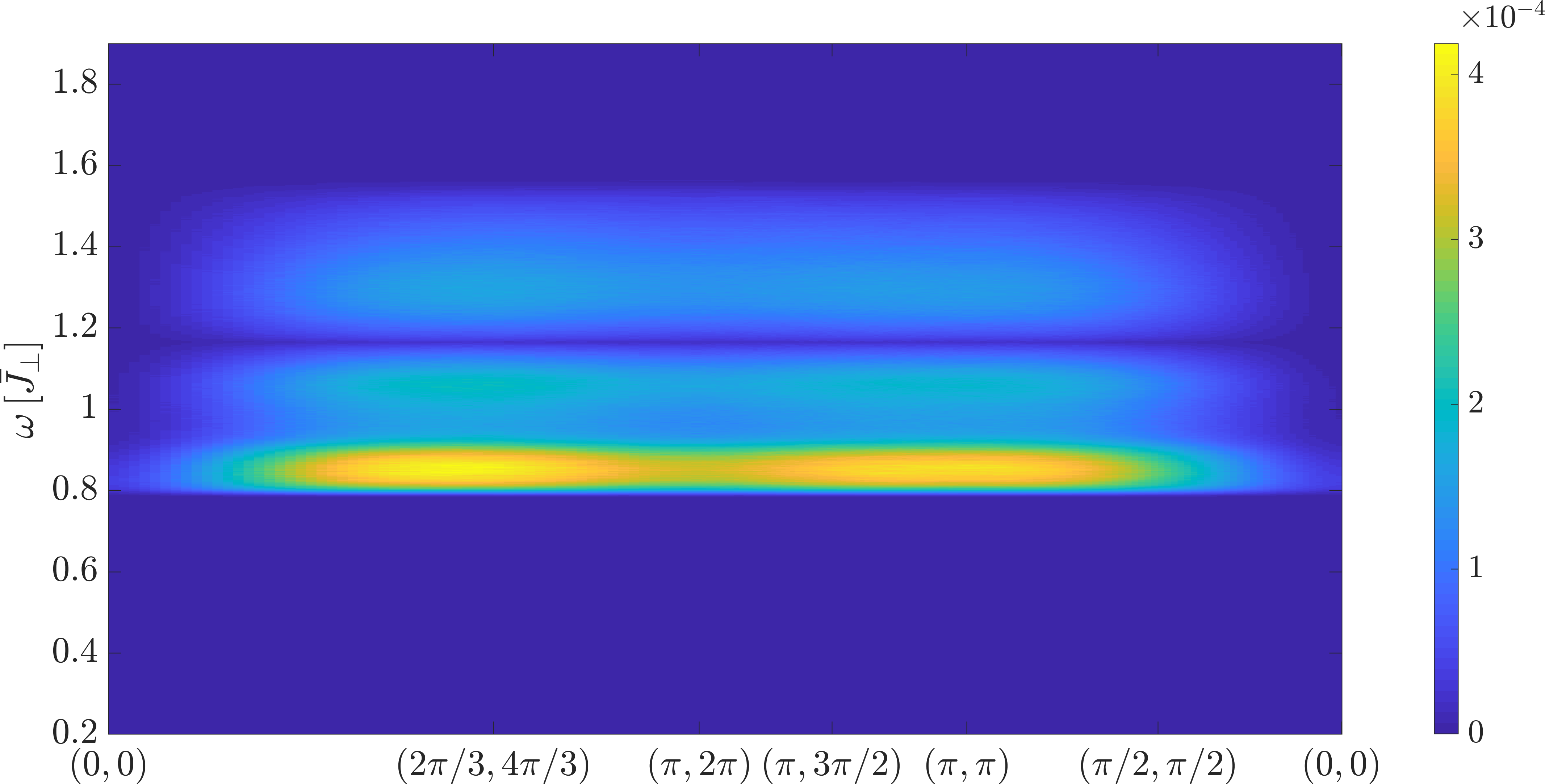}
		\label{fig:Ng2}
	\end{subfigure}

	\caption[]{
		The DSF $\mathcal{S}_+(k,\omega)$ is shown for bimodal inter-dimer disorder, $p=0.5$ and the corresponding disorder configurations of the main body of the paper in (a) for the square, in (b) for the triangular and  in (c) for the kagome lattice.
	}
	\label{fig:Leg_Plus}
\end{figure}

\begin{figure}[H]
	\centering
	\begin{subfigure}[b]{.9\columnwidth}
		\caption{}		
		\includegraphics[width=1\linewidth]{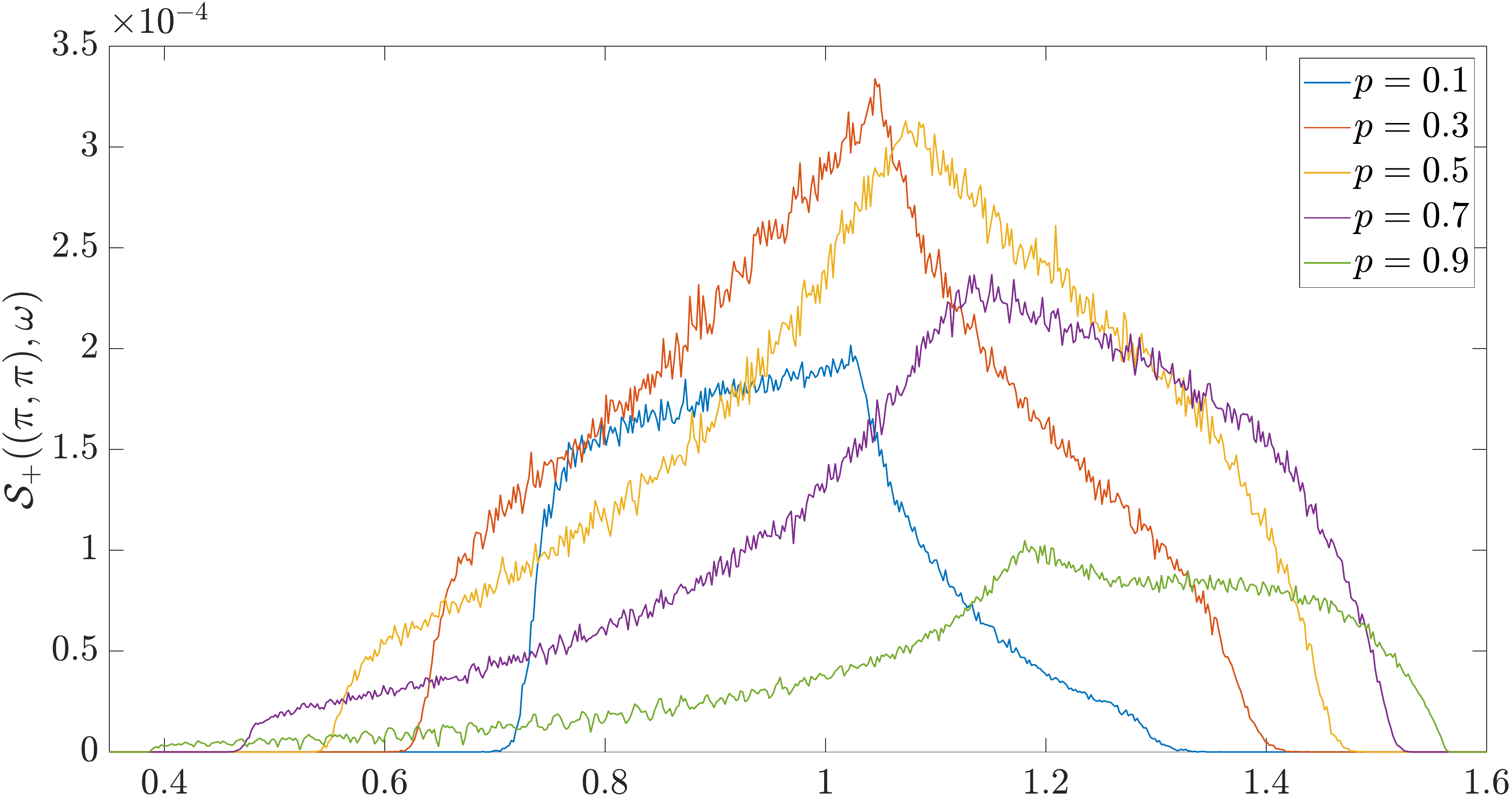}
		\label{fig:Ng1} 
	\end{subfigure}
	
	\begin{subfigure}[b]{.9\columnwidth}
		\caption{}		
		\includegraphics[width=1\linewidth]{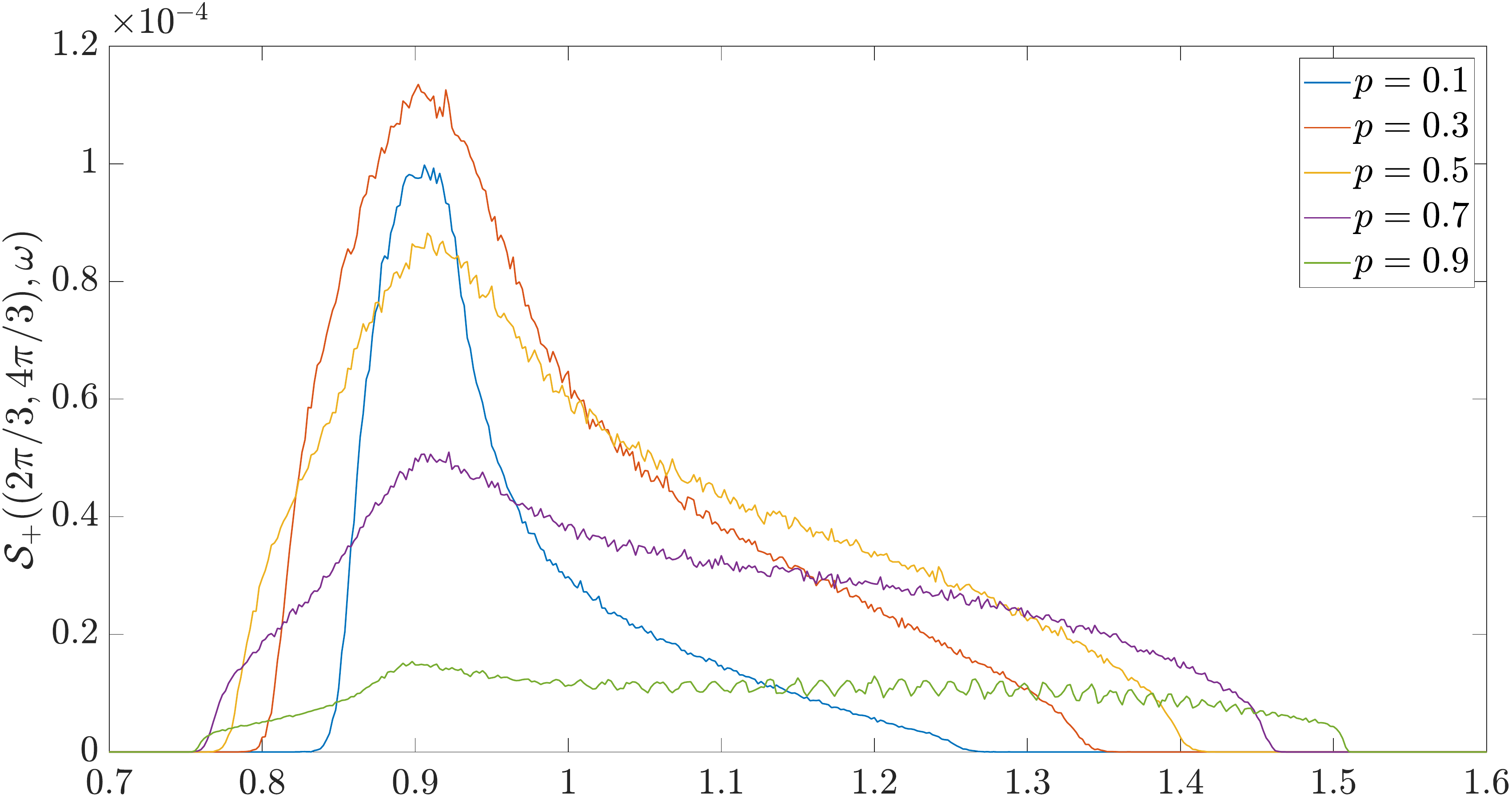}
		\label{fig:Ng2}
	\end{subfigure}
\begin{subfigure}[b]{.9\columnwidth}
	\caption{}	
	\includegraphics[width=1\linewidth]{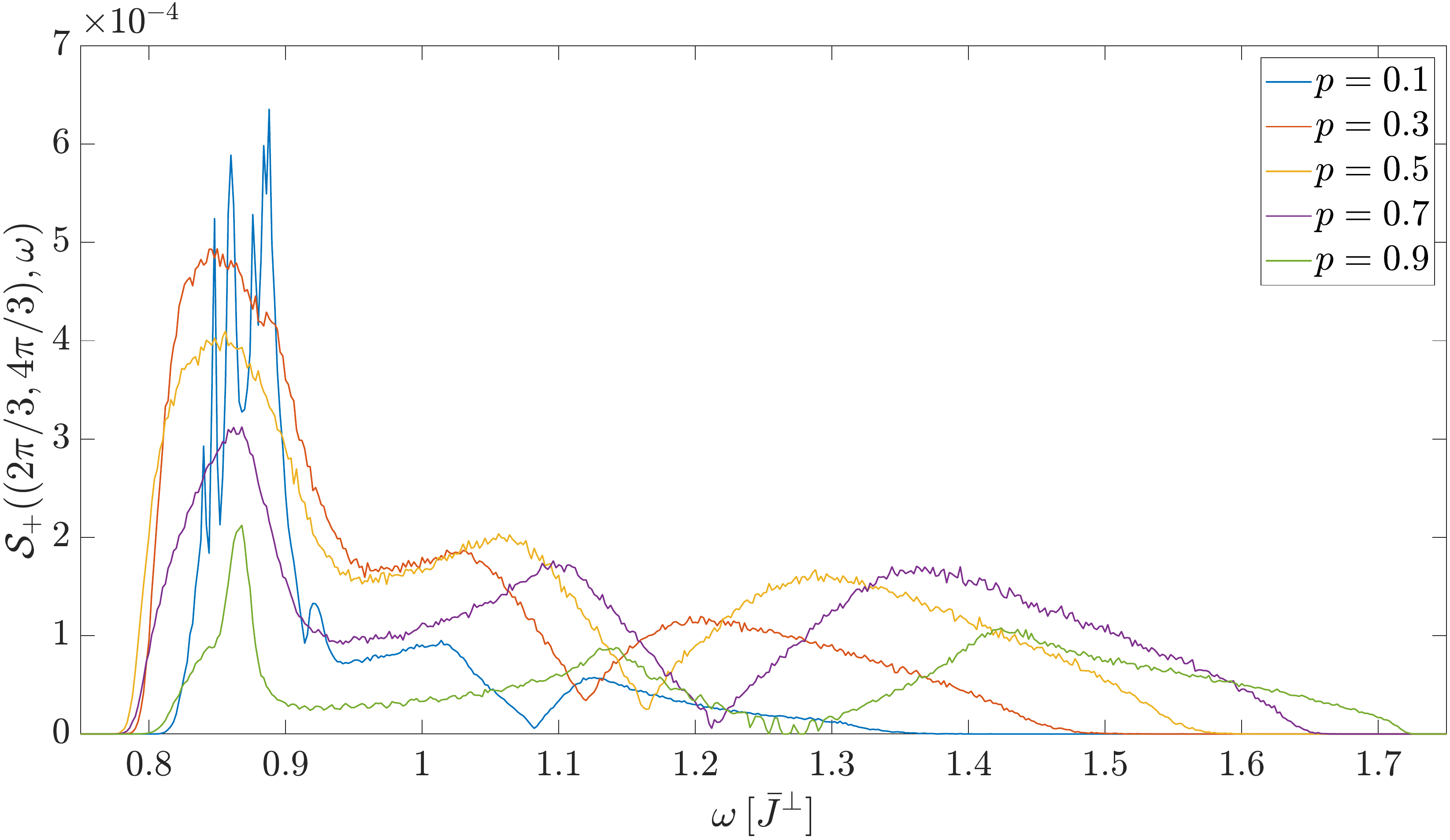}
	\label{fig:Ng2}
\end{subfigure}

	\caption[]{
		The DSF $\mathcal{S}_+(k,\omega)$ is shown for bimodal inter-dimer disorder, $p=0.1,0.3,0.5,0.7,0.9$ and the corresponding disorder configurations of the main body of the paper in (a) for the square, in (b) for the triangular lattice and in (c) for the kagome lattice. The momentum $\vec{k}$ is the gap momentum in the corresponding lattices, i.e. $\vec{k}=(\pi,\pi)$ for the square, $\vec{k}=(2\pi/3,4\pi/3)$ for the triangular and $\vec{k}=(2\pi/3,4\pi/3)$ for the kagome lattice.
	}
	\label{fig:Leg_Plus_k_p}
\end{figure}

\end{appendix}

\end{document}